\documentclass[11pt,letterpaper]{article}
\pdfoutput=1

\usepackage{fancyhdr}
\usepackage{amsmath,amssymb}
\usepackage{graphicx}               
\usepackage{tabularx}
\usepackage{url,cite}
\usepackage[x11names]{xcolor}
   \definecolor{db}{RGB}{0,0,221}
\usepackage[colorlinks=true,urlcolor=blue,linkcolor=blue,citecolor=db]{hyperref}

\usepackage{mathabx}
\usepackage[T1]{fontenc}
\usepackage{textcomp}
\usepackage{pifont}

\usepackage{caption}
\newcommand{\nc}{\newcommand}

\nc{\beq}{\begin{equation}}
\nc{\eeq}{\end{equation}}
\nc{\beqa}{\begin{eqnarray}}
\nc{\eeqa}{\end{eqnarray}}
\nc{\bea}{\begin{eqnarray}}
\nc{\eea}{\end{eqnarray}}
\nc{\ra}{\rightarrow}
\nc{\lsim}{\begin{array}{c}\,\sim\vspace{-21pt}\\< \end{array}}
\nc{\gsim}{\begin{array}{c}\sim\vspace{-21pt}\\> \end{array}}
\nc{\Tr}{{\rm Tr}}
\nc{\slsh}{\slash\hspace*{-0.22cm}}

\def\be{\begin{equation}}
\def\ee{\end{equation}}
\def\bea{\begin{eqnarray}}
\def\eea{\end{eqnarray}}
\def\bit{\begin{itemize}}
\def\eit{\end{itemize}}

\def\to{\rightarrow}

\newcommand{\mini}[2]{\begin{minipage}{#1}\centering #2\end{minipage}}

\title{
\vspace*{-2.3cm}
\begin{flushright}
\normalsize{
  }
\end{flushright}
\vspace{1.5cm}
\Large
\textbf{
Galactic Neutrinos in the TeV to PeV Range
}
\vspace*{1.0cm}
}

\author{\normalsize \bf Markus Ahlers,$^{a,b}$ Yang Bai,$^{a}$ Vernon Barger,$^{a}$ and Ran Lu$^{a}$
\vspace{5mm}
\\
$^{a}$ \normalsize\emph{Department of Physics, University of Wisconsin-Madison, Madison, WI 53706, USA}  \vspace{1mm} \\
$^{b}$ \normalsize\emph{Wisconsin IceCube Particle Astrophysics Center, Madison, WI 53703, USA}
}

\date{}

\begin{document}
\setcounter{page}{0}
\maketitle

\vspace*{1cm}
\begin{abstract}
\normalsize{
We study the contribution of Galactic sources to the flux of astrophysical neutrinos recently observed by the IceCube Collaboration. We show that in the simplest model of homogeneous and isotropic cosmic ray diffusion in the Milky Way the Galactic diffuse neutrino emission consistent with $\gamma$-ray ({\it Fermi}-LAT) and cosmic ray data (KASCADE, KASCADE-Grande and CREAM) is expected to account for only 4\%$-$8\% of the IceCube flux above 60~TeV. Direct neutrino emission from cosmic ray-gas ($pp$) interactions in the sources would require an unusually large average opacity above 0.01. On the other hand, we find that the IceCube events already probe Galactic neutrino scenarios via the distribution of event arrival directions. Based on the latter, we show that most Galactic scenarios can only have a limited contribution to the astrophysical signal: diffuse Galactic emission ($\lesssim50$\%), quasi-diffuse emission of neutrino sources ($\lesssim65$\%), extended diffuse emission from the {\it Fermi Bubbles} ($\lesssim25$\%) or unidentified TeV $\gamma$-ray sources ($\lesssim25$\%). The arguments discussed here leave, at present, dark matter decay unconstrained.
}
\end{abstract}

\thispagestyle{empty}
\newpage

\setcounter{page}{1}

\baselineskip16pt

\section{Introduction}\label{sec:intro}

Galactic cosmic rays (CRs) interact with gas and radiation during the acceleration in their sources or during diffusion in the Galactic medium. Neutral pions produced in this interaction promptly decay into pairs of $\gamma$-rays and are one component of the observed $\gamma$-ray emission of the Galaxy. A significant component of the Galactic diffuse $\gamma$-ray emission in the 100~MeV to 100~GeV range observed with EGRET~\cite{Hunter:1997} and {\it Fermi}-LAT~\cite{FermiLAT:2012aa} can be attributed to CR interactions with gas and is expected to become increasingly important towards higher energies. Direct evidence of hadronic interactions in Galactic sources could be inferred via the observation of low-energy cutoffs in $\gamma$-ray emission spectra related to the production threshold of neutral pions~\cite{Ackermann:2013wqa}.

The production of pions in the same CR interactions and the subsequent decay via $\pi^+\to\mu^+\nu_\mu$ followed by $\mu^+ \to e^+\nu_e\bar\nu_\mu$ (and the charge-conjugate process) predicts a flux of high-energy neutrinos. The mean energy of a secondary neutrino produced in the production and decay of charged pions is typically of the order of 5\% of the initial CR nucleon energy. Galactic CRs are believed to dominate the observed CR spectrum at least up to the CR {\it knee}, a spectral break at about $3-4$~PeV.  Therefore, for proton-dominated CR spectra the Galactic emission of $\gamma$-rays and neutrinos is expected to extend to at least an energy of a few 100~TeV and possibly even higher. 

\begin{figure*}[p]\centering
\includegraphics[width=0.8\textwidth]{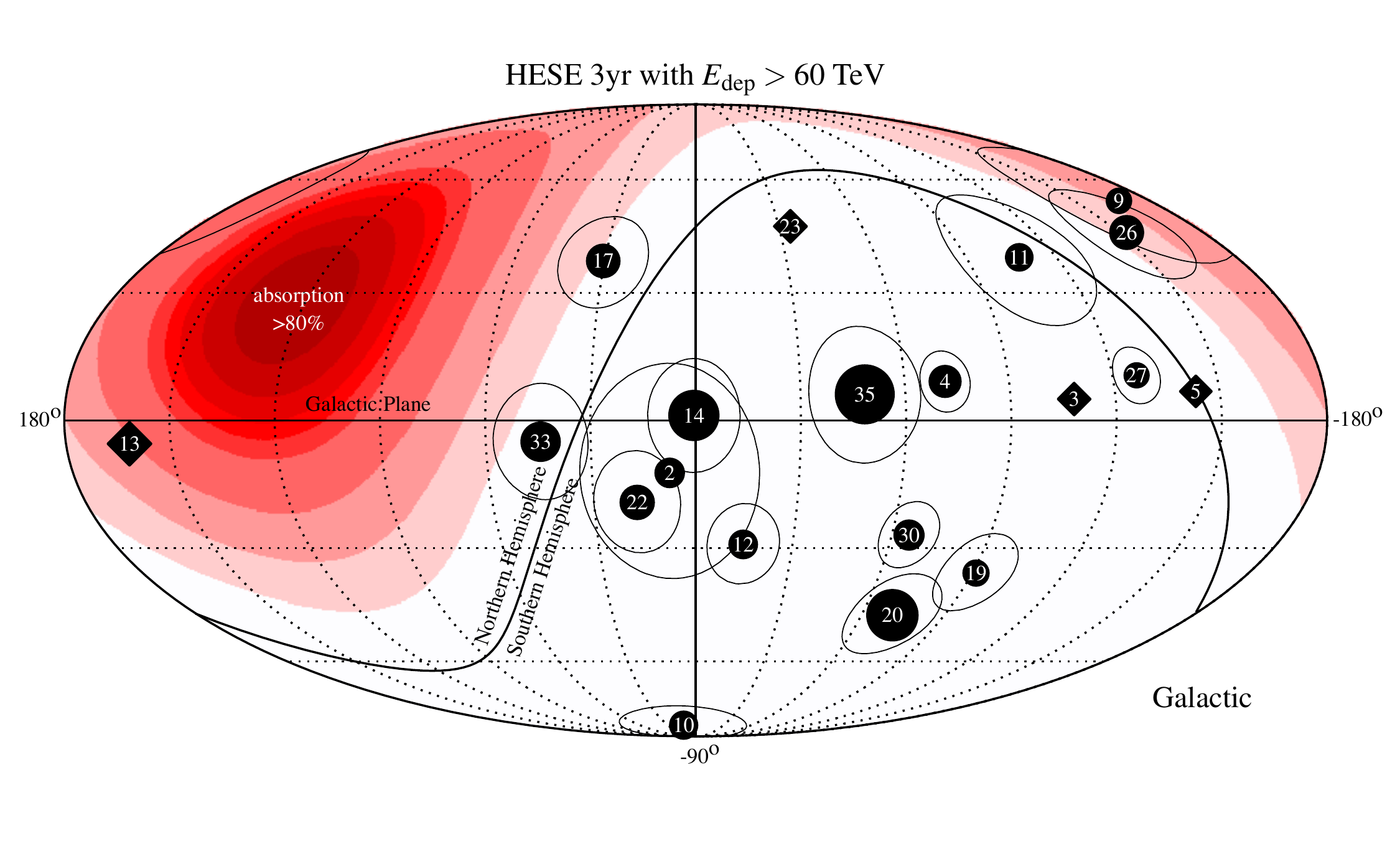}
\vspace{-0.5cm}
\caption{Mollweide projection of the arrival directions of IceCube events from Ref.~\cite{Aartsen:2014gkd} with deposited energies $E_{\rm dep}>60$~TeV in Galactic coordinates. The events are classified as tracks (diamonds) and cascades (filled circles). We use the same event numbers as in Ref.~\cite{Aartsen:2014gkd}. The relative detected energies of the events are indicated by the areas of the symbols. The thin lines around the arrival direction of the cascade events indicate the systematic uncertainty of the reconstruction. The red shaded region shows the minimal ($E_\nu = 60$~TeV) absorption of the neutrino flux due to scattering in the Earth in $\log_{10}$-steps of $0.1$.}\label{fig:HESE3yr}
\vspace{0.5cm}
\includegraphics[width=0.48\textwidth]{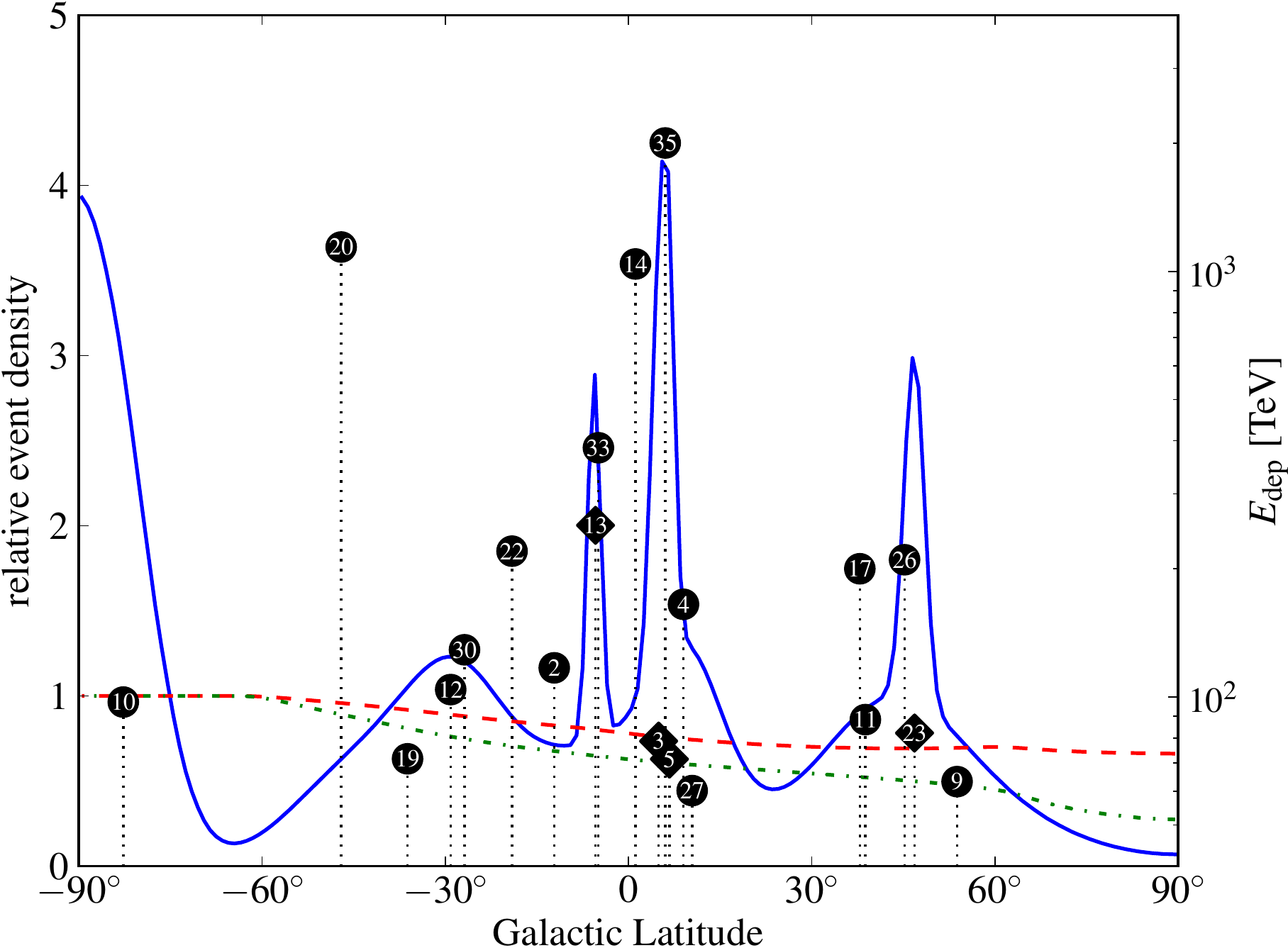}\hfill
\includegraphics[width=0.48\textwidth]{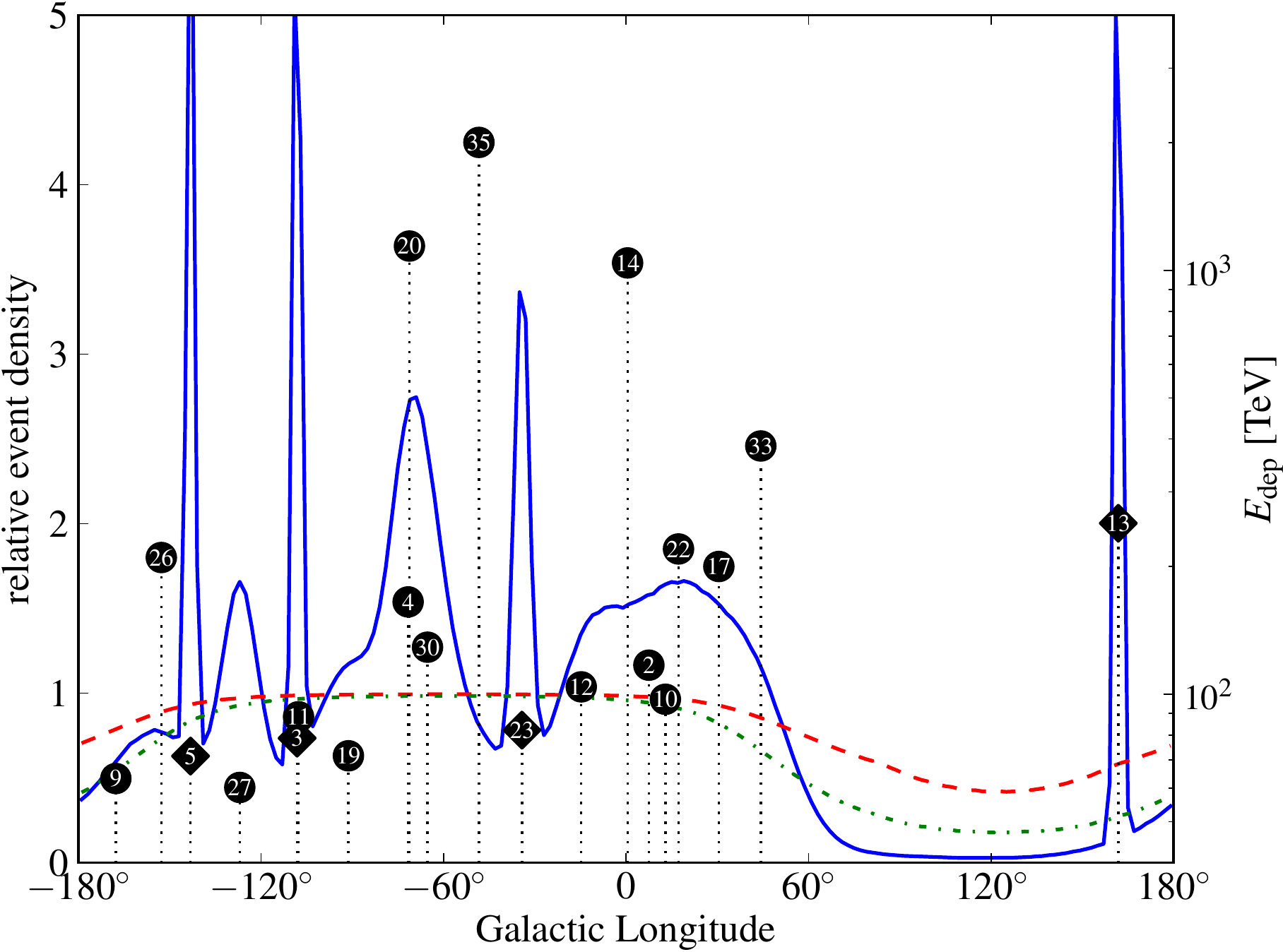}
\caption{The Galactic latitude (left) and longitude (right) distribution of the HESE 3yr events with $E_{\rm dep}>60$~TeV. The blue solid lines show projections of the sum of probability distribution of event arrival directions normalized to an isotropic distribution. The red dashed and green dot-dashed lines show the expected distribution of an isotropic flux of 60~TeV and 1~PeV neutrinos, respectively, taking into account Earth absorption effects. The vertical positions of diamonds and circles indicate the deposited energy (right axis) of tracks and cascades, respectively. The dotted vertical lines indicate the central fit of the event location. 
\label{fig:HESE_lat_long} }
\end{figure*}

Interestingly, the IceCube Collaboration has recently identified a flux of high-energy neutrinos in the TeV-PeV energy range~\cite{Aartsen:2013jdh,Aartsen:2014gkd}. The analysis is based on high-energy starting events (HESE), {\it i.e.}~events with neutrino interaction vertex inside the detector, as opposed to signals ranging into the detector. Figure \ref{fig:HESE3yr} shows the arrival direction of a subset of HESE events with deposited energy of $E_{\rm dep}>60$~TeV in Galactic coordinates. This high-energy sub-sample shows a significant contribution of astrophysical neutrinos at the $5.7\sigma$ level. The best-fit isotropic $E^{-2}$ power-law flux is at the level of 
$E_\nu^2{\rm d}N/{\rm d}E_\nu = (0.95 \pm 0.3) \times 10^{-8}\,{\rm GeV}\,{\rm cm}^{-2}\,{\rm s}^{-1}\,{\rm sr}^{-1}$ per flavor. 

The topologies of the HESE events are classified in terms of {\it tracks} and {\it cascades}, depending on whether the neutrino interaction produced a muon track inside the detector or just a nearly spherical emission pattern at its interaction vertex, respectively. In Fig.~\ref{fig:HESE3yr} these different event topologies are shown as diamonds and filled circles, respectively. The area of the symbols scales with the deposited energy of the event. The most energetic events are three PeV cascades (14, 20 \& 35). At these high energies, absorption of neutrinos in the Northern Hemisphere due to charged and neutral current interactions as they pass through the Earth becomes relevant. As an illustration, the red-shaded area shows the minimal Earth absorption of neutrinos in the sample in $\log_{10}$-steps of $0.1$.

The IceCube Collaboration has analyzed a possible association of HESE events with known Galactic point-sources and a diffuse neutrino emission along the Galactic Plane; Galactic emission could not be significantly established~\cite{Aartsen:2014gkd}. On the other hand, an isotropic distribution of event arrival directions is expected for an extragalactic source population. As an illustration, the two plots of Fig.~\ref{fig:HESE_lat_long} show the HESE event distribution corresponding to events already shown in Fig.~\ref{fig:HESE3yr}, but now projected onto Galactic latitude (left) and longitude (right). In this plot we indicate the deposited energy via the height of the symbols, corresponding to the scale of the right axis. A strong emission from {\it e.g.}~the Galactic Plane could be visible via an event cluster in the center of the latitude plot. However, the presently limited event statistics and the systematic uncertainties of the neutrino arrival directions make an identification of this Galactic structure less straightforward. In other words, the absence of a strong correlation of neutrino events with Galactic neutrino emission templates does not necessarily imply that a Galactic origin is already ruled out. 
 
Indeed, some authors have suggested that a significant contribution of this flux could have a Galactic origin (see Ref.~\cite{Anchordoqui:2013dnh} for a recent review). Possible sources are unidentified Galactic PeV sources~\cite{Fox:2013oza,Gonzalez-Garcia:2013iha} or microquasars~\cite{Anchordoqui:2014rca}, pulsar wind nebulae~\cite{Padovani:2014bha}, extended Galactic structures like the {\it Fermi Bubbles}~\cite{Razzaque:2013uoa,Ahlers:2013xia,Lunardini:2013gva,Lunardini:2015laa}, the Galactic Halo~\cite{Taylor:2014hya} or Sagittarius A$^*$~\cite{Bai:2014kba}. A possible association with a hard diffuse Galactic $\gamma$-ray emission~\cite{Neronov:2013lza,Guo:2014laa} has also been investigated as well as a contribution of neutrino emission from decaying heavy dark matter~\cite{Feldstein:2013kka,Esmaili:2013gha,Bai:2013nga,Bhattacharya:2014vwa,Esmaili:2014rma,Cherry:2014xra,Murase:2015gea}.  A guaranteed flux of Galactic neutrinos is provided by the diffuse emission of Galactic CRs~\cite{Stecker:1978ah,Domokos:1991tt,Berezinsky:1992wr,Bertsch:1993,Ingelman:1996md,Evoli:2007iy,Ahlers:2013xia,Joshi:2013aua,Kachelriess:2014oma}. 

In this study we will discuss the expected high-energy emission of Galactic neutrinos with a special emphasize on its morphology. We provide estimates of the diffuse neutrino emission of the Galaxy in the TeV to PeV energy range utilizing the {\tt GALPROP} propagation code~\cite{Strong:1998pw}. Our results are consistent with the hadronic sub-TeV $\gamma$-ray emission inferred by the {\it Fermi} Collaboration~\cite{FermiLAT:2012aa}. We will estimate the uncertainties on the overall flux from uncertainties of the CR spectrum in the {\it knee} region. The inferred emission rate density of Galactic CR sources can then be utilized to estimate the direct emission of $\gamma$-rays and neutrinos, which can be parametrized by the opacity of the sources with respect to CR-gas interactions. We then try to estimate via an unbinned maximum likelihood test the sensitivity and discovery potential of the IceCube observation for Galactic emission.

Our paper is organized as follows. We first calculate the Galactic diffuse neutrino flux in section~\ref{sec:diffuse}, based on the {\tt GALPROP} propagation code and with both power-law and exponential function fit to the CR spectra. In section~\ref{sec:source}, we discuss the fluxes and morphologies of point-like and extended sources. We then perform an anisotropy test of Galactic emission and calculate the sensitivity and 90\% upper limit on the Galactic fraction for different neutrino source candidates in section~\ref{sec:morphology}. Finally, we conclude in section~\ref{sec:conclusion}. Supplementary material for our study is provided in the appendices. In Appendix~\ref{app1}, we compare our Galactic $\gamma$-ray distributions with results of the {\it Fermi} Collaboration. We discuss the dark matter decay spectra in Appendix~\ref{appDM}, the background and signal samplings in Appendix~\ref{app2} and the likelihood function in Appendix~\ref{app3}. In Appendix~\ref{app4} we discuss the extragalactic neutrino emission associated with other galaxies similar to the Milky Way.
 
\section{Galactic Diffuse Neutrino Flux}\label{sec:diffuse}

{\it A guaranteed} component of the diffuse $\gamma$-ray emission of the Galaxy is due to CR interactions with the interstellar medium. Hadronic interactions with gas such as $pp \rightarrow pp + \pi^0$ produce neutral pions that decay into two $\gamma$-rays. The associated process $pp \rightarrow pn + \pi^+$ can generate charged pions, which then decay into neutrinos.
Therefore, the observation of diffuse Galactic $\gamma$-ray emission in combination with the local measurement of CRs determines the flux and morphology of diffuse Galactic neutrino emission~\cite{Stecker:1978ah,Domokos:1991tt,Berezinsky:1992wr,Bertsch:1993,Ingelman:1996md,Evoli:2007iy,Ahlers:2013xia,Joshi:2013aua,Kachelriess:2014oma}. 

\begin{figure}[t]\centering
\includegraphics[width=0.48\textwidth,clip=true,viewport= 10 0 540 400]{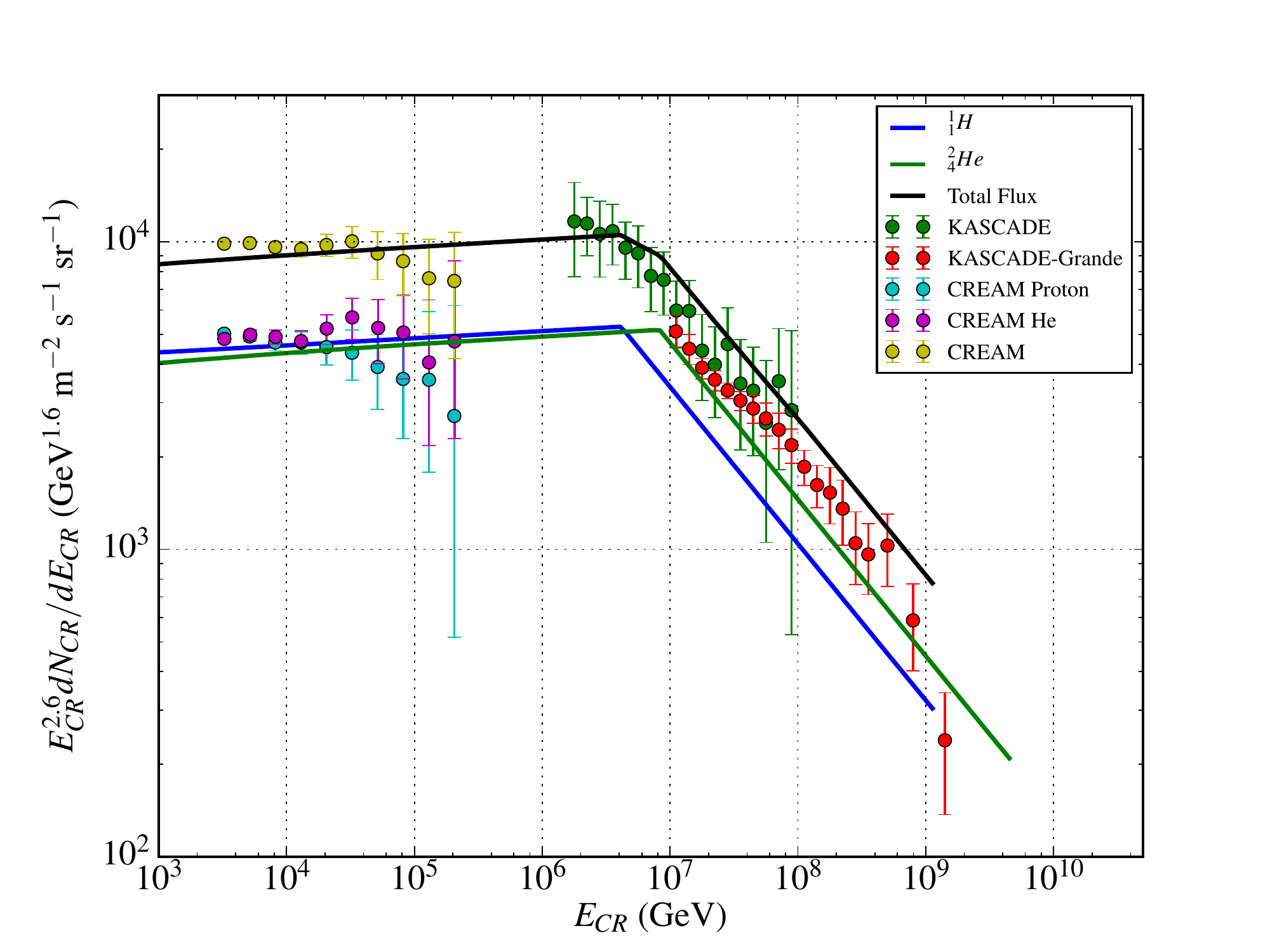}
\includegraphics[width=0.48\textwidth,clip=true,viewport= 10 0 540 400]{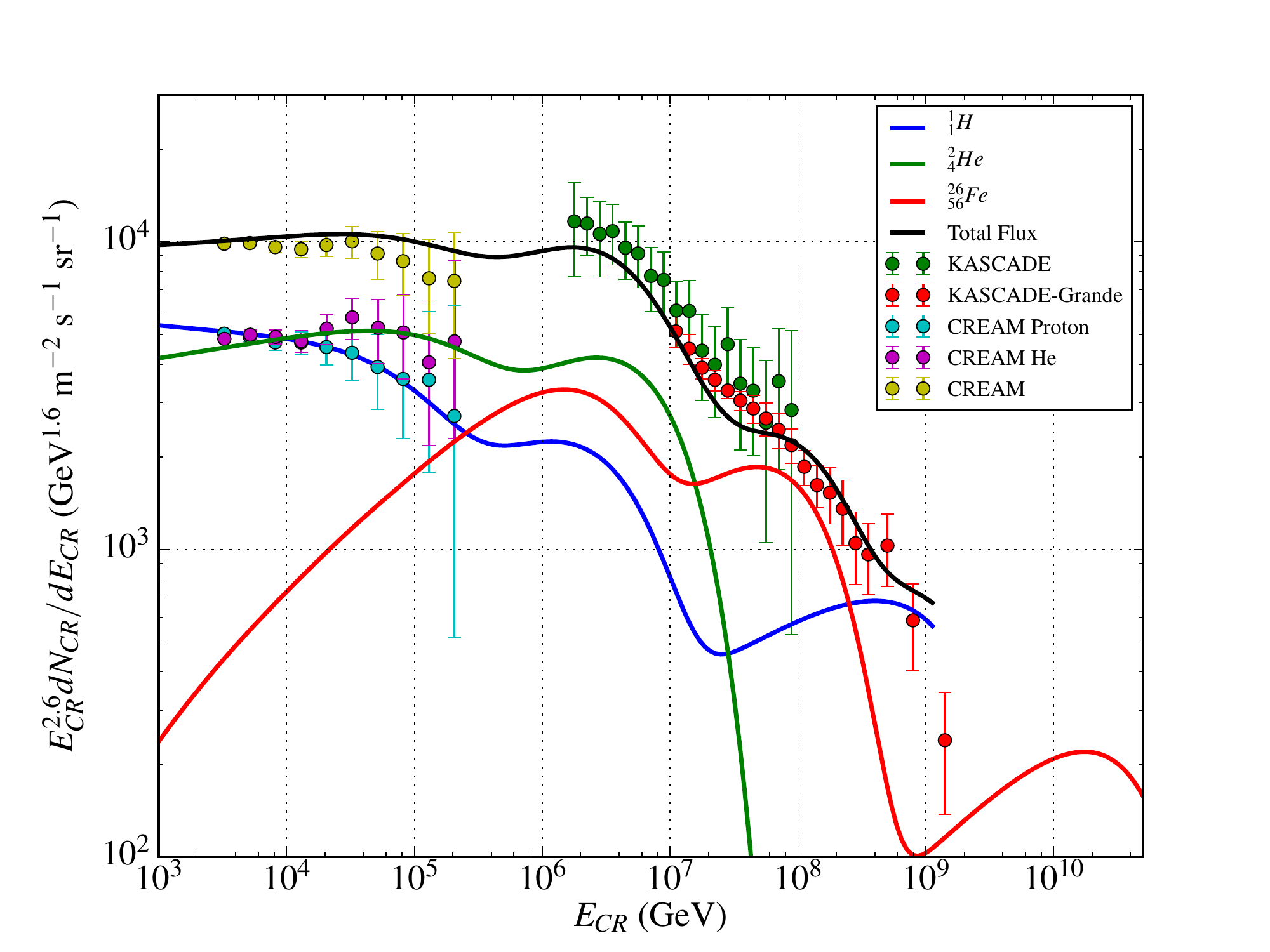}
\caption{{\bf Left panel:} a broken-power-law fit to the CR spectrum including only proton and helium contributions. {\bf Right panel:} a more sophisticated fit with exponential functions from Ref.~\cite{Gaisser:2013bla}, where the iron flux is important around the knee. 
\label{fig:cosmic-ray-fit} }
\end{figure}

To calculate the secondary diffuse neutrino flux with an energy above 1 TeV, it is important to understand the primary CR flux, especially at energies around the CR {\it knee} at a few PeV. This feature could be related to the maximum rigidity achievable in Galactic accelerators or a change in the rigidity-scaling of diffusion. In both cases we expect a strong dependence on the chemical composition of Galactic CRs.
In this study, we will adopt two Galactic CR models. The first one assumes a CR composition of proton and Helium following a broken-power-law spectra that we fit to the CREAM~\cite{CREAM}, KASCADE~\cite{KASCADE} and KASCADE-Grande~\cite{KASCADE-Grande} data. 
Our best-fit CR spectrum follows
\begin{equation}\label{eq:broken-power-fit}
E_A^2\frac{{\rm d}N_{A}}{{\rm d}E_{A}} \simeq a_A\begin{cases}\left(\frac{E_{A}}{E^*_{A}}\right)^{-0.58} &E_{A} \leq E^*_{A} \\ 
\left(\frac{E_{A}}{E^*_{A}}\right)^{-1.1}&E_{A}>E^*_{A}  \end{cases}  \,, 
\end{equation}
with $A=p, {\rm He}$, $a_p = 0.77\,\mbox{GeV}\,\mbox{m}^{-2}\,\mbox{s}^{-1}\, \mbox{sr}^{-1}$ and $E^*_{p} = 4.1\,\mbox{PeV}$ as well as $a_{\rm He} = 0.39\,\mbox{GeV}\,\mbox{m}^{-2}\,\mbox{s}^{-1}\, \mbox{sr}^{-1}$ and $E^*_{\rm He} = 8.2\,\mbox{PeV}$\,.
In the left panel of Fig.~\ref{fig:cosmic-ray-fit}, one can see that our power-law fit in good agreement with to the observed CR data up to the {\it knee} region. 
Note, that CR diffusion in the Galactic medium softens the observable spectra.
As a consistency check of our implementation of the CR injection spectrum Eq.~(\ref{eq:broken-power-fit}) into {\tt GALPROP} one can compare the energy dependence of the resulting CR flux to analytic expectations.
The difference between the observed CR power $E^{-\beta}$ and the injected power $E^{-\alpha}$ is expected to follow the relation $\beta = \alpha + \delta$, where $\delta$ is the rigidity-dependence of the diffusion coefficient. In this study, we assume diffusion of CRs in a Kolmogorov-like magnetic turbulence that predicts $\delta =1/3$. The numerical prediction of {\tt GALPROP} shown in the plot reproduces this expecting rigidity-scaling well.

As a second CR model we adopt the best-fit result of a study by Gaisser {\it et al.}~\cite{Gaisser:2013bla} modeling the CR spectrum as a superposition of three populations with individual contributions of different mass groups that have a common exponential rigidity cutoff~\cite{Hillas:2005cs}. 
As in the first CR model, the proton and helium components still provide the dominant contribution to the neutrino flux. The CR spectrum is given by a parametric form in Ref.~\cite{Gaisser:2013bla} [Table II and Eq.(3) therein] and is shown in the right panel of Fig.~\ref{fig:cosmic-ray-fit}. The parametrized spectra are given by the sum over three populations,
\begin{equation}\label{eq:global-fit}
E_A^2\frac{{\rm d}N_{A}}{{\rm d}E_A} =  \sum_{i=1}^3a_{A,i}\left(\frac{E_A}{E^*_{A,i}}\right)^{2-\Gamma_{A,i}}\exp\left(-\frac{E_A}{E^*_{A,i}}\right)\,,
\end{equation}
with normalization $a_{p} =$ (0.34, 0.020, 0.00032) $\mbox{GeV}\,\mbox{m}^{-2}\,\mbox{s}^{-1}\mbox{sr}^{-1}$, cutoff energies $E^*_{p} =$ (4, 30, 2000)~PeV and spectral indices $\Gamma_{p} =$ (2.66,2.4,2.4) for the three proton populations and $a_{{\rm He}} =$ (0.35, 0.015, 0.00025) $\mbox{GeV}\,\mbox{m}^{-2}\,\mbox{s}^{-1}\mbox{sr}^{-1}$, $E^*_{{\rm He}} = 2E^*_{p}$ and $\Gamma_{{\rm He}} =$ (2.58, 2.4, 2.4) for helium. The two CR models have different powers below the {\it knee} and the power-law model provides a harder spectrum, {\it i.e.}~larger CR fluxes. In the low energy part below 100 TeV, neutrino fluxes follow a simple power law behavior and have a power of $-2.54 (-2.69)$ close to the proton CR fluxes in Eq.~(\ref{eq:broken-power-fit}) [Eq.~(\ref{eq:global-fit})]. 

\begin{figure}[t]
\centering
\includegraphics[width=0.48\textwidth,clip=true,viewport= 0 0 540 400]{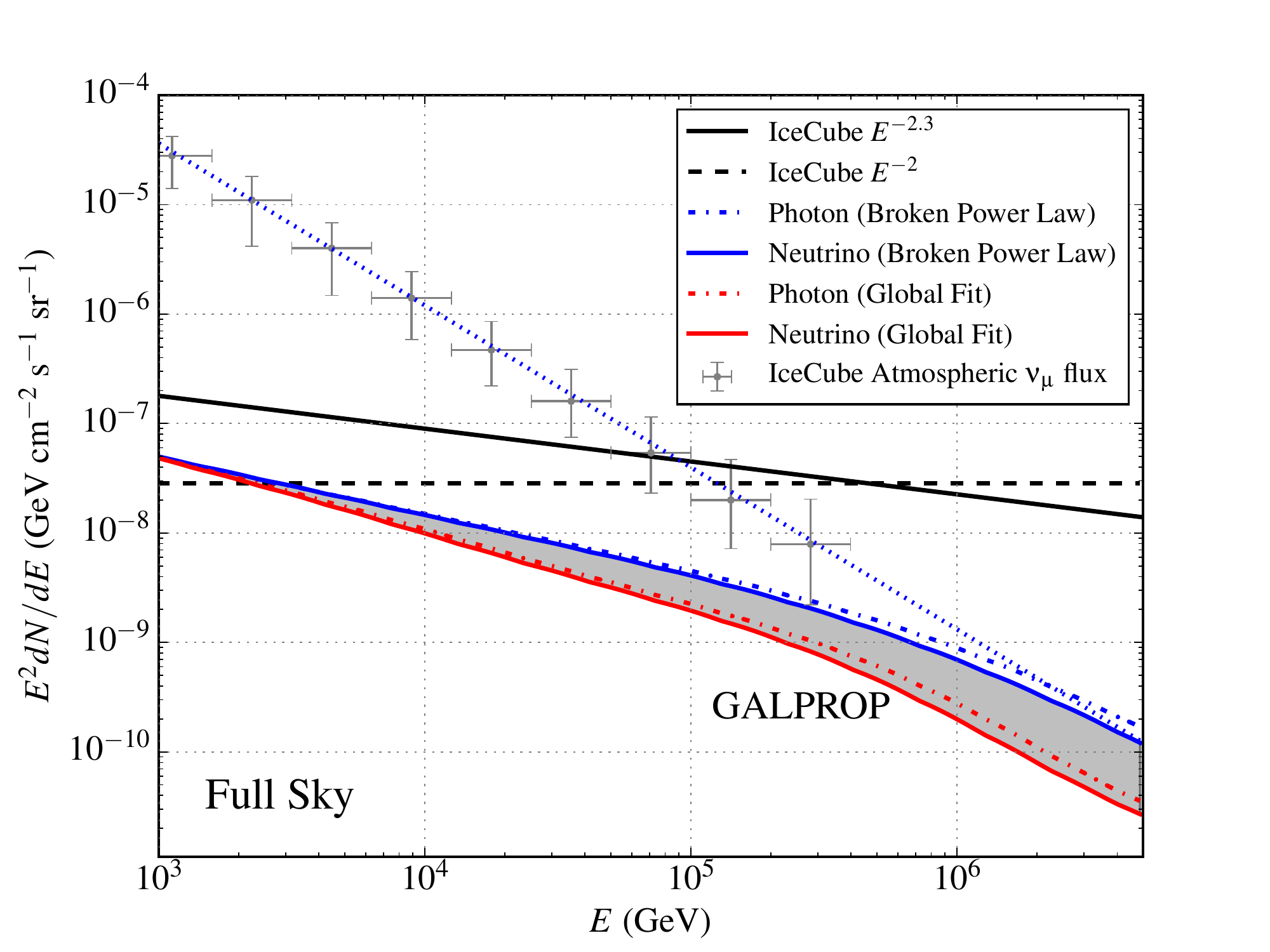}
\includegraphics[width=0.48\textwidth,clip=true,viewport= 0 0 540 400]{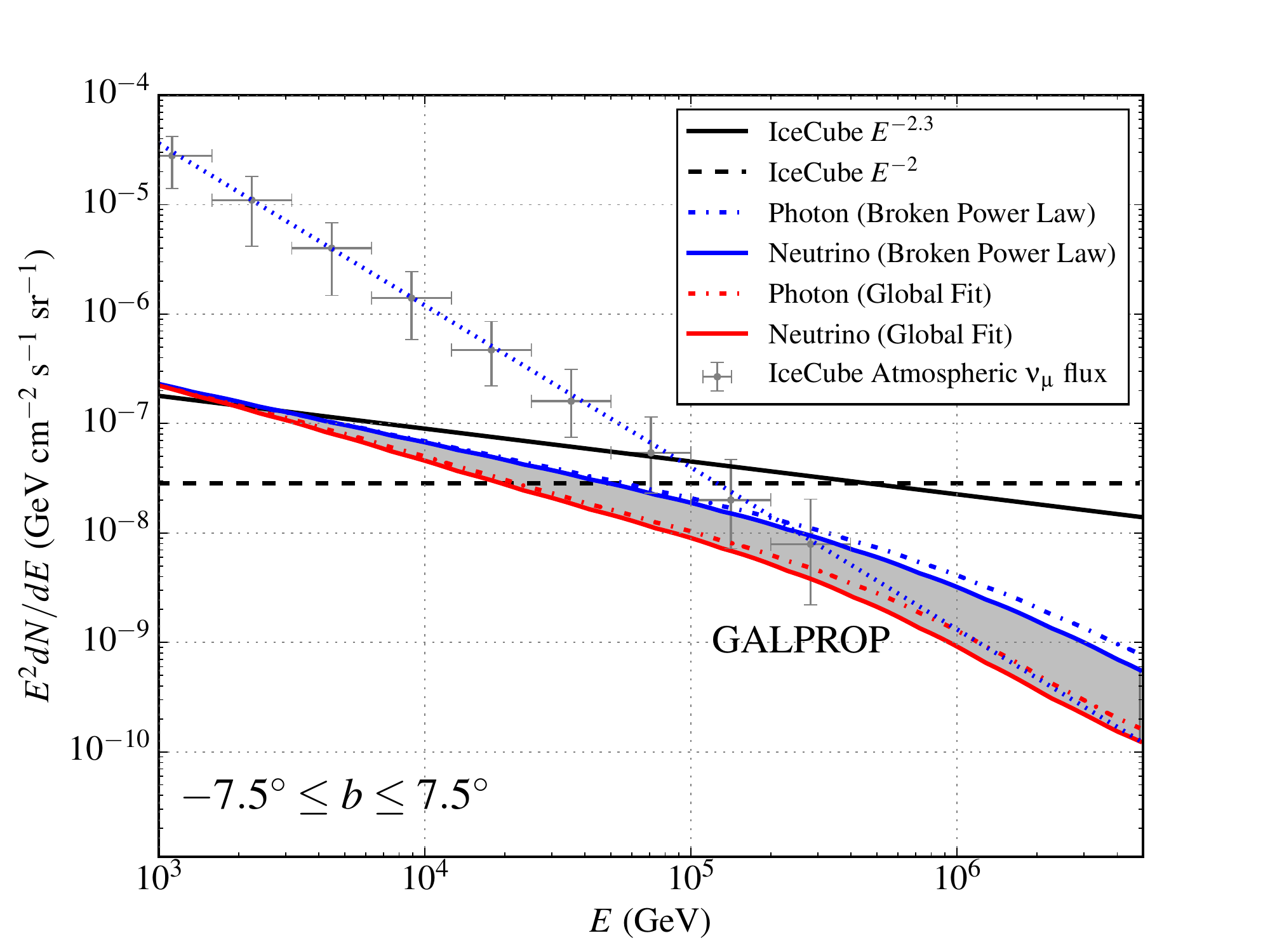}
\caption{ {\bf Left panel:} the differential neutrino and $\gamma$-ray spectra averaged over all the sky, based on the {\tt GALPROP} $^S S ^Z 4 ^R 20 ^T 150 ^C5$ diffuse model. Also shown is the IceCube best-fit and all-flavor spectrum flux: $E_\nu^2 {\rm d}N/{\rm d}{E_\nu} = 4.5 \times 10^{-8} (E_\nu/100~\mbox{TeV})^{-0.3}\,\mbox{GeV}\,\mbox{cm}^{-2}\mbox{s}^{-1}\mbox{sr}^{-1}$~\cite{Aartsen:2014gkd}. For the $\gamma$-ray spectrum, absorbing effects become significant only above 10 TeV due to the CMB absorbing effects~\cite{Moskalenko:2005ng}. We also show the IceCube measured atmospherical muon neutrino background flux in the green and cross points~\cite{Aartsen:2014qna}. {\bf Right panel:} the same as the left one but for the Galactic Plane region with $|b| \leq 7.5^\circ$.
\label{fig:flux-ratio} }
\end{figure}

In the left panel of Fig.~\ref{fig:flux-ratio}, we show our predictions for the all-sky-averaged diffuse neutrino fluxes for the two different CR models. We also show the best-fit power-law flux observed by IceCube~\cite{Aartsen:2014gkd}, $E_\nu^2 {\rm d}N/{\rm d}E_\nu$ $= 4.5 \times 10^{-8} (E_\nu/100~\mbox{TeV})^{-0.3}\,\mbox{GeV}\,\mbox{cm}^{-2}\mbox{s}^{-1}\mbox{sr}^{-1}$ summed over neutrino flavors. The diffuse neutrinos account for 10\% (5\%) for the broken-power-law (global) fit with the sky-averaged flux and $E_{\nu/\bar{\nu}} \ge 30$~TeV (see Refs.~\cite{Blum:2014ewa,Fong:2014bsa,Chen:2014gxa,Palomares-Ruiz:2015mka,Kamada:2015era} for other spectral analysis). For  $E_{\nu/\bar{\nu}} \ge 60$~TeV, the diffuse neutrinos account for a slightly smaller fraction of 8\% (4\%) for the broken-power-law (global) fit. Compared to the atmospherical muon neutrino background measured by IceCube~\cite{Aartsen:2014qna} (see also \cite{Gondolo:1995fq,Honda:2006qj,Enberg:2008te} for theoretical calculations), the Galactic diffuse neutrino flux dominates over the Galactic diffuse spectrum below PeV. However, this comparison is based on neutrino fluxes averaged over the whole sky. As we know from the density distribution indicated in Fig.~\ref{fig:pi-zero-latitude} of Appendix~\ref{app1}, the Galactic diffuse neutrinos mainly come from the Galactic Plane. In the right panel of Fig.~\ref{fig:flux-ratio}, we show the comparison for the neutrino differential fluxes around the Galactic Plane region with $|b| \leq 7.5^\circ$~\cite{Aartsen:2014qna}. The Galactic diffuse neutrinos can have a flux over the atmospheric one for energy above  $\sim 300$~TeV. We also note that the IceCube HESE analysis (see Fig.~\ref{fig:HESE3yr}) has reported a reduced atmospherical neutrino background and decmonstrated an enhanced ability to measure Galactic diffuse neutrinos~\cite{Gaisser:2014bja}.

Before we move on to the next section, a few remarks are in order. Our analysis of Galactic diffuse gamma-ray and neutrino emission with the {\tt GALPROP} code follows the standard approximation of homogenous and isotropic CR diffusion in the Milky Way. It also assumes that the observed CR spectrum can be approximated as a steady-state solution of a smooth distribution of continuous sources. All these assumptions can only be considered as a first-order approximation. Deviations from these assumptions can decrease or enhance the diffuse emission. For instance, it is well known that diffusion along the regular component of the Galactic magnetic field is stronger compared to orthogonal directions, {\it e.g.}~\cite{Casse:2001be}. This can lead to local variations of the CR flux normalization compared to the Galactic average~\cite{Effenberger:2012jc}. A similar effect can be observed for models accounting for non-azimuthally symmetric source distributions including the Galactic spiral arms. This has been studied with three-dimensional propagation codes like {\tt DRAGON}~\cite{Evoli:2008dv,Gaggero:2013rya} and {\tt PICARD}~\cite{Kissmann:2014sia,Werner:2014sya}. It was also shown that non-homogenous Galactic diffusion implemented in the propagation code {\tt DRAGON}~\cite{Evoli:2008dv} can enhance the hadronic gamma-ray and neutrino emissions in the multi-TeV region~\cite{Gaggero:2015xza}. Finally, it was argued that time-dependent and local CR injection episodes could also result in locally observed CR spectra that are softer than the Galactic average~\cite{Neronov:2013lza}. This harder average emission would increase the diffuse emission in the multi-TeV region after normalization to {\it Fermi} data.

\section{Neutrino Fluxes from Point-like and Extended Sources}\label{sec:source}
Point-like sources in the Milky Way may also contribute to the observed neutrino flux. Since there is presently no evidence for a correlation of the IceCube HESE events with Galactic point-like sources~\cite{Aartsen:2014cva} we do not expect a strong contribution of individual Galactic sources, but rather look for hints in the data for their combined (stacked) emission. The most likely candidates of this neutrino emission are the sources that contribute the bulk of Galactic CRs. Candidate sources include supernova remnants (SNRs) and pulsar wind nebulae (PWNe).

To first order, the observed density $n_{\rm CR}$ of CRs  can be related to the time and volume-averaged CR emission rate of the Galaxy $\langle Q_{\rm CR}\rangle(E)$ via the energy-dependent diffuse escape time $\tau_{\rm esc}(E)$ as $n_{\rm CR} \simeq \tau_{\rm esc}\langle Q_{\rm CR}\rangle(E)$ (see {\it e.g.}~Ref.~\cite{Blasi:2011fi}). The escape time from the Galactic Plane with half height $H$ can be estimated as $\tau_{\rm esc}\simeq H^2/D(E)$ for a diffusion coefficient $D(E)$. The initial spectrum of CRs is therefore softened by the energy dependence of $D(E)$.

The fit to the observed CR data via {\tt GALPROP} presented in the previous section can also be utilized to determine the {\it in situ} contribution of the sources of Galactic CRs more precisely. From the {\tt GALPROP} fit we know the time-averaged differential source emission rate of Galactic CR sources $Q_{\rm CR}(r,z,E)$ for a given source distribution and propagation model. The difference to the diffuse neutrino emission of Galactic CRs is that here we assume that secondary neutrinos are directly produced at the source location via interaction with gas in the source environment, {\it e.g.}~Refs.~\cite{Berezhko:2000vy,Berezhko:2004qb}. This is quantified via the optical thickness $\tau_{pp}$. So, one can directly relate the source emission rate of neutrinos (per flavor) and CRs as
\beqa
E_\nu^2Q_\nu(r,z,E_\nu) \simeq \frac{1}{6}\left(1-e^{-\kappa_{\rm ine}\tau_{pp}}\right)E_p^2 Q_{\rm CR}(r,z,E_p) \,.  \label{eq:neutrino-rate-from-proton}
\eeqa
with inelasticity $\kappa_{\rm ine}\simeq 0.5$. The overall factor of $1/6$ is a rough estimation of the total neutrino energy over the total proton energy accounting for the per flavor emission ($1/3$), the total neutrino energy fraction in the charge pion decay  ($3/4$) and for the charged pion fraction in $pp$ collisions ($2/3$). Furthermore, for the individual neutrino energy, one can use the approximate relation $E_\nu \simeq E_p / 20$ with a factor of $1/4$ for the neutrino (each flavor) energy in the pion decay and a factor of $1/5$ for the average energy of the pion produced in $pp$ interactions. 

For a source with dynamical time-scale $t_{\rm dyn}$ and average gas density $n_{\rm gas}$ the opacity is given as $\tau_{pp} \simeq ct_{\rm dyn}n_{\rm gas}\sigma_{pp}$ for an inelastic scattering cross section $\sigma_{pp}\simeq 34 +1.88L +0.25L^2$~mb for $L=\ln(E_p/{\rm TeV})$~\cite{Kelner:2006tc,Block:2011vz}. We estimate the dynamical time-scale as $10^4$~yrs corresponding to the beginning of the snowplow phase, where the acceleration efficiency in typical supernova remnants are expected to cease~\cite{Blondin1998}. We therefore estimate the optical thickness, following Refs.~\cite{Berezhko:2000vy,Berezhko:2004qb} as
\begin{equation}\label{eq:opacity}
\tau_{pp} \simeq 3\times10^{-4} \left(\frac{t_{\rm dyn}}{10^4~{\rm yr}}\right)\left(\frac{n_{\rm gas}}{1\,{\rm cm}^{-3}}\right)\,.
\end{equation}

For a given class of sources, the approximate Galactic source distribution is typically expressed in cylindrical Galacto-centric coordinates $\rho(r, z)$, which is related to spherical Helio-centric coordinates $(s,\ell,b)$ as
$r^2 = s^2\cos^2b +R_\odot^2-2\,sR_\odot\cos\ell\cos b$ and $z = s\,\sin b$. The distance from the Sun to the Galactic Center is $R_\odot \simeq 8.5$~kpc. To keep track of the source distribution in the skymap, it is convenient to introduce the normalized ``$J$''-factor after integrating along the line of sight
\beqa
{J(b,\ell)} =\frac{1}{4\pi}\frac{1}{N_{\rm s}}\int {\rm d}s \rho\left[r(s,\ell,b),z(s,\ell,b) \right] \,.
\label{eq:J-factor-definition}
\eeqa
with the total number of sources $N_{\rm s} = 2\pi\int{\rm d}z \int {\rm d}r\,r\,\rho(r,z)$. The total $J$-factor is defined to be $\widehat{J} \equiv \int d\Omega J(b, \ell) = \int {\rm d}b\,{\rm d}\ell  \cos{b} \,J(b, \ell)$. We can then express the all-sky-averaged neutrino diffuse neutrino flux via the total neutrino spectral emission rate $\widehat{Q}_\nu$ as
\beqa
E_\nu^2\phi_\nu(E_\nu) =\frac{1}{4\pi}\int {\rm d}\Omega J(b,\ell)E_\nu^2\widehat{Q}_\nu(E_\nu)  = \frac{\widehat{J}}{4\pi}E_\nu^2\widehat{Q}_\nu(E_\nu)  \,.
\label{eq:source-generated-neutrino}
\eeqa
%

\subsection{SNRs and PWNe}
In the following, we study two CR source densities that are expected to approximate the distributions of SNRs and PWNe in our Milky Way. For SNRs we use the parametrized distribution from Case {\it et al.} in~\cite{Case:1998qg}, while for the PWNe, we take the pulsar distribution from Lorimer {\it et al.} in~\cite{Lorimer:2006qs}. Both can be described by the following function
\beqa\label{eq:Case}
\rho(r,z) = \rho_\odot\left(\frac{r}{R_\odot}\right)^\alpha\exp\left(-\beta\frac{r-R_\odot}{R_\odot}\right)\exp\left(-\frac{|z|}{h}\right) \,,
\eeqa
with $\alpha=2$, $\beta=3.53$ and $h=0.181$~kpc ({\it Case}, SNR) and $\alpha=1.93$, $\beta=5.06$ and $h=0.181$~kpc ({\it Lorimer}, pulsars). Both models give $\widehat{J}\simeq (16.2~{\rm kpc})^{-2}$. 

\begin{figure*}[t]
\begin{center}
\includegraphics[width=0.6\textwidth]{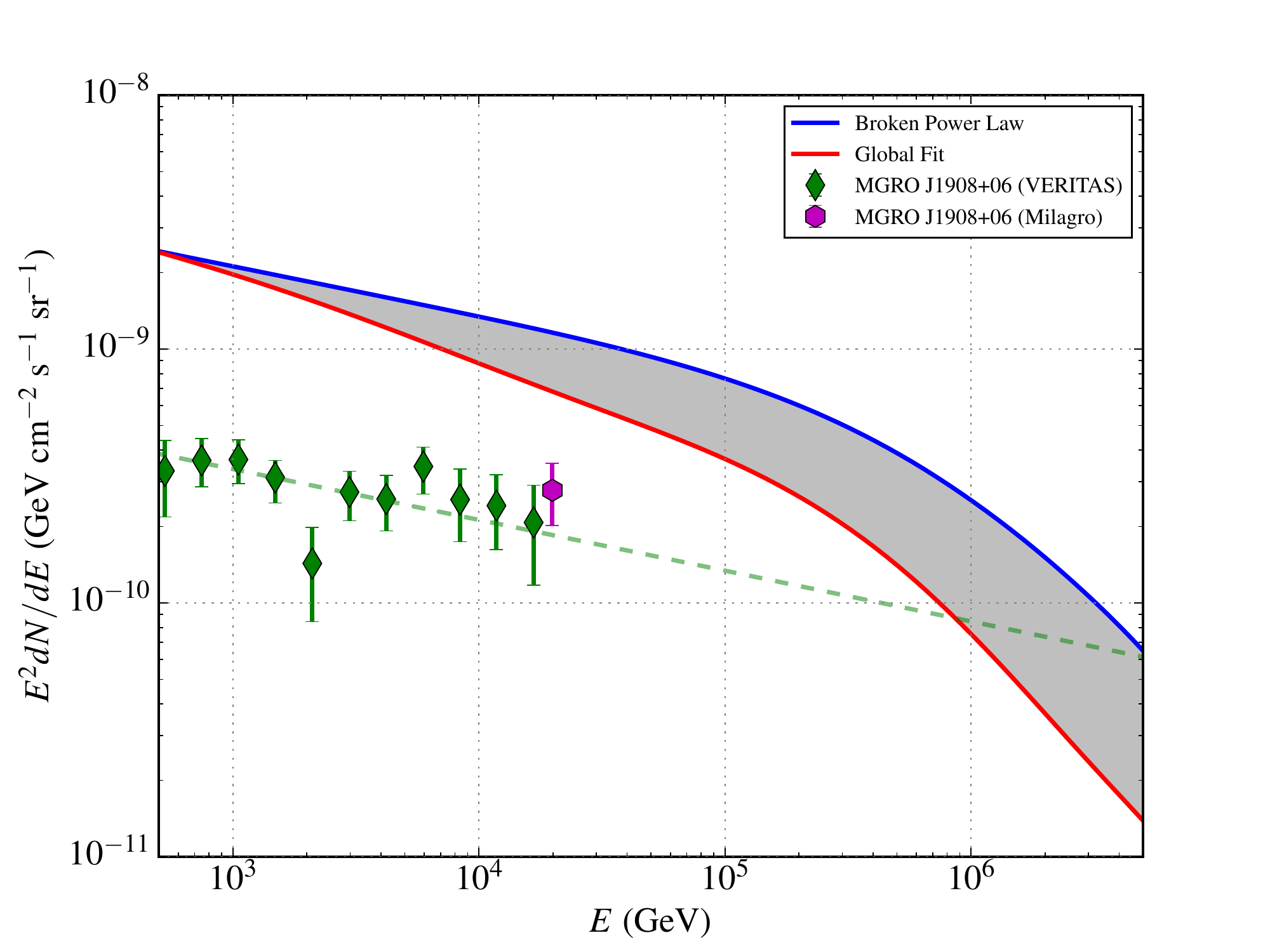}\hfill
\caption{The data points are the $\gamma$-ray flux for one representative UnID TeV source: MGRO J1908+06~\cite{Abdo:2007ad,Aliu:2014xra}, after multiplied by a factor of $1/4\pi$. The solid lines are calculated neutrino fluxes directly from the supernovae remnant sources.
}
\label{fig:sourcespectrum}
\end{center}
\end{figure*}

In Fig.~\ref{fig:sourcespectrum} we show the estimated all-sky-averaged neutrino flux from CR sources in our Galaxy assuming a distribution following that of SNRs. We assume an average optical thickness of the sources of $\tau_{pp}\simeq 3\times10^{-4}$ following the SNR benchmark values shown in Eq.~(\ref{eq:opacity}). Note that the flux level is proportional to the source opacity as long as $\tau_{pp}\ll 1$ and saturates for $\tau_{pp}\gg 1$. To have the flux not overshoot the observed IceCube flux, the averaged opacity is required to be $\langle \tau_{pp}\rangle \lesssim 0.01$. The overall normalization is determined from the {\tt GALPROP} fit to CR data assuming two different fitted CR spectra in Eq.~(\ref{eq:broken-power-fit}) and Eq.~(\ref{eq:global-fit}) resulting in a total Galactic CR power of about $(3-5)\times10^{40}$~erg/s. This is consistent with a Galactic supernova rate of 3~${\rm yr}^{-1}$ if a few percent of the kinetic energy of supernova ejecta of typically $10^{51}$~erg is converted into CRs. Since the spectrum is harder than the diffuse emission by the diffusion parameter $\delta=1/3$, the source spectrum can dominate at high energies unless the opacity of the sources becomes too small. 

\subsection{Unidentified TeV Sources}
Other than the cumulative contribution of SNRs and PWNe, we also consider the contribution from unidentified TeV  (UnID TeV) $\gamma$-ray sources~\cite{Fox:2013oza}, which may originate in hypernova (superluminous supernova) remnants of our Galaxy~\cite{Ioka:2009dh} and contribute to high energy neutrinos from hadronic $pp$ interactions. For instance ({\it cf.}~data points in Fig.~\ref{fig:sourcespectrum}), one of the bright sources, MGRO J1908+06, observed first by the Milagro collaboration, has the differential $\gamma$-ray flux of, $E^2 \, {\rm d}N/{\rm d}E = 1.3 \times 10^{-9} (E/\mbox{1~GeV})^{-0.20}\,\mbox{GeV}\mbox{cm}^{-2}\mbox{s}^{-1}\mbox{sr}^{-1}$ (after multiplying the source flux by $1/4\pi$), from a more precise measurement by the VERITAS collaboration~\cite{Aliu:2014xra} which is just one order of magnitude below the IceCube neutrino flux. Two of the UnID TeV sources studied in~\cite{Fox:2013oza} have since been identified (HESS J1018-589 \& HESS J1837-069) whereas one new unidentified source has been announced (VER J2019+368). Consequently, we have updated the UnID source catalogue from the tabulated list provided in Ref.~\cite{Fox:2013oza}.

\subsection{Fermi Bubbles}
Beyond the point-like sources, there are also a few candidates of extended neutrino sources. The {\it Fermi Bubbles}~\cite{Su:2010qj} are two large structures extending to latitudes of about $55^\circ$ above and below the Galactic Plane. The precise mechanism of the $\gamma$-ray emission is unknown and, both, hadronic~\cite{Crocker:2010dg} and leptonic scenarios~\cite{Mertsch:2011es,Yang:2013kca} have been proposed in the literature as an explanation. A possible connection to the IceCube HESE flux has been pointed out in Refs.~\cite{Razzaque:2013uoa,Ahlers:2013xia,Lunardini:2013gva,Lunardini:2015laa}. The {\it Fermi} Collaboration has also performed a dedicated search to understand the energy spectrum and morphology of the {\it Fermi Bubbles}~\cite{Fermi-LAT:2014sfa}. For the hadronic origin, a power-law CR proton spectrum with an exponential cutoff at about $15$~TeV provides a better fit than a power-law spectrum without a cutoff~\cite{Fermi-LAT:2014sfa}. The corresponding cutoff in the neutrino emission would then be below TeV. However, the significance of this cutoff is weak and it is possible that the emission continues with a softer power towards higher energies and then contributes to the IceCube data above 10~TeV~\cite{Fermi-LAT:2014sfa}. Future $\gamma$-ray observations with CTA~\cite{Doro:2012xx,Pierre:2014tra} and HAWC~\cite{Abeysekara:2013tza,Abeysekara:2014ffg} in the $(0.1-100)$~TeV range will be able to identify the spectrum~\cite{Ahlers:2013xia,Lunardini:2015laa}.

\subsection{Decaying Dark Matter}
Another more exotic scenario that has been proposed as a source of the IceCube flux is decaying dark matter with a dark matter mass of ${\cal O}(5~\mbox{PeV})$ and a lifetime of ${\cal O}(10^{28}~\mbox{s})$~\cite{Feldstein:2013kka,Esmaili:2013gha,Bai:2013nga,Cherry:2014xra,Murase:2015gea}. The contributions of Galactic emission from the Galactic dark matter halo and an extragalactic isotropic emission are of similar magnitude~\cite{Esmaili:2013gha,Ema:2013nda,Ema:2014ufa,Murase:2015gea}. The relative intensity of the Galactic emission is again calculated from Eq.~(\ref{eq:J-factor-definition}). For the dark matter density distribution in our galaxy, we use the isotropic Einasto profile~\cite{Graham:2006ae} with 
\beqa\label{eq:Einasto}
\rho_{\rm DM}(r) = \rho_{\odot} \, \exp\left({-\frac{2}{\beta} \left[ \left( \frac{r}{r_s}  \right)^{\beta} - \left( \frac{r_\odot}{r_s} \right)^{\beta}    \right]   }\right) \,,
\eeqa
with $\rho_\odot=0.4~{\rm GeV}/{\rm cm}^3$, $r_s = 20$~kpc, $r_\odot=8.5$~kpc and $\beta = 0.17$. The total dark matter mass in the Milky Way is then approximately $M^{\rm total}_{\rm DM}\simeq2.7\times10^{12}~M_\odot$.

Note, that 
the contribution from extragalactic dark matter will be mostly isotropic, except for close-by galaxies and galaxy clusters~\cite{Murase:2015gea}. This isotropic background and the very extended Galactic diffuse neutrino emission will make it difficult to identify the emission by anisotropy studies alone. However, the dark matter decay spectrum can be identified by spectral properties, like strong line-features and mass thresholds that are unexpected in astrophysical scenarios. Two representative dark matter decay spectra are shown in Fig.~\ref{fig:DM-photon} of Appendix~\ref{appDM}.

\begin{figure*}[p]
\begin{center}
\includegraphics[width=0.5\textwidth]{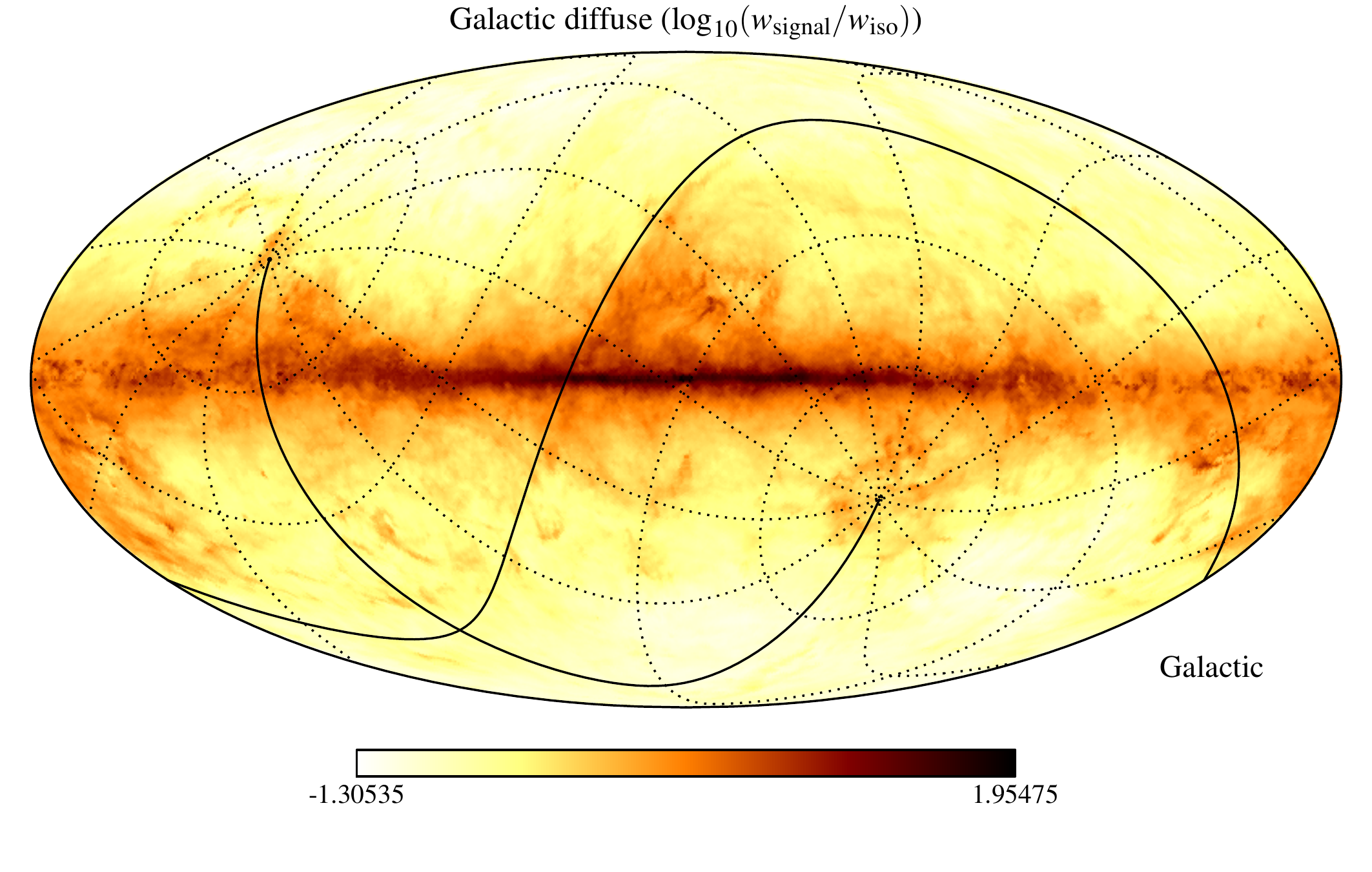}\includegraphics[width=0.5\textwidth]{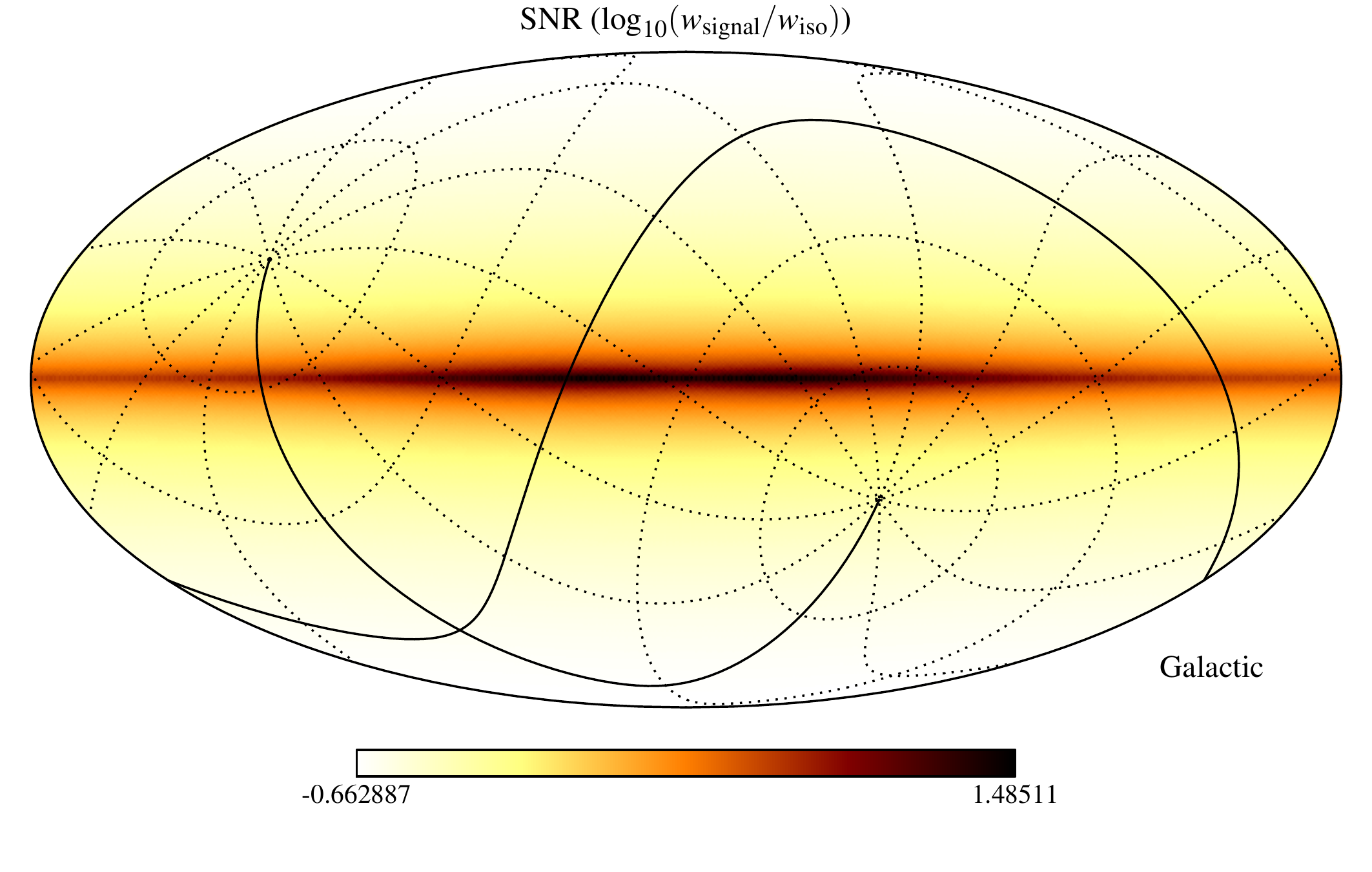}\\[-0.1cm]
\includegraphics[width=0.5\textwidth]{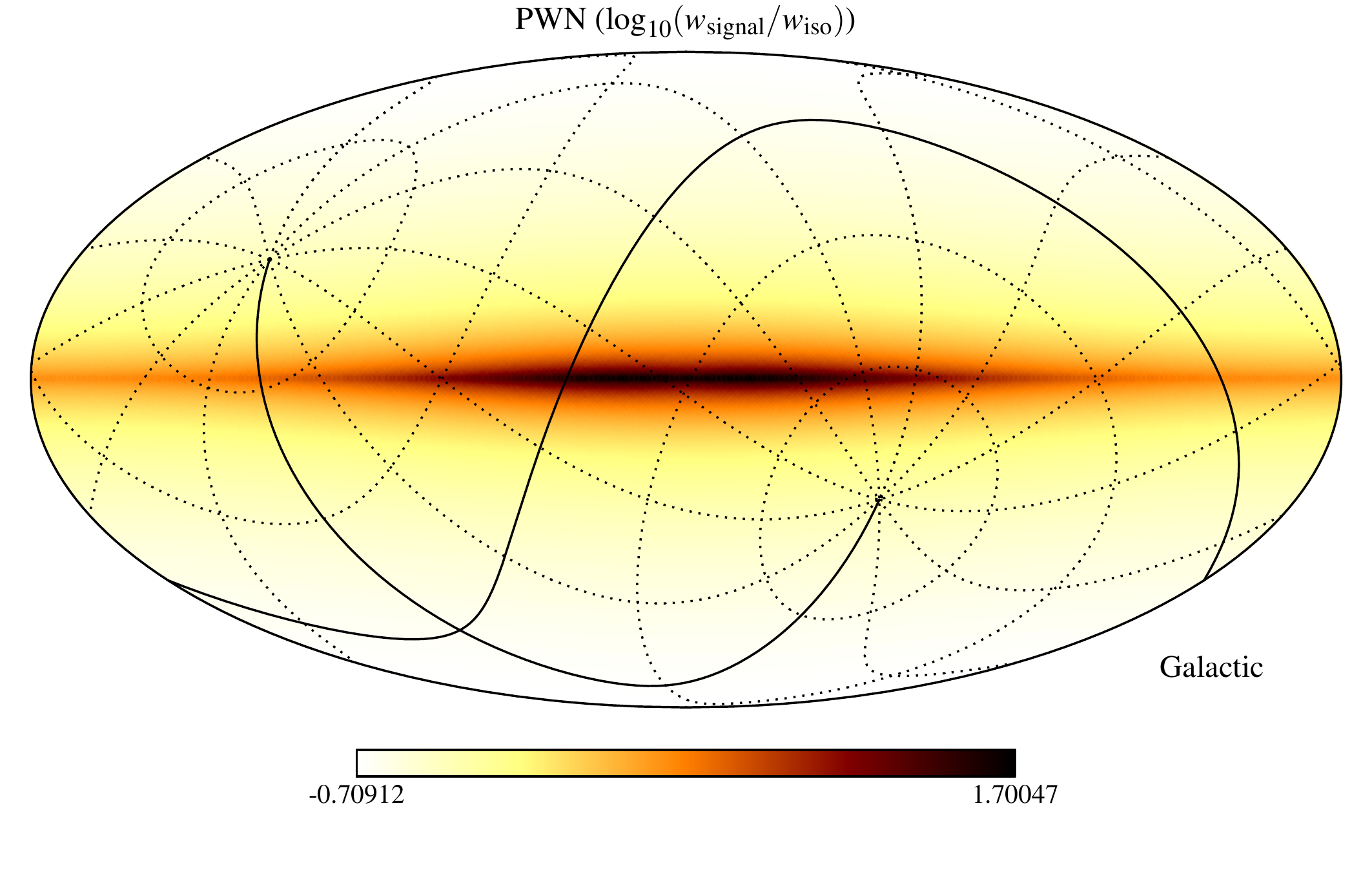}\includegraphics[width=0.5\textwidth]{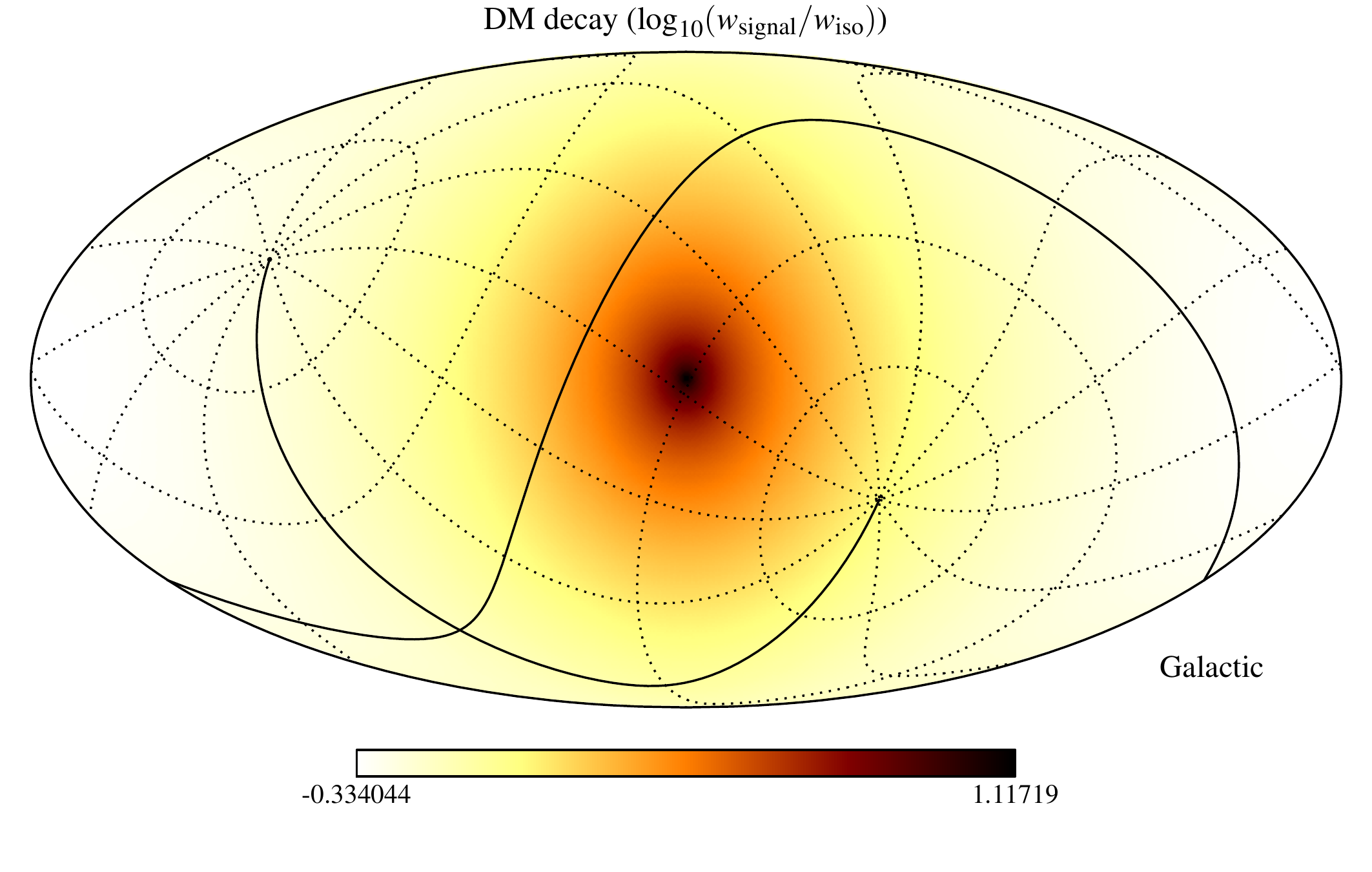}\\[-0.3cm]
\includegraphics[width=0.5\textwidth]{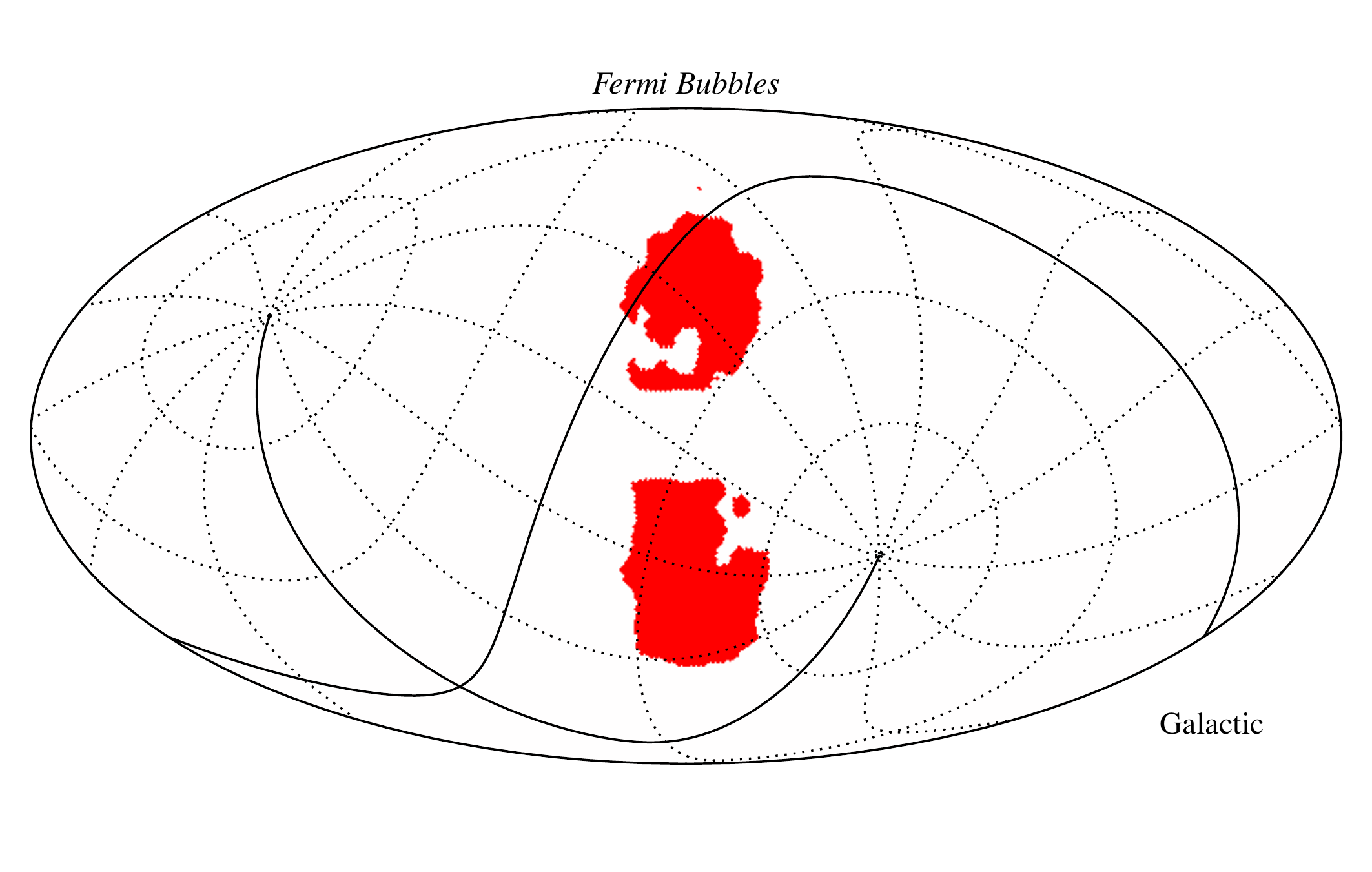}\includegraphics[width=0.5\textwidth]{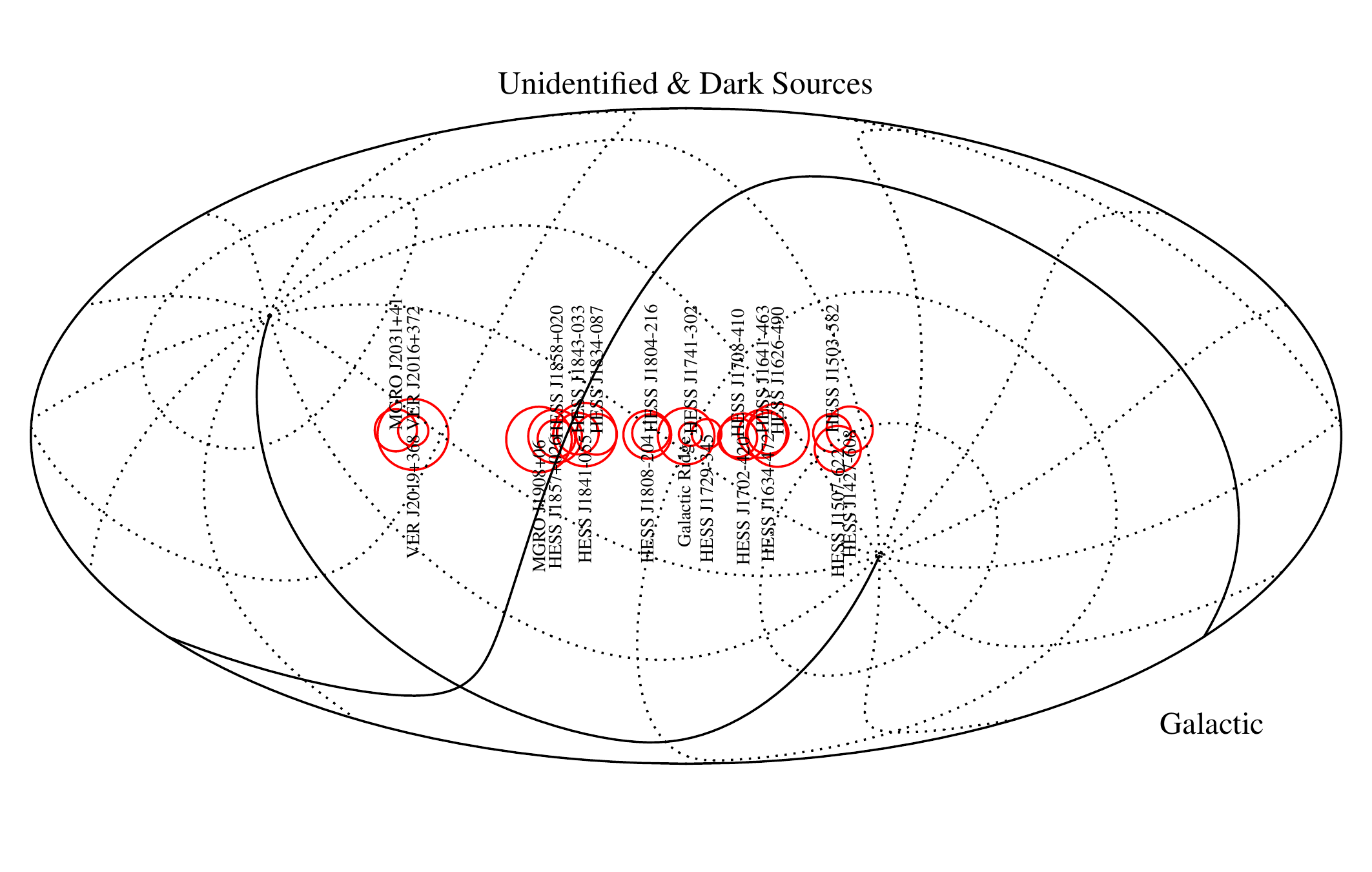}
\caption{\label{templates} Six Mollweide projections of Galactic emission templates used in the statistical analysis. From left to right and top to bottom, they are Galactic diffuse ($E_\nu = 10$~TeV), supernovae remnants (Case), pulsar wind nebulae (Lorimer), decaying dark matter, {\it Fermi Bubbles} and unidentified and dark TeV $\gamma$-ray sources. The mesh indicates the equatorial coordinate system with right ascension $\alpha=0^\circ$ and declination $\delta=0^\circ$ indicated as solid lines. Note that at the location of IceCube declination and zenith angles have the simple relation $\theta_{\rm zen} = \pi/2+\delta$. 
}\label{fig:morphology}
\end{center}
\end{figure*}

\section{Anisotropy Test of Galactic Neutrino Emission}\label{sec:morphology}
In general, Galactic neutrino emission is expected to be anisotropic. The diffuse emission from CR propagation is required to follow the integrated column density along the line-of-sight. This emission is structured due to the higher gas concentration towards the Galactic Plane (GP) and Galactic Center, but it also has high-latitude fine-structure due to molecular gas clouds. Emission from CR-gas interactions in the vicinity of CR sources is also correlated with the Galactic Plane. In the following we will consider two source density distributions already introduced in the previous section, approximating supernova remnants (Case~\cite{Case:1998qg}) and pulsar wind nebulae (Lorimer~\cite{Lorimer:2006qs}). On the other hand, a hypothetical signal from the decay of heavy dark matter (DM) is expected to follow the DM density distribution producing an extended emission around the Galactic Center. As a template we use the Einasto profile~\cite{Graham:2006ae}. To test for extended diffuse neutrino emission from the {\it Fermi Bubbles}, we use the gamma ray emission template provided by the {\it Fermi} Collaboration~\cite{Fermi-LAT:2014sfa}. The collective emission from Galactic hypernovae~\cite{Fox:2013oza} is already covered by a test of emission along the Galactic plane. In addition to that we also include the unidentified and dark TeV $\gamma$-ray sources that have been summarized in Ref.~\cite{Fox:2013oza} which we update with recent observations. In Fig.~\ref{fig:morphology}, we show the geometric distributions of all six Galactic emission templates in the Galactic coordinate. For Galactic diffuse, supernovae remnants (Case), pulsar wind nebulae (Lorimer) and decaying dark matter templates, we show their flux density distributions with respect to an isotropic distribution. 

Correlations of the neutrino arrival direction with Galactic structures can be tested using an unbinned maximum likelihood test (see {\it e.g.}~\cite{Agashe:2014kda}). The IceCube Collaboration has tested emission along the Galactic Plane via a {\it top-hat} template, $w_{\rm GP}(\Omega) = {\Theta(b_{\rm GP}-|b|)}/({4\pi\sin b_{\rm GP}})$, where $\pm b_{\rm GP}$ is the Galactic lattitude width of the emission region. A scan over $b_{\rm GP}$ revealed a maximal excess over background at $b_{\rm GP}=7.5^\circ$, with post-trial significance of only 2.8\%~\cite{Aartsen:2014gkd}.

\begin{figure*}[p]
\mini{0.5\textwidth}{Samples for 3yr HESE ($E_{\rm dep}>60$~TeV)}\mini{0.5\textwidth}{Samples for 3yr Up-going $\nu_\mu$ ($E_\mu>100$~TeV)}\\
\includegraphics[width=0.5\textwidth]{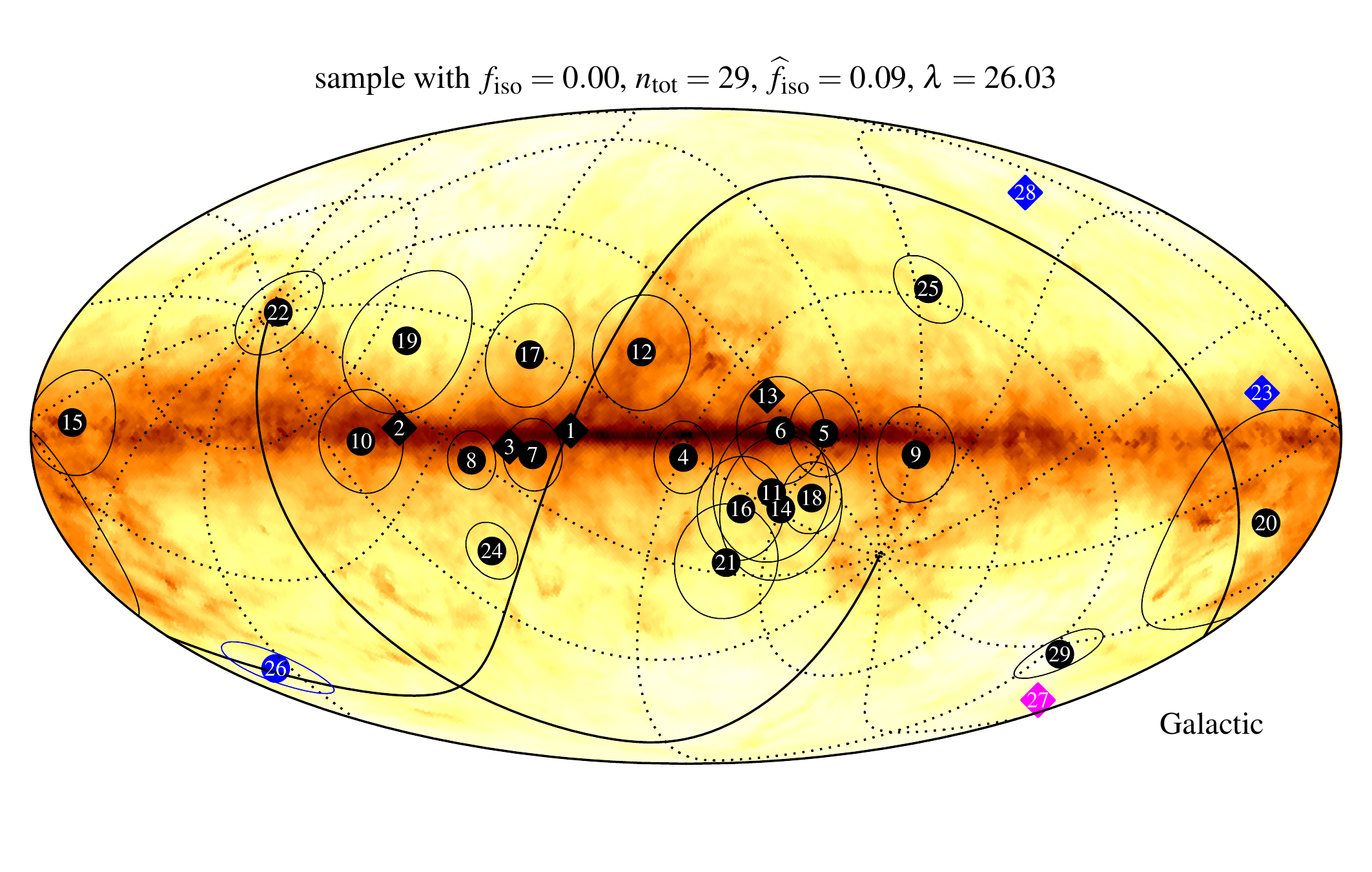}\hfill\includegraphics[width=0.5\textwidth]{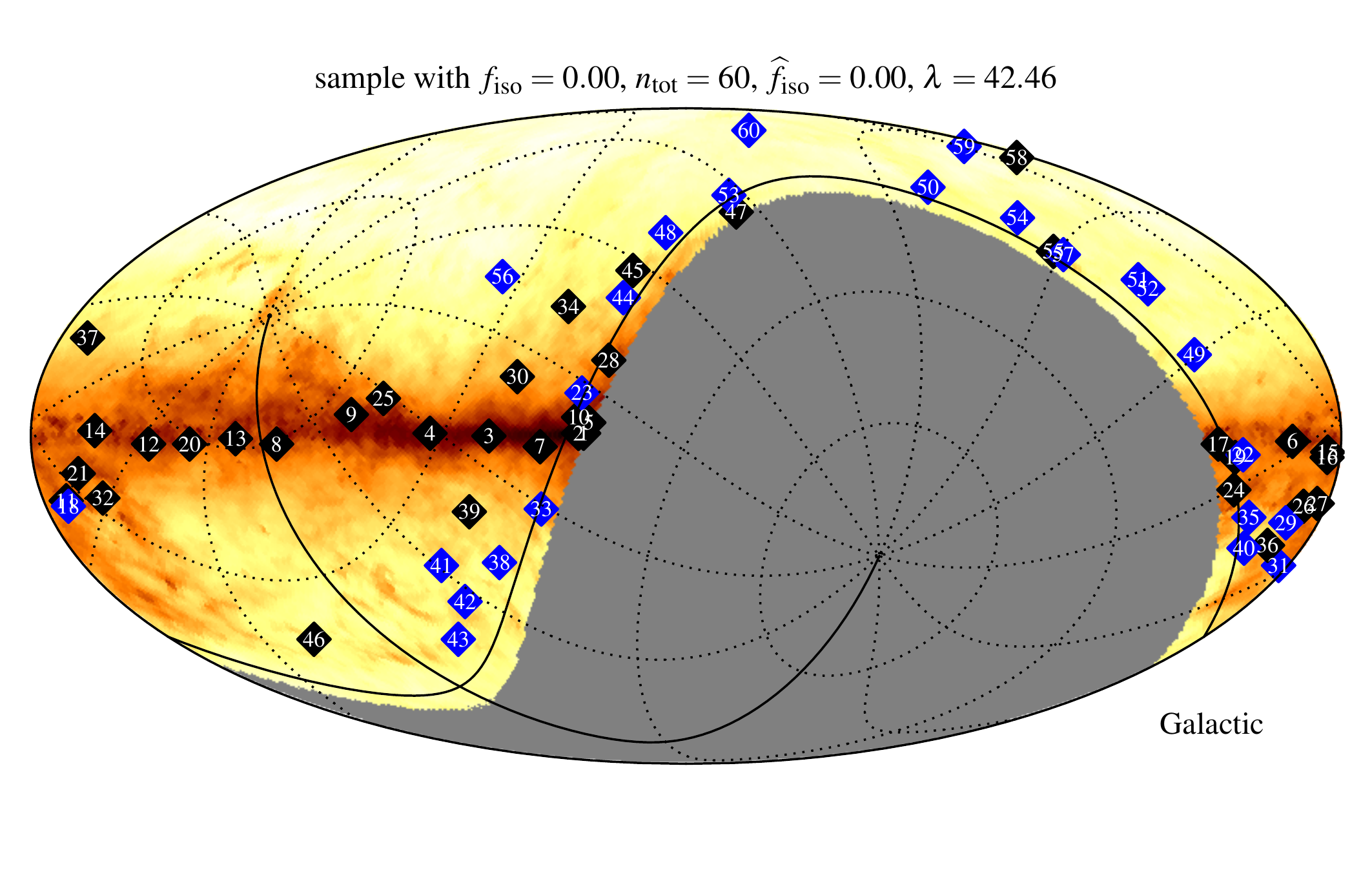}\\[-0.5cm]
\includegraphics[width=0.5\textwidth]{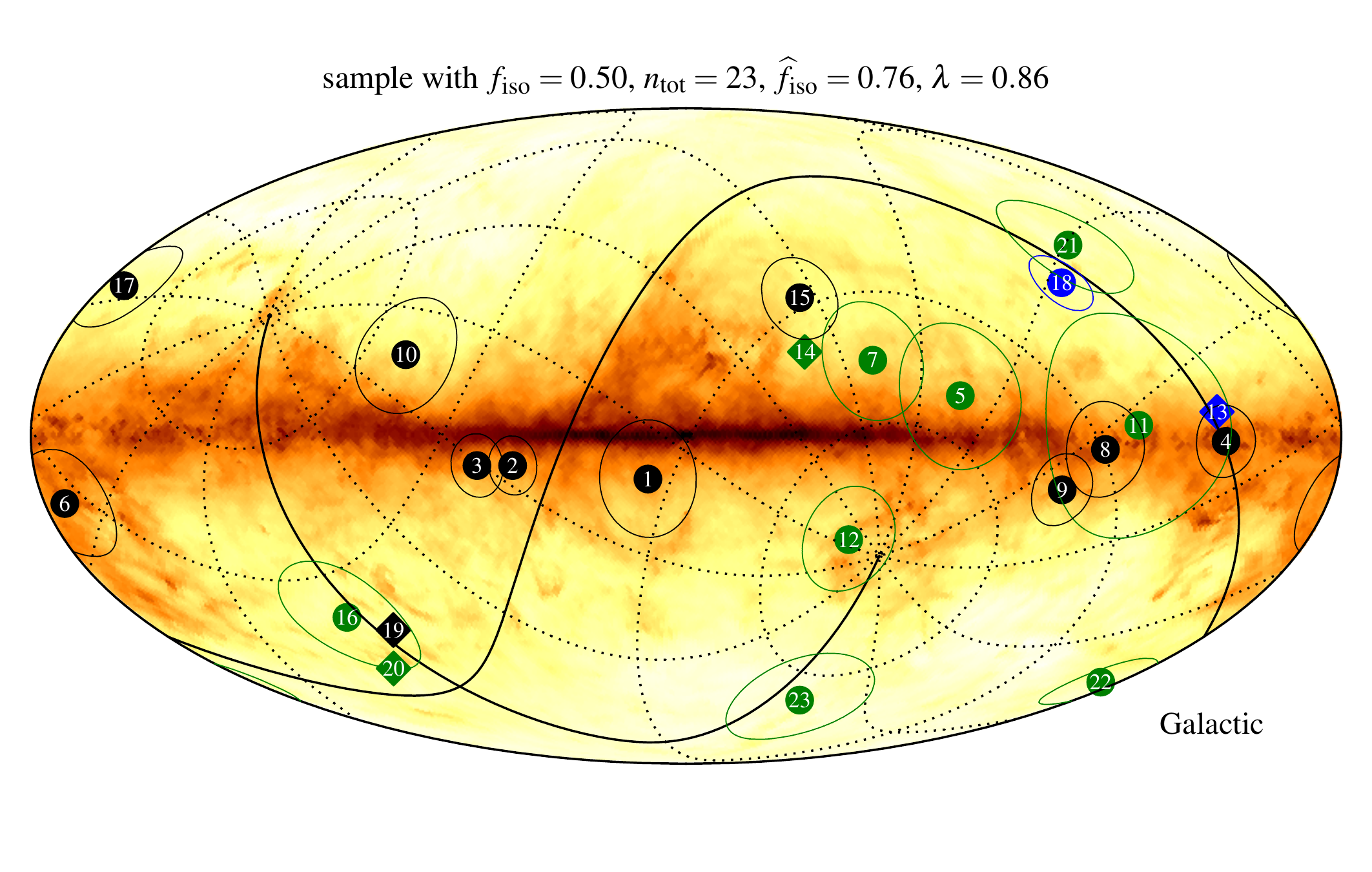}\hfill\includegraphics[width=0.5\textwidth]{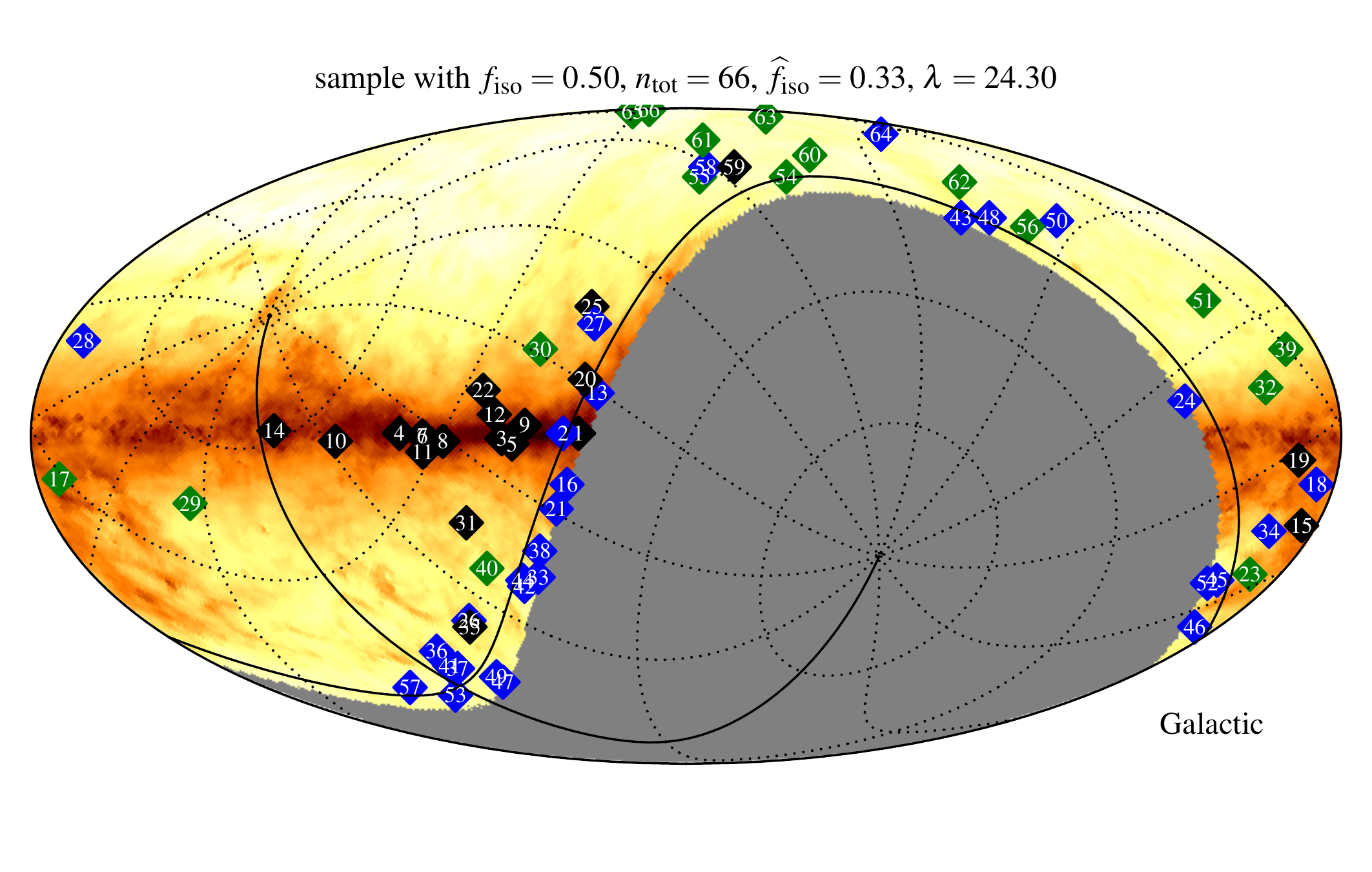}\\[-0.5cm]
\includegraphics[width=0.5\textwidth]{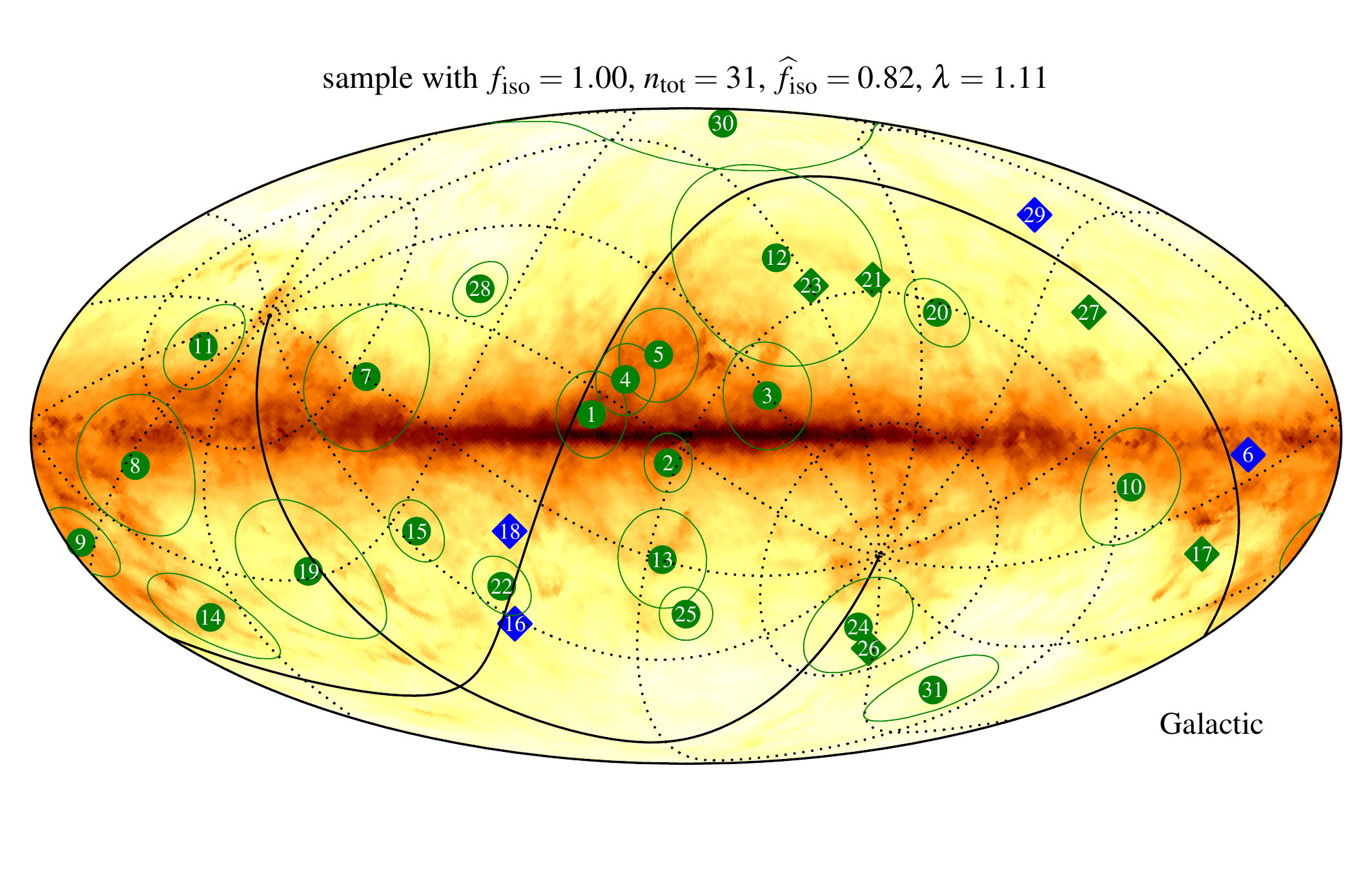}\hfill\includegraphics[width=0.5\textwidth]{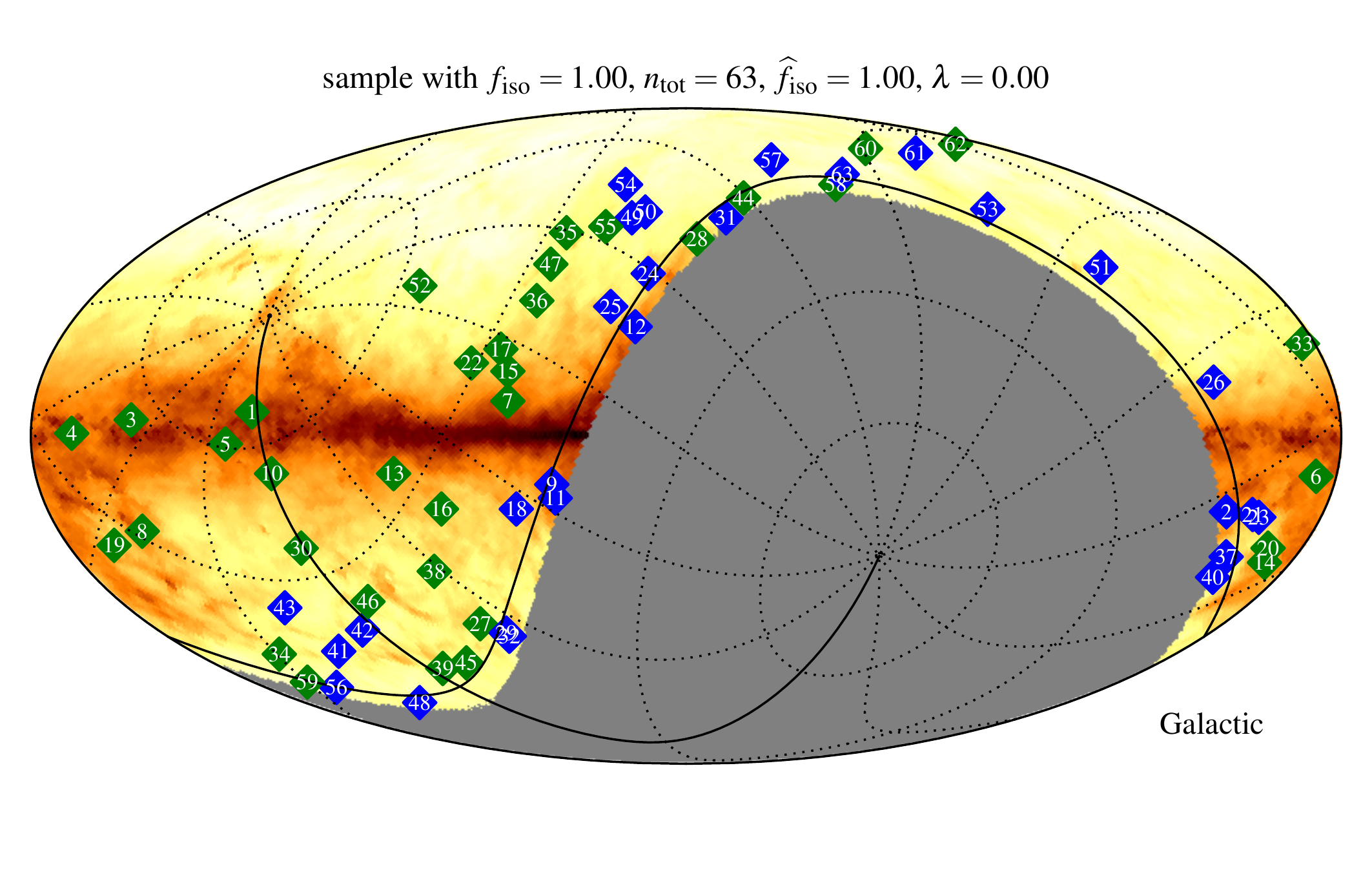}\\[-0.2cm]
\mini{\textwidth}{\small{\Large$\diamond/\circ$} : Galactic $\nu$ | \textcolor{Green4}{\Large$\diamond/\circ$} : isotropic $\nu$ | \textcolor{Blue1}{\Large$\diamond/\circ$} : atmospheric $\nu$ | \textcolor{Magenta2}{\Large$\diamond/\circ$} : atmospheric $\mu$}\\
\caption[]{Event samples with decreasing Galactic diffuse fraction $f_{\rm iso}$ of $0$ (top), $0.5$ (middle) and $1$ (bottom) for the HESE 3yr study with $E_{\rm dep}>60$~TeV (left panels) and a classical up-going $\nu_\mu$ search with $E_\mu>100$~TeV in the same period (right panels). In the sample maps the simulated events are composed of Galactic neutrinos (black), isotropic neutrinos (green), conventional atmospheric neutrinos (blue) and atmospheric muons (magenta). Tracks are indicated as diamonds and cascades as filled circles. The angular uncertainty (sampled from data) is indicated as circles. The numbers in all maps follow the relative signal strength of the events for a Galactic diffuse origin starting from strong to weak. Each map indicate total number of events $n_{\rm tot}$, the maximum likelihood ratio $\lambda$ and the corresponding maximum fraction $\widehat{f}_{\rm iso}$.
\label{samplemaps} 
}
\end{figure*}

Here, we want to improve on previous tests in several ways. Firstly, the likelihood ratio in our statistical test accounts for the intrinsic anisotropy of atmospheric backgrounds and Galactic emission as well as the expected anisotropy due to Earth absorption and detector acceptance. This intrinsic dependence on zenith angle does not cancel in the likelihood ratio and is not accounted for in the IceCube analysis. The likelihood ratio with respect to the background hypothesis (a full isotropic astrophysical signal) can be written in the form
\begin{equation}\label{eq:LH}
\frac{\mathcal{L}(f_{\rm iso})}{\mathcal{L}(1)} = \prod_{j}{\left[\frac{\mu^{\rm sig}_j(f_{\rm iso})+\mu_j^{\rm bgr}(f_{\rm iso})}{\mu_j^{\rm bgr}(1)}\right]}\,,
\end{equation}
where the signal and background expectation values $\mu^{\rm sig}_j$ and $\mu^{\rm bgr}_j$, respectively, for the $j$'th event depend on the fraction $f_{\rm iso}$ of the total number of expected events from the best-fit $E^{-2.3}$ spectrum that is due to an isotropic emission (see discussions in Appendices~\ref{app2} and \ref{app3}). The complement $f_{\rm Gal} = 1-f_{\rm iso}$ is the event fraction associated with Galactic sources. The maximum log-likelihood (LLH) ratio 
\begin{equation}\label{lambda}
\lambda = 2\ln \frac{\mathcal{L}(\widehat{f}_{\rm iso})}{\mathcal{L}(1)}\,.
\end{equation}
with maximum point $\widehat{f}_{\rm iso}$ is then defined as our test statistic (TS), which will be used to estimate the sensitivity of the IceCube Observatory for Galactic emission and the significance of emission in the HESE analysis and a classical up-going $\nu_\mu$ search.

\begin{figure*}[p]
\centering
\includegraphics[width=0.45\textwidth]{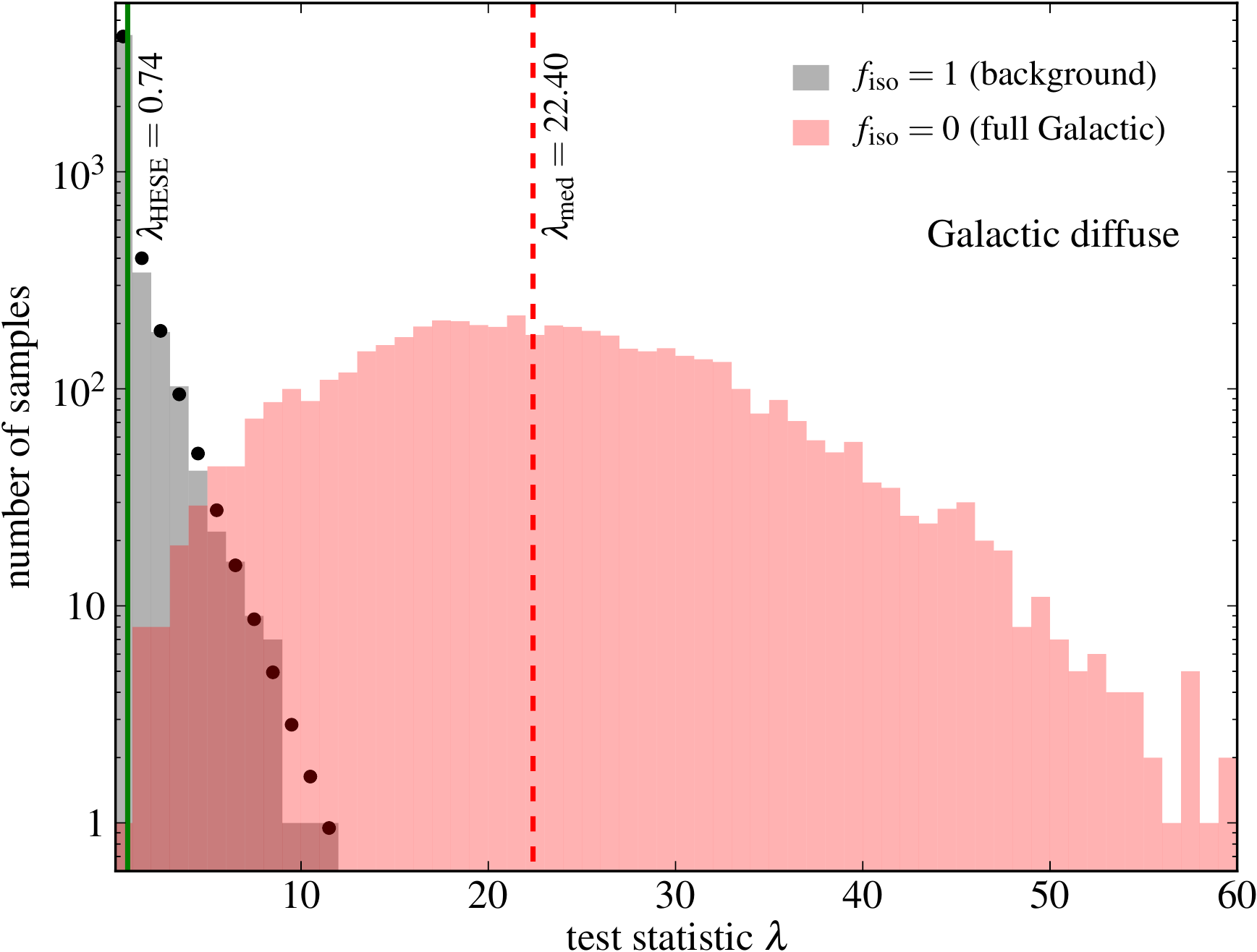}\hspace{0.5cm}\includegraphics[width=0.45\textwidth]{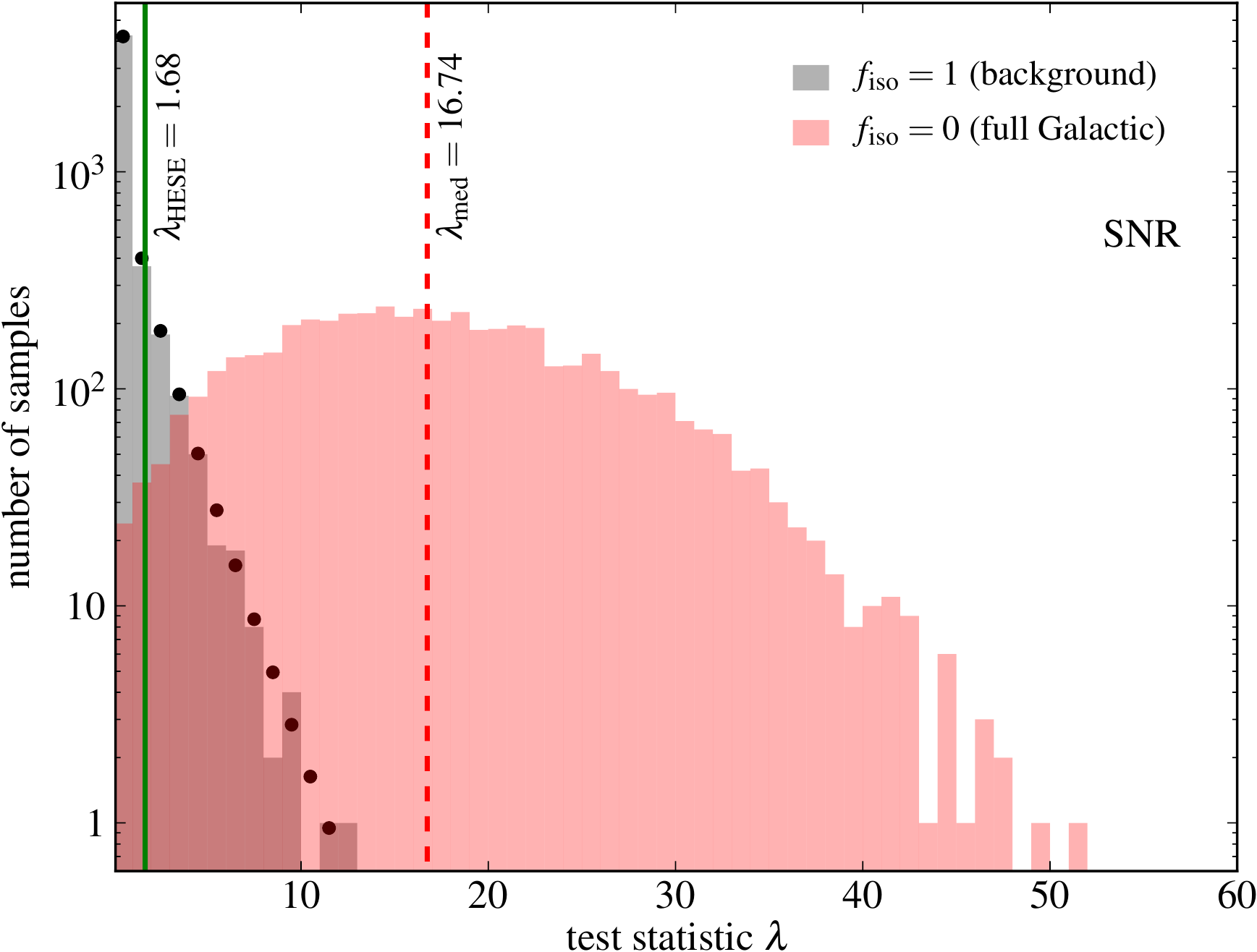}\\[0.5cm]
\includegraphics[width=0.45\textwidth]{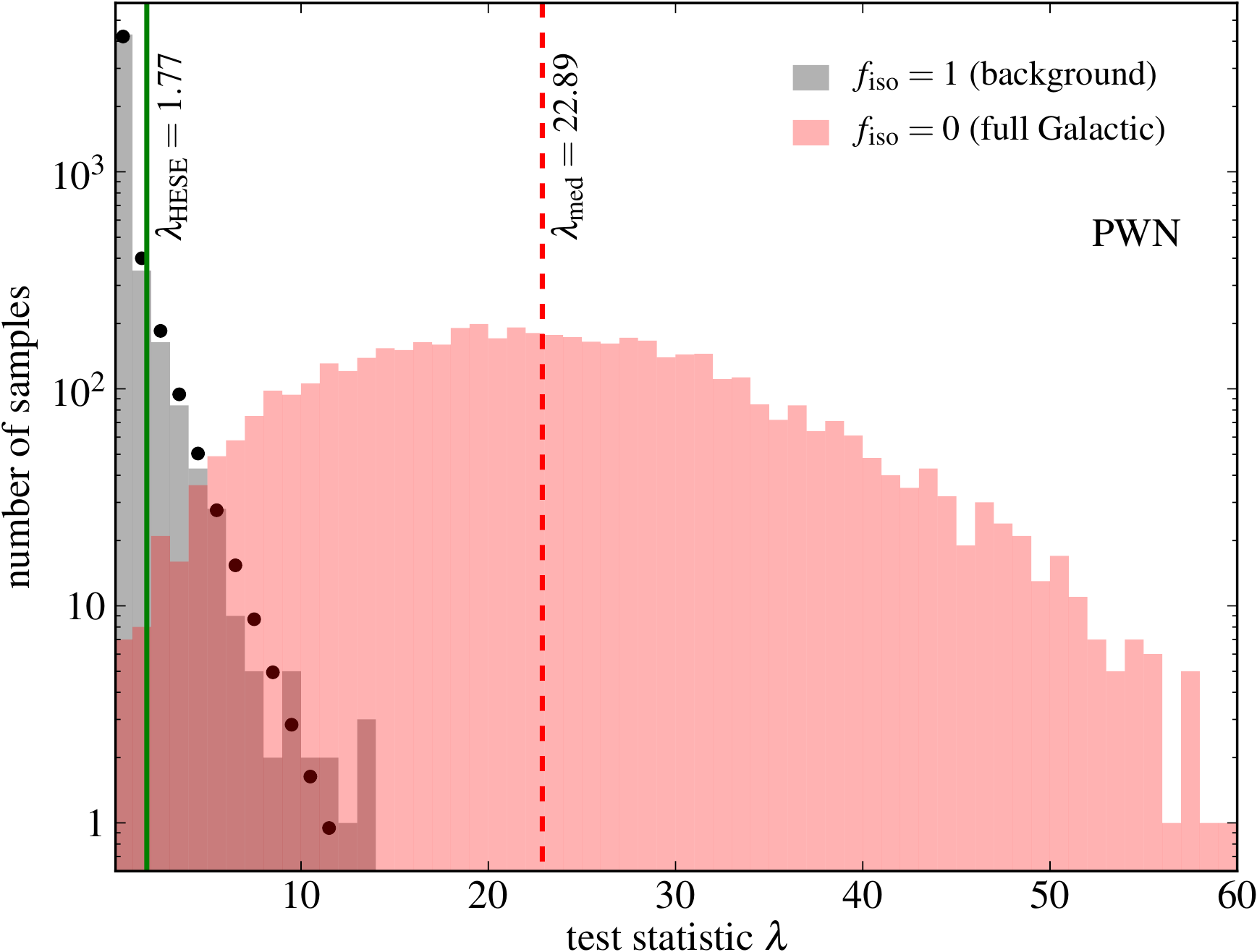}\hspace{0.5cm}\includegraphics[width=0.45\textwidth]{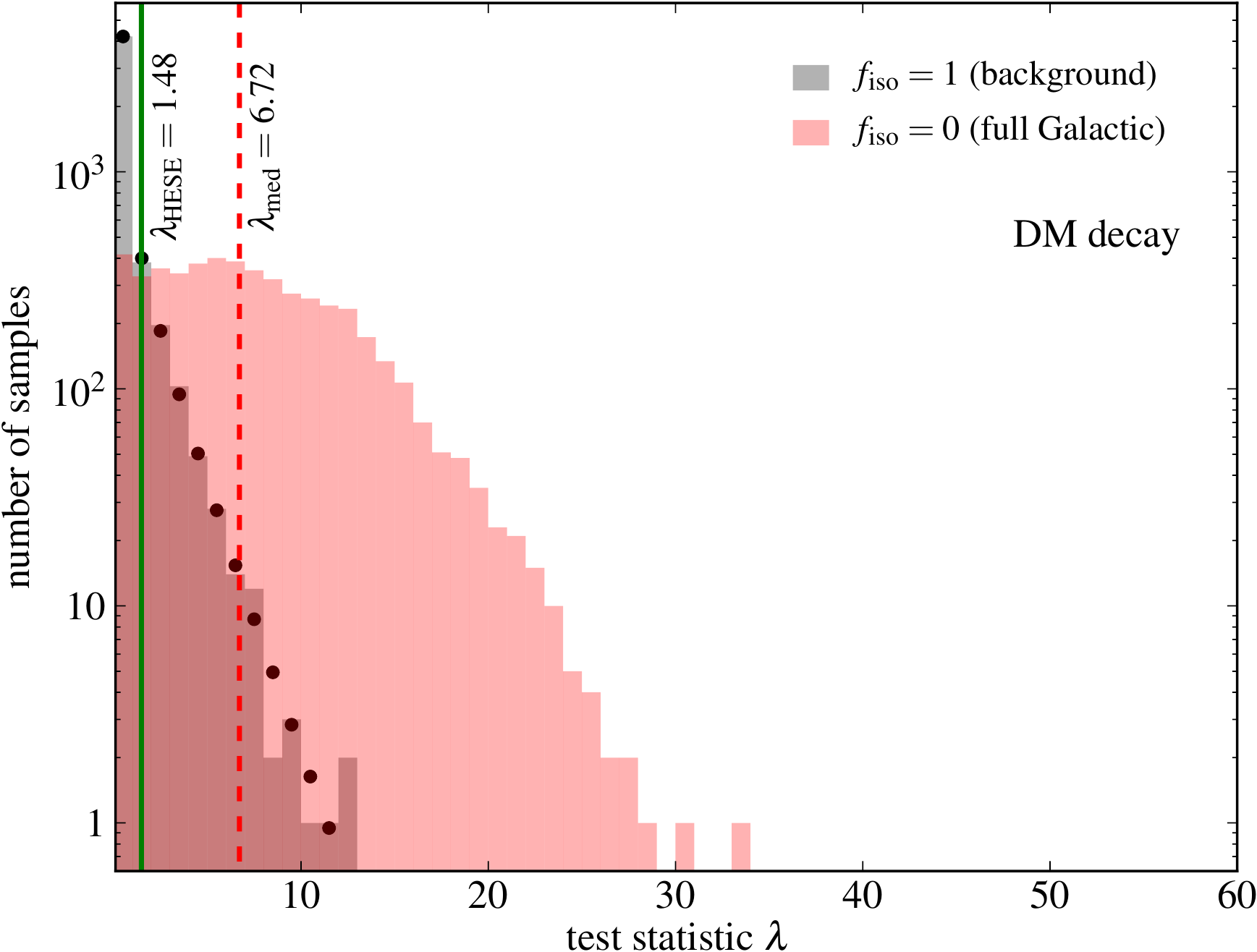}\\[0.5cm]
\includegraphics[width=0.45\textwidth]{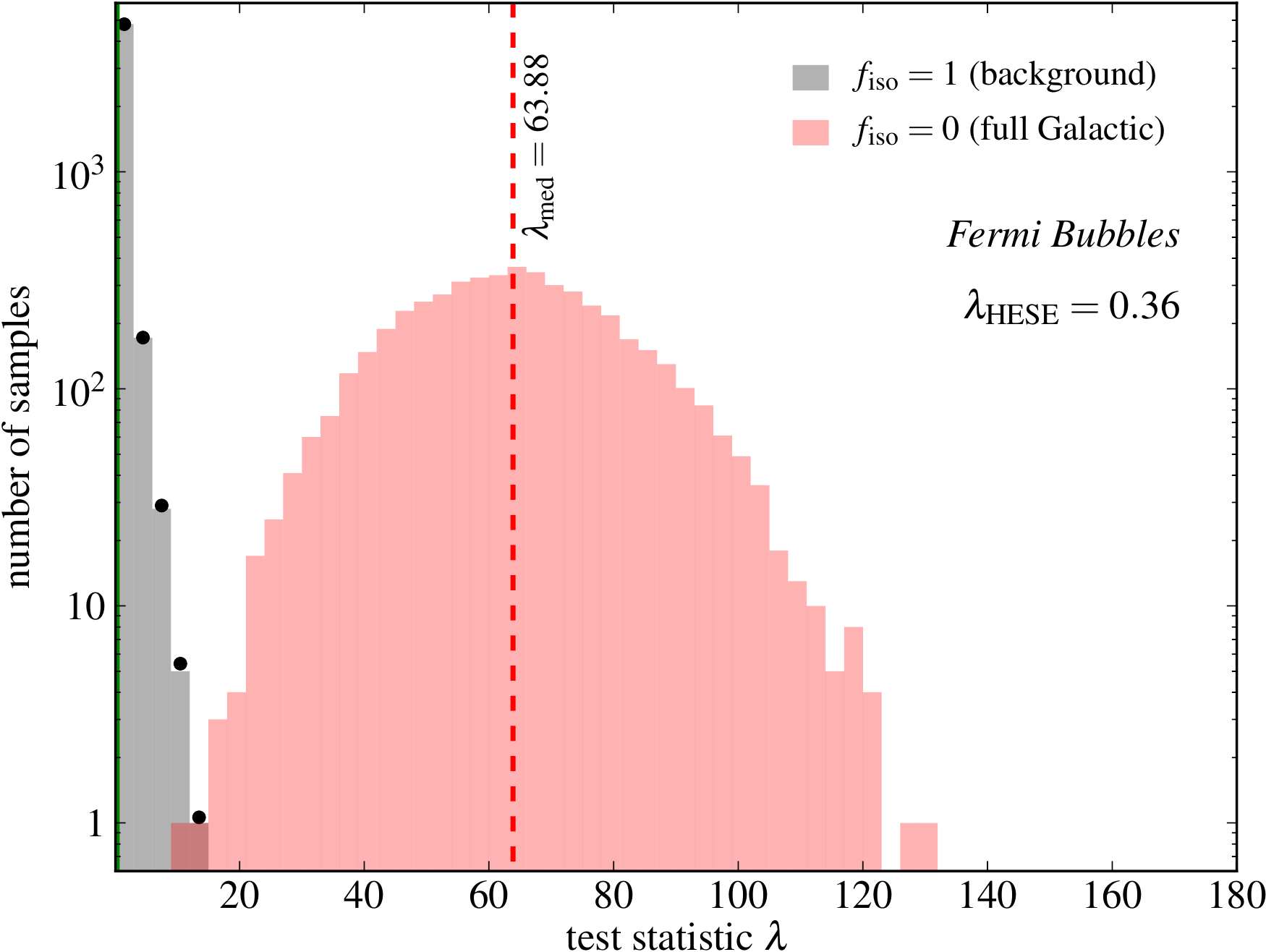}\hspace{0.5cm}\includegraphics[width=0.45\textwidth]{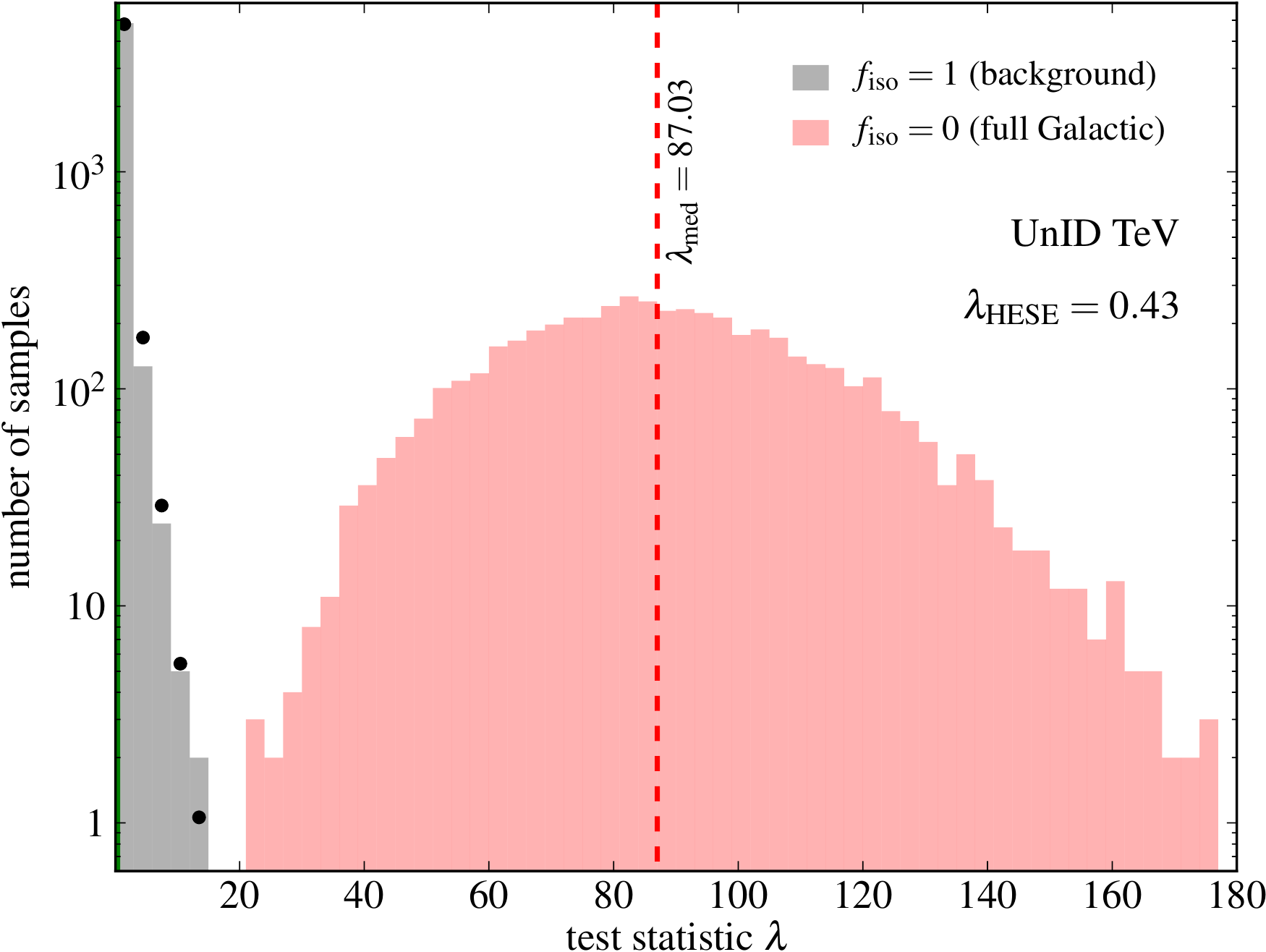}
\caption[]{Test Statistic (TS) distributions of 5000 Monte Carlo samples for the HESE 3yr study. For each model the histograms show background (gray) and 100\% Galactic emission (red). The vertical red dashed line shows the median TS of simulated Galactic emission and the vertical green solid line the observed value $\lambda_{\rm HESE}$. The black points indicate the bin-wise contribution of the expected distribution following $[\delta(\lambda)+\chi_1^2(\lambda)]/2$~\cite{Cowan:2010js}. 
}\label{TSdis}
\end{figure*}

For the sensitivity and significance studies of Galactic emission in the HESE data and the classical up-going $\nu_\mu$ data we generate event maps following the prescription outlined in Appendix~\ref{app2}. The left plots in Fig.~\ref{samplemaps} show examples of samples for the full signal case ($f_{\rm iso}=0$, top), a partial distribution ($f_{\rm iso}=0.5$, middle) and the background case ($f_{\rm iso}=1$, bottom). The title of the plots indicate the test statistics value $\lambda$ and the value $\widehat{f}_{\rm iso}$ at maximum. 

\begin{figure*}[t]
\includegraphics[height=0.38\linewidth]{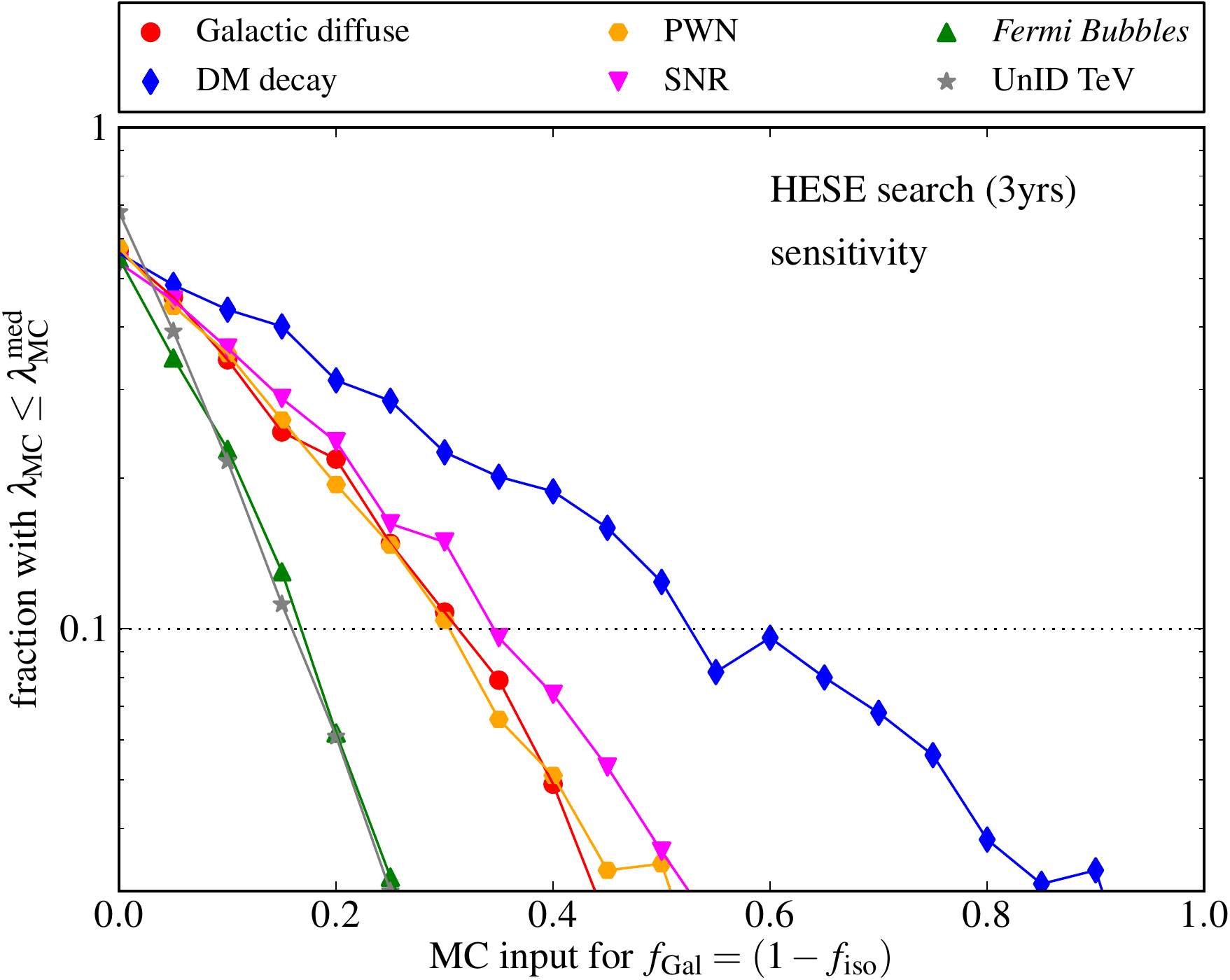}\hfill\includegraphics[height=0.38\linewidth]{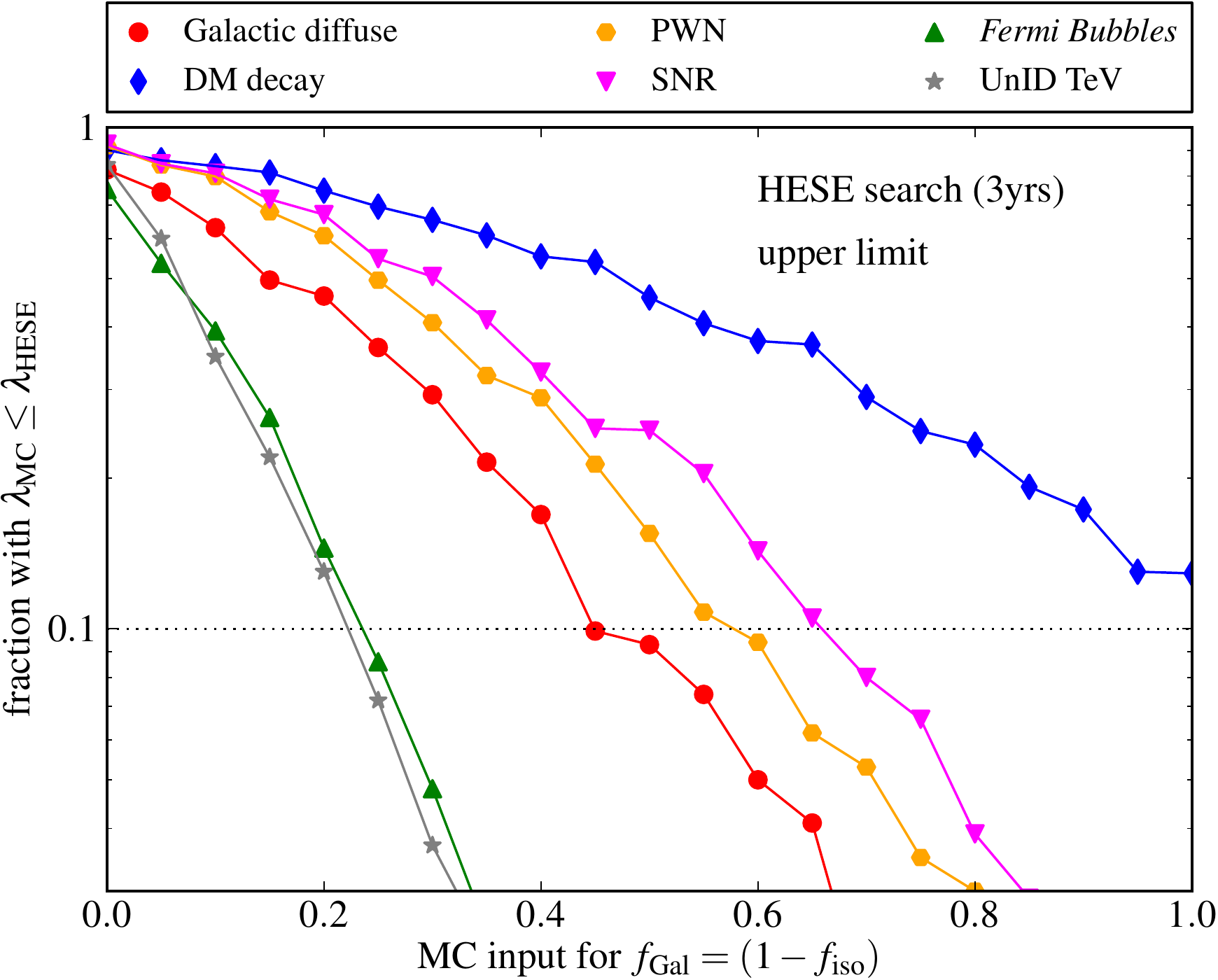}
\caption[]{Sensitivity (left panel) and 90\% upper limit (right panel) on the Galactic fraction $f_{\rm Gal} = 1-f_{\rm iso}$ for six different Galactic emission templates.}\label{fig:scan}
\end{figure*}

\begin{figure*}[t]
\includegraphics[height=0.38\linewidth]{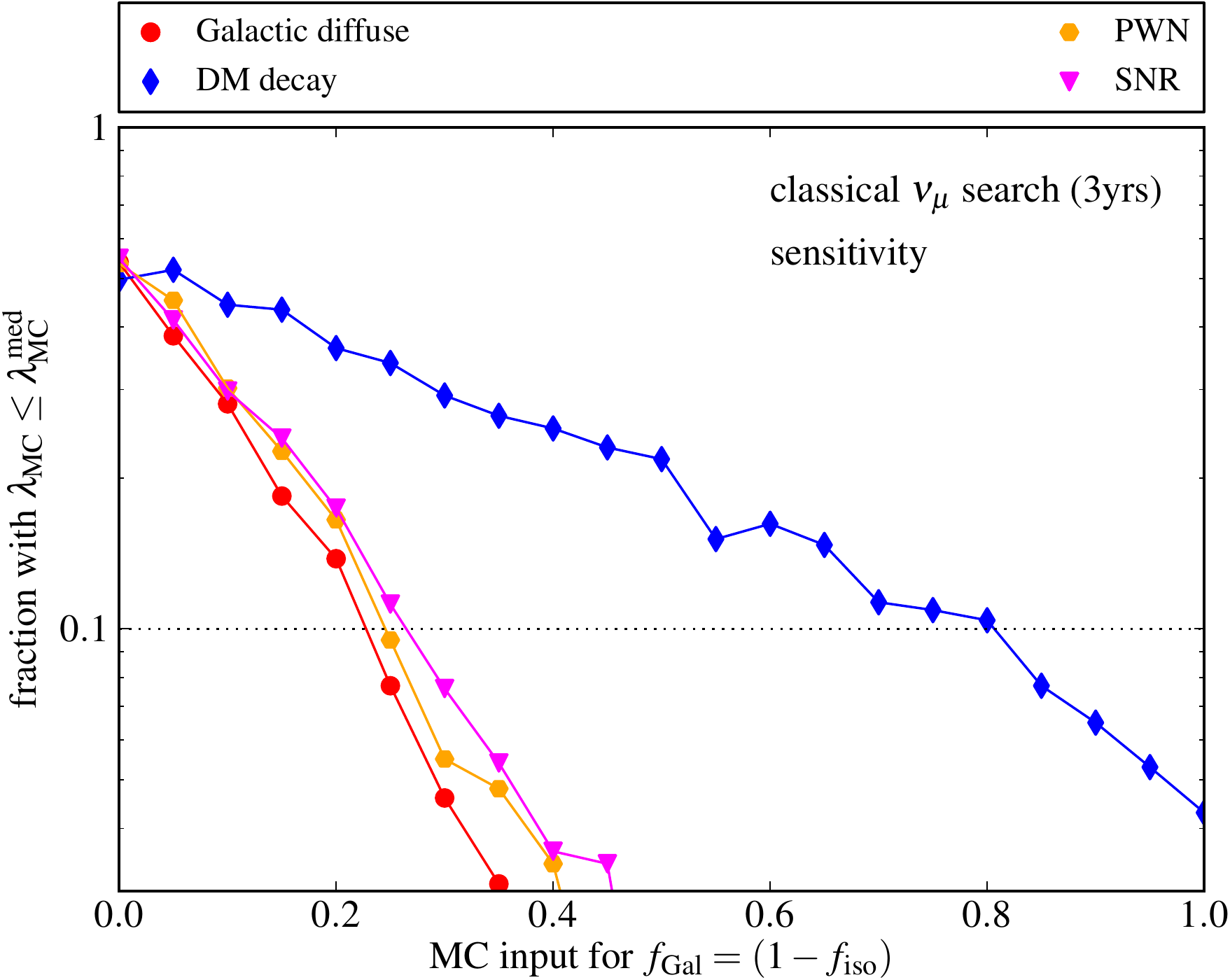}
\hfill\includegraphics[height=0.38\linewidth]{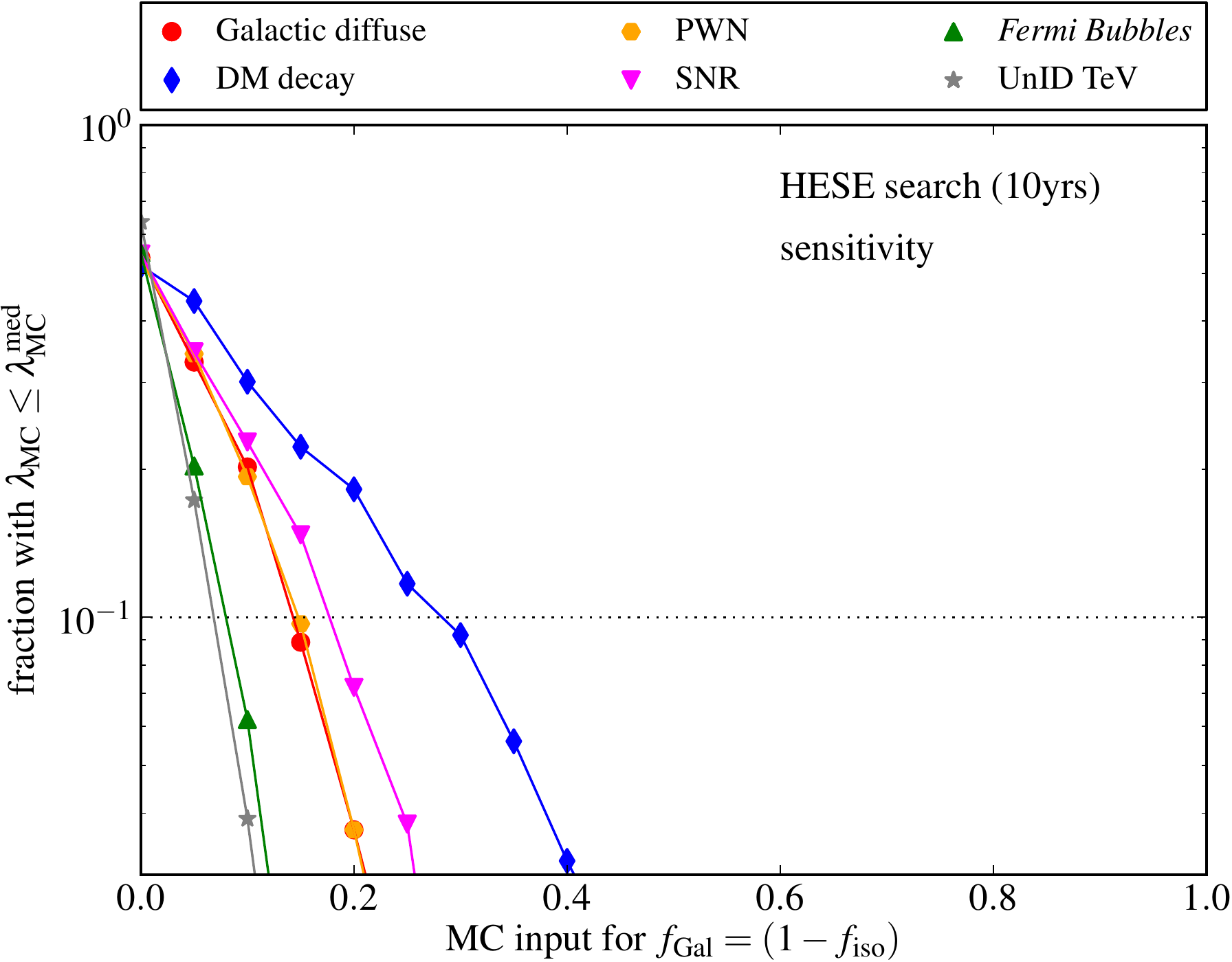}
\caption[]{Sensitivity of the search of up-going $\nu_\mu$ above muon energies of $100$~TeV in three years (left) and the sensitivity of the HESE search after 10 years of observation (right). In the case of the classical up-going $\nu_\mu$ search we do not include the templates of {\it Fermi Bubbles} and unidentified or dark TeV $\gamma$-ray sources, since the overlap with the Northern Hemisphere is small.}\label{fig:scan2}
\end{figure*}

Fig.~\ref{TSdis} shows the TS distribution of 5000 samples generated for the background case ($f_{\rm iso}=1$) and a full contribution ($f_{\rm iso}=0$). The test statistics of the HESE 3yr data are indicated as vertical green lines and summarized as the second column of Table~\ref{tab:results}. For all six emission models we find that the data are consistent with the background distributions, {\it i.e.}~an isotropic arrival distribution of the astrophysical signal. This is consistent with previous findings of the IceCube Collaboration~\cite{Aartsen:2014gkd}. The highest excess above background is obtained with the PWN distribution with a chance probability of $9$\%.

The medians of the TS distributions (vertical red dashed lines) are well separated from the background distribution in most cases. This indicates that the non-observation of significant Galactic anisotropies in the HESE 3yr data (vertical green line) is constraining the contribution of most Galactic emission scenarios. The 90\% sensitivity to a Galactic fraction is defined as the minimum isotropic fraction $f_{\rm iso}$ where 90\% of the signal samples have a TS larger than the median TS of the background distribution~\cite{Braun:2008bg}. For the likelihood defined via Eq.~(\ref{eq:LH}) the median for all six Galactic emission templates is 0. We estimate the sensitivity via a scan over $f_{\rm iso}$ with $\Delta f_{\rm iso} = 0.05$ steps and show our results in Fig.~\ref{fig:scan} and the sixth column of Table~\ref{tab:results}.

All scenarios considered in this paper have been suggested before to explain a part of the HESE data. Therefore our study does not correspond to a ``blind'' analysis and it is no surprise that all models show an up-fluctuation above background. The 90\% upper limit from this analysis corresponds to the maximum Galactic fraction where 90\% of the signal samples have a TS larger than the value observed in the data. We derive this upper limit via a scan over Galactic contributions $f_{\rm Gal}=1-f_{\rm iso}$ and show the result in the right panel of Fig.~\ref{fig:scan} and the fifth column of Table~\ref{tab:results}. Interestingly, for the case of Galactic dark matter decay we can not obtain a (non-trivial) upper limit on the Galactic fraction based on the HESE 3yr data and hence we leave the entry empty in Table~\ref{tab:results}.

With our method we can also estimate IceCube sensitivities for the various Galactic scenarios with increased statistics and with alternative up-going muon neutrino searches. For the classical up-going ($\theta_{\rm zen}>85^\circ$) $\nu_\mu +\bar{\nu}_\mu$ search we assume a total astrophysical event rate of 10 events per year and a total conventional atmospheric background rate of 10 events per year. This approximates the expected integrated event rate above muon energies of $100$~TeV~\cite{Karle_Arlington}, comparable to the cut of $60$~TeV in deposited energy used for the HESE search. The details of the expected signal and background distributions are described in Appendices~\ref{app2} and \ref{app3}. The right plots in Fig.~\ref{samplemaps} show examples of samples for the full signal case ($f_{\rm iso}=0$, top), a partial distribution ($f_{\rm iso}=0.5$, middle) and the background case ($f_{\rm iso}=1$, bottom) over a period of three years. Note that this approach is not sensitive to contributions which dominate in the Southern Hemisphere and therefore we leave out the {\it Fermi Bubbles} and UnID sources in this study.

The left plot in Fig.~\ref{fig:scan2} shows the result of the sensitivity scan for the muon neutrino search and the last column of Table~\ref{tab:results} summarizes the results. Note that the sensitivities are not much different compared to the HESE 3yr search for the four scenarios, although the muon search offers a better angular resolution of the order of $0.5^\circ$. The reason for this lies in the fact that the templates themselves correspond to extended emission such that the improved pointing precision does not matter that much. The right plot in Fig.~\ref{fig:scan2} shows the result of the sensitivity scan for 10 years of HESE observations and the sensitivity level is also summarized in the seventh column of Table~\ref{tab:results}. The sensitivity compared to the three years data improves by about a factor of two in this period.
\begin{table*}[t]
\centering
\renewcommand*{\arraystretch}{1.5}
\begin{tabular}{c|cccc|ccc}
\hline\hline
&\multicolumn{4}{c|}{HESE 3yr observation}&\multicolumn{3}{c}{sensitivity for $f_{\rm Gal}$${}^\star$}\\[0.2cm]
\mini{3cm}{template}&
\mini{1.2cm}{$\lambda$}&
\mini{1.5cm}{$p$-value${}^*$}&
\mini{1.2cm}{$\widehat{f}_{\rm Gal}$${}^\star$}&
\mini{1.2cm}{$f_{\rm Gal}^{90\%}$${}^\star$}&
\mini{1.5cm}{HESE 3\,yr}&
\mini{1.5cm}{HESE 10\,yr}&
\mini{1.5cm}{Northern $\nu_\mu$ 3\,yr}\\[0.2cm]
\hline
Galactic diffuse $\nu$ ${}^\#$
&0.74&0.19&0.19&0.50&
0.30&0.15&0.25\\
SNR~\cite{Case:1998qg}
&1.68&0.10&0.34&0.65&
0.35&0.20&0.30\\
PWN~\cite{Lorimer:2006qs}
&1.77&0.09&0.30&0.60&
0.30&0.15&0.25\\
DM decay~\cite{Graham:2006ae}
&1.48&0.11&0.46&--&
0.60&0.30&0.85\\
{\it Fermi Bubbles}~\cite{Fermi-LAT:2014sfa}
&0.36 &0.27&0.07&0.25&
0.20&0.10&--\\
UnID TeV~\cite{Fox:2013oza}
&0.43 &0.25&0.07&0.25&
0.20&0.10&--\\
\hline\hline
\end{tabular}
\flushleft \scriptsize ${}^\#$ \begin{minipage}[t]{0.95\linewidth}The emission template is using {\tt GALPROP}. We estimate the systematic uncertainty of $f_{\rm Gal}$ from the diffusion model to be at the level of $\pm10$\%.\end{minipage}
\vspace{-0.2cm}
\flushleft{\scriptsize ${}^*$ \begin{minipage}[t]{0.95\linewidth}The $p$-value is calculated from $\lambda$ assuming a background distribution $[\delta(\lambda)+\chi^2_1(\lambda)]/2$.\end{minipage} }
\vspace{-0.4cm}
\flushleft{\scriptsize ${}^\star$ \begin{minipage}[t]{0.95\linewidth}The Galactic fraction is defined as $f_{\rm Gal} = 1-f_{\rm iso}$.\end{minipage}}
\caption[]{Sensitivity and 90\% C.L.~lower limits of a Galactic fraction in the HESE data above 60~TeV. The first two columns shows the TS and maximum point $\widehat{n}_s$ using the IceCube approach via Eq.~(\ref{eq:LH}).}\label{tab:results}
\end{table*}

Note that our result is not a full replacement of an IceCube analysis. Several steps of this analysis can be improved, in particular the zenith and energy dependence of the events. We expect that a dedicated IceCube analysis will improve the sensitivity of the analysis by a factor of a few. In particular, for very high energy neutrinos the classical muon neutrino is also sensitive to emission in the Southern Hemisphere, although at a much lower level~\cite{Aartsen:2014cva}. A strong Galactic contribution can also alter the best-fit value of the astrophysical contribution which requires a simultaneous fit in the first place. 

%
\section{Discussion and Conclusions}\label{sec:conclusion}
We have studied the contributions of extended Galactic TeV-PeV neutrino emission sources in relation to the IceCube observations. A guaranteed contribution to Galactic emission is from CR propagations and interactions in the Galactic medium. We have studied the corresponding diffuse emission of gamma-rays and neutrinos with the numerical cosmic ray propagation code {\tt GALPROP}. In our calculations we have assumed that the locally observed CR flux corresponds to the steady-state solution of the diffusion-convection equation with a homogeneous and isotropic diffusion coefficient. We found that under these assumptions the expected Galactic diffuse neutrino emission that is consistent with $\gamma$-ray ({\it Fermi}-LAT) and CR data (KASCADE, KASCADE-Grande and CREAM) can only contribute 4\%-8\% above 60~TeV. We also derived the neutrino emission from the Galactic CR sources parametrized by the opacity of the sources to CR-gas interactions. We showed that the same CR model assumptions require an unusually large average opacity of $\langle \tau_{pp}\rangle \simeq 0.01$ to reach the flux level of IceCube integrated over the whole sky.

However, a more realistic study of CR injection and diffusion can alter these results, for instance time-dependent and inhomogeneous sources as well as inhomogeneous and/or anisotropic diffusion in the Galaxy. For instance, it was argued that remnant CRs from a locally enhanced CR emission episode could soften the CR spectra observed today compared to the global Galactic average~\cite{Neronov:2013lza}. Anisotropic diffusion in our local environment would similarly introduce local fluctuations from the Galactic average leading to, both, enhancements or reductions of the expected all-sky hadronic gamma-ray and neutrino emission~\cite{Effenberger:2012jc}. A similar effect has been discussed in the context of more realistic source distributions following the Galactic Arms~\cite{Kissmann:2014sia,Werner:2014sya}. The combined effect can reach to local enhancement and deficits of the CR spectrum at the level of $0.5-3.0$. Note, however, that these studies do not account for the {\it Fermi} gamma-ray data to constrain their model parameters and the actual range might be somewhat smaller. Finally, inhomogeneous diffusion could enhance the local concentration of CRs~\cite{Evoli:2008dv}. It was argued in Ref.~\cite{Gaggero:2015xza} that this would also increase the multi-TeV hadronic emission of the steady-state solution in the inner Galaxy by a factor of two in better agreement with {\it Fermi} and {\it Milagro} data.

An independent limit on Galactic neutrino emission that does not rely on the overall normalization scale of CR injection can be derived from the absence of correlations of neutrino arrival directions with Galactic emission templates. We studied the published IceCube data with respect to Galactic diffuse neutrino emission, emission from Galactic source populations like SNRs or PWNe, emission from the {\it Fermi Bubbles} or unidentified/dark {\rm TeV} $\gamma$-ray sources and a dark matter halo. We did not find a strong statistical excess in the 3yr HESE data above background expectations.  This conclusion is consistent with previous studies of the IceCube data. However, we showed that most of these scenarios are already constrained by the 3yr HESE data. The contribution at the 90\% confidence level is limited to $\lesssim50$\% for diffuse Galactic emission, $\lesssim65$\% for quasi-diffuse emission of neutrino sources and $\lesssim25$\% for extended diffuse emission from the {\it Fermi Bubbles} or unidentified TeV $\gamma$-ray sources. Interestingly, the emission of PeV dark matter decay in our Galactic center is presently unconstrained by the data.

We also estimated the sensitivity of IceCube for these emission scenarios with 10 years of HESE data and with 3 years of a classical up-going muon neutrino search. We estimate that the classical muon neutrino search in the same time period (three years) and energy range (muon energies above 100~TeV) has a similar sensitivity to extended diffuse Galactic emission in the Northern Hemisphere. The sensitivity of the HESE data after ten years of observation is expected to increase by a factor of two, not accounting for an up-fluctuation of the test statistic in the first three years.

Our analysis only relies on published IceCube data and can not be considered a full replacement of a dedicated IceCube study. We expect that our sensitivity estimates and upper limits are conservative and can be improved for the same data set using additional event information. In particular, our maximum likelihood analysis does not account for the energy dependence of individual events and can be improved by IceCube: if the Galactic emission spectrum responsible for the 60~TeV to 3~PeV data shows strong spectral variations with respect to the isotropic emission, a statistical separation between Galactic and extragalactic components can be much stronger. In particular, this is expected for dark matter decay, since narrow spectral features ({\it e.g.}~line features) of close-by Galactic dark matter are expected to be smoothed out for extragalactic contributions due to the redshift distribution.

Finally, let us conclude with a few additional remarks. The data-driven limit on Galactic diffuse neutrino emission is based on a template derived with {\tt GALPROP}, assuming an azimuthal distribution of CR sources and homogeneous isotropic diffusion. As noted earlier, the template does not depend on the absolute normalization, but it is expected to vary for models that predict {\it relative} enhancements of local CR distributions~\cite{Neronov:2013lza,Effenberger:2012jc,Werner:2014sya,Gaggero:2015xza}. This introduces a systematic uncertainty for the limit on Galactic diffuse neutrino emission. However, due to the dominance of cascade events in the IceCube data, which have a poor angular resolution of more than 10~degrees, we don't expect that the statistical analysis of the HESE data has a very strong dependence on the diffusion and source model.

One possibility to estimate the effect of local CR enhancements by source and diffusion models is by a comparison of the statistical outcome for the SNR and PWN templates. The relative ratio of the corresponding azimuthal source distributions (normalized to the same number of sources) is comparable to the variation of $0.5-3.0$ shown in detailed studies~\cite{Werner:2014sya} for non-azimuthal source distributions. The resulting templates shown in Fig.~\ref{fig:morphology} differ in the maximum-to-minimum ratio by a factor two. However, from Fig.~\ref{fig:scan} and Tab.~\ref{tab:results} we can see that the different statistical results for the SNR and PWN emission templates are within $\pm10$~\% and we take this as an estimate for the systematic error associated to the modeling of Galactic CR emission and diffusion.

A strong neutrino emission from our own Milky Way also implies an extragalactic contribution from similar galaxies. In Appendix~\ref{app4} we show that this extragalactic contribution is in general expected to be not as significant as the Galactic emission. The situation is different for the case of PeV dark matter decay: Here, the extragalactic flux is  comparable to the Galactic flux. In addition, any strong spectral features from particle physics, that are possible for the local Galactic emission in this scenario (see Appendix~\ref{appDM}), are expected to be smoothed out due to integration over red-shift.

Hadronic production of TeV-PeV neutrino sources will inevitably predict TeV-PeV $\gamma$-rays~\cite{Ahlers:2013xia}, which can be observed by large-scale $\gamma$-ray telescopes like HAWC~\cite{Abeysekara:2013tza,Abeysekara:2014ffg,Lunardini:2015laa} or CTA~\cite{Doro:2012xx,Pierre:2014tra}. In particular, the observation of PeV $\gamma$-rays would correspond to a {\it smoking gun} of Galactic contributions due to the small absorption length of about 10~kpc via $e^+e^-$ production in the CMB~\cite{Prodanovic:2006bq,Gupta:2013xfa,Ahlers:2013xia,Anchordoqui:2013qsi}. For $pp$ interactions, the relation between the $\gamma$-ray and neutrino fluxes is shown in Fig.~\ref{fig:flux-ratio}, which has small dependence on the primary CR index. For decaying dark matter models, a wide range of dark matter models predict the associated $\gamma$-ray flux above 10\% of the observed IceCube neutrino flux as shown in~Fig.~\ref{fig:DM-spectrum-ratio} in Appendix~\ref{appDM}. This is within the reach of HAWC-100 or HAWC-300 with one year of data~\cite{Abeysekara:2013tza}.

In our study we have simulated Galactic and extragalactic (isotropic) event distributions assuming an equal neutrino flavor ratio with $(f_e : f_\mu : f_\tau )_\oplus \approx (1:1:1)_\oplus$ which corresponds to an initial flavor combination $(1:2:0)_{\rm s}$ (pion decay) at the astrophysical sources, after flavor oscillations~\cite{Learned:1994wg,Athar:2000yw}. This is consistent with IceCube observations~\cite{Aartsen:2014gkd,Aartsen:2015ivb,Palomares-Ruiz:2015mka,Palladino:2015zua}. Future observations could be used to distinguish different explanations for the observed IceCube events~\cite{Beacom:2002vi,Majumdar:2006px,Baerwald:2012kc,Laha:2013lka,Barger:2014iua}, in particular $(0:1:0)_{\rm s}$ (muon-damped pion decay),  $(1:0:0)_{\rm s}$ (neutron decay) and  $(0:0:1)_{\rm s}$ (flavor-dependent dark matter decay). This fact makes our morphology studies of prime interest.

\vspace{3mm}
\noindent{\bf Acknowledgments:} We thank Luis Anchordoqui, Dario Grasso, Gu{\dh}laugur J\'ohannesson, Claudio Kopper and Kohta Murase for informative discussions. This work is supported by the U. S. Department of Energy under the contract DE-FG-02-95ER40896 and by the National Science Foundation under grants OPP-0236449 and PHY-0236449.

\begin{appendix}

\section{Galactic Diffuse Gamma-Ray Emission}\label{app1}

The recent measurement of the diffuse Galactic $\gamma$-ray emission by {\it Fermi}-LAT~\cite{FermiLAT:2012aa} covers an energy range from 200~MeV to 100 GeV. While we are interested in neutrino energies above 1 TeV, we need to have a specific model to perform an extrapolation. Similar to the model used by the {\it Fermi} Collaboration, we use the {\tt GALPROP} code~\cite{Moskalenko:1997gh,Strong:1998fr,Strong:2004de} to calculate the $\gamma$-ray and neutrino spectra as well as their morphology. 

\begin{figure*}\centering
\includegraphics[width=0.5\textwidth,clip=true,viewport= 10 0 540 400]{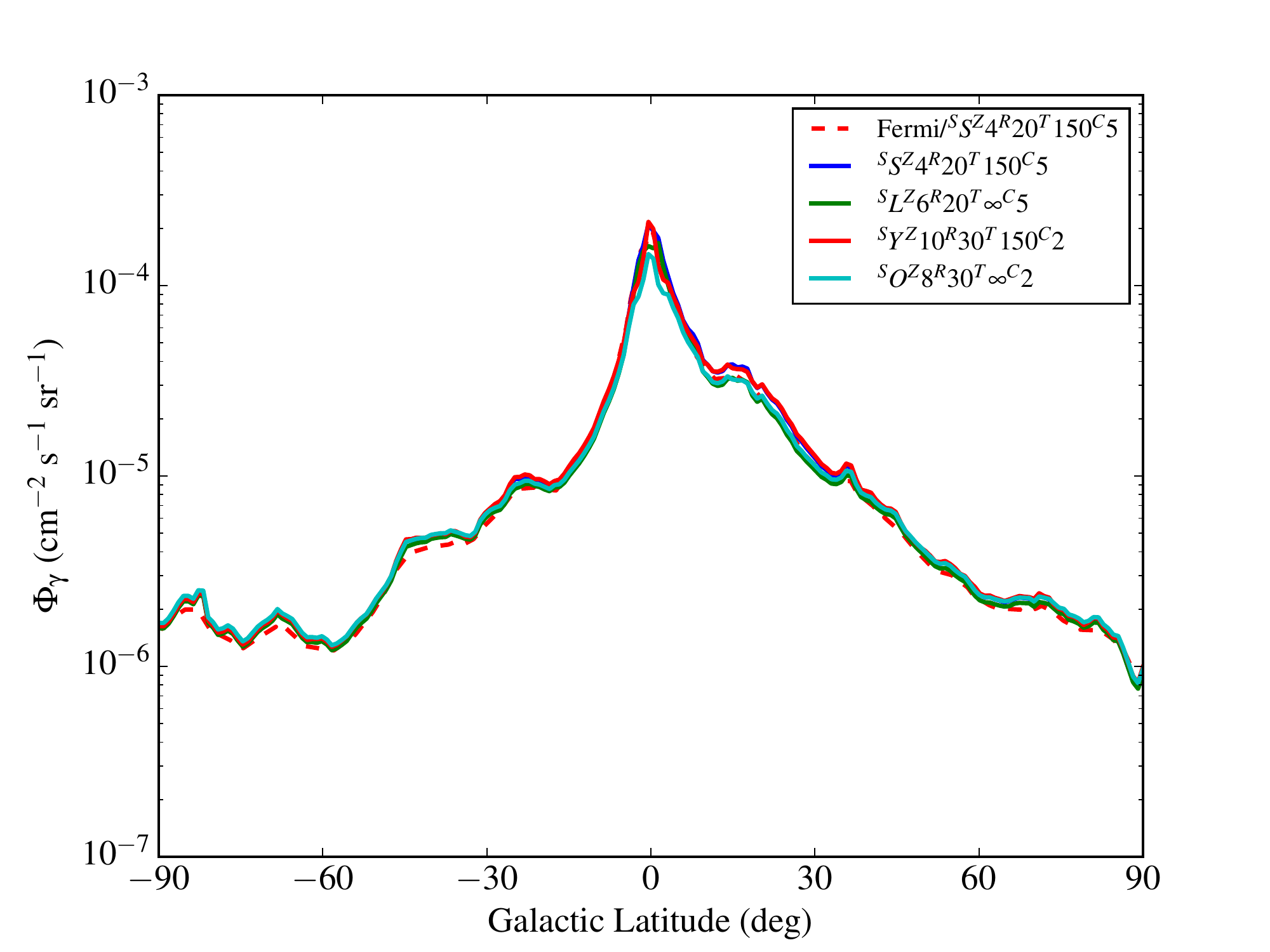}\hfill
\includegraphics[width=0.5\textwidth,clip=true,viewport= 10 0 540 400]{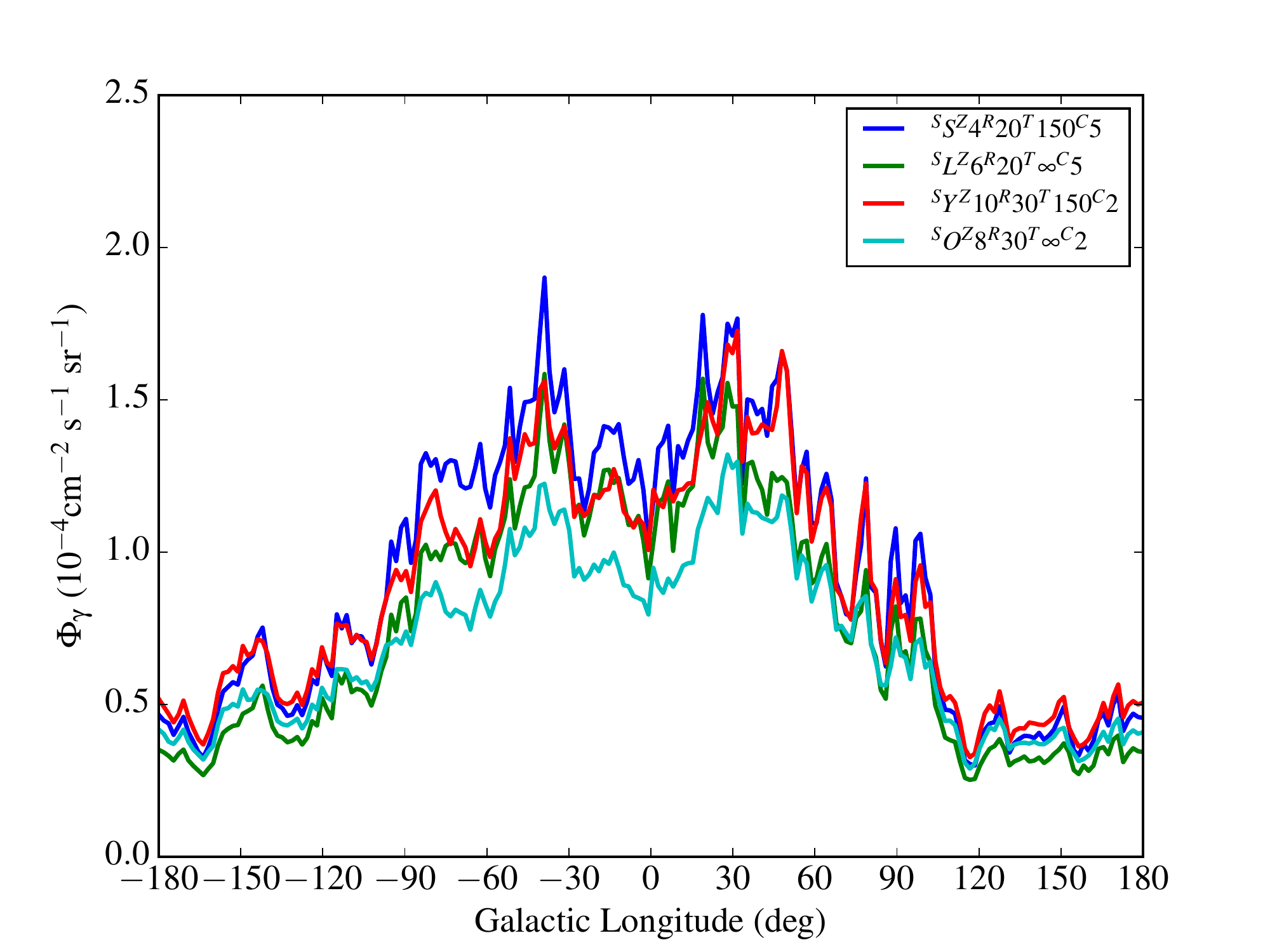}
\caption{{\bf Left panel:} a comparison of diffuse $\gamma$-ray distributions in Galactic latitude in the inner part of the galaxy with $-30^\circ\leq l \leq 30^\circ$ and $200~\mbox{MeV} \leq E_\gamma \leq 1.6$~GeV for four different {\tt GALPROP} models:  $^S S ^Z 4 ^R 20 ^T 150 ^C5$, $^S L ^Z 6 ^R 20 ^T \infty ^C5$, $^S Y ^Z 10 ^R 30 ^T 150 ^C2$ and $^S O ^Z 8 ^R 30 ^T \infty ^C2$. The current  black and solid line is from our simulation. The red and dashed line is the H{\small I} gas part from {\it Fermi}-LAT (the top panel of Fig.~20 in~\cite{FermiLAT:2012aa}). The energy range is from 200 MeV to 1.6 GeV. {\bf Right panel:} a comparison of diffuse $\gamma$-ray distributions in longitudinal latitude for $-5^\circ \leq b \leq 5^\circ$ and $200~\mbox{MeV} \leq E_\gamma \leq 1.6$~GeV.
\label{fig:pi-zero-latitude} }
\end{figure*}

As a starting point, we first reproduce the $\pi^0$ background of the H{\small I} gas part in Ref.~\cite{FermiLAT:2012aa}. Our corresponding results are shown in Fig.~\ref{fig:pi-zero-latitude}. Note that for the numerical calculations in this paper we obtain the total Galactic diffuse neutrino flux from both the H{\small I} and molecular gas components. Fig.~\ref{fig:pi-zero-latitude} shows the spatial distributions of $\gamma$-ray flux for four different {\tt GALPROP} models:  $^S S ^Z 4 ^R 20 ^T 150 ^C5$, $^S L ^Z 6 ^R 20 ^T \infty ^C5$, $^S Y ^Z 10 ^R 30 ^T 150 ^C2$ and $^S O ^Z 8 ^R 30 ^T \infty ^C2$. We use the same short hand notation $^S X ^Z z_h ^R R_h ^T T_S ^Cc$  as {\it Fermi}-LAT~\cite{FermiLAT:2012aa} where $X$ is the first letter of the CR source distribution (S: SNR~\cite{Case:1998qg}, L: Lorimer~\cite{Lorimer:2006qs}, Y: Yusifov~\cite{Yusifov:2004fr}, O: OB stars~\cite{Bronfman:2000tw}); $z_h$ and $R_h$ are the vertical and radial size of the diffusion region given in units of kpc; $T_S$ is the spin temperature in units of Kelvin; $c$ is the $E(B-V)$ magnitude cut accounting for high-extinction regions in the determination of the gas-to-dust ratio. As one can see from Fig.~\ref{fig:pi-zero-latitude}, our simulated results can match those from {\it Fermi}-LAT very well.


We use {\tt GALPROPv54\_r2504}~\cite{GALPROPv54} to model the propagation of the CRs in the Galaxy and calculate the photon and neutrino diffused emission from the
interactions between the CRs and the interstellar medium. For the
neutrino flux, we use the energy spectrum and the production cross section
calculated in Ref.~\cite{Huang:2007wk}. To be consistent and for the $\gamma$-ray flux calculation, we also use the photon energy spectrum and the production cross section in Ref.~\cite{Huang:2007wk}. Our study requires the  calculation of Galactic CRs above 100~TeV, slightly beyond the recommended energy range of the {\tt GALPROP} code. However, since the morphology of secondary neutrino and photon fluxes only depend on the gas target distribution and since we account for uncertainties of the CR spectrum in the {\it knee} region the additional systematic uncertainties from this extrapolation is not expected to be dominant.

\section{Neutrino and Photon Spectra from Dark Matter Decay}\label{appDM}

\begin{figure*}[p]\centering
\includegraphics[width=0.5\textwidth,clip=true,viewport= 10 0 540 420]{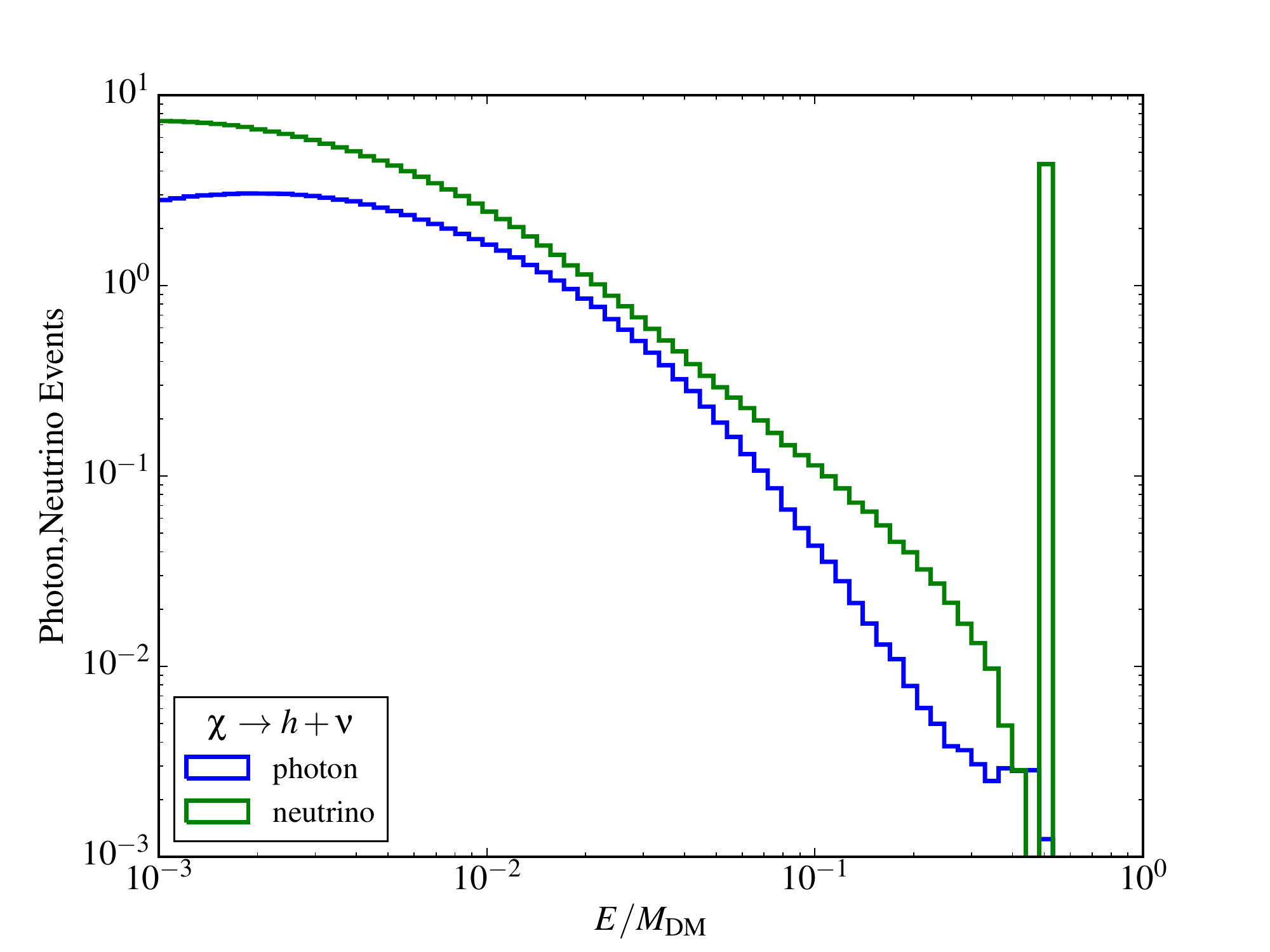}\hfill
\includegraphics[width=0.5\textwidth,clip=true,viewport= 10 0 540 420]
{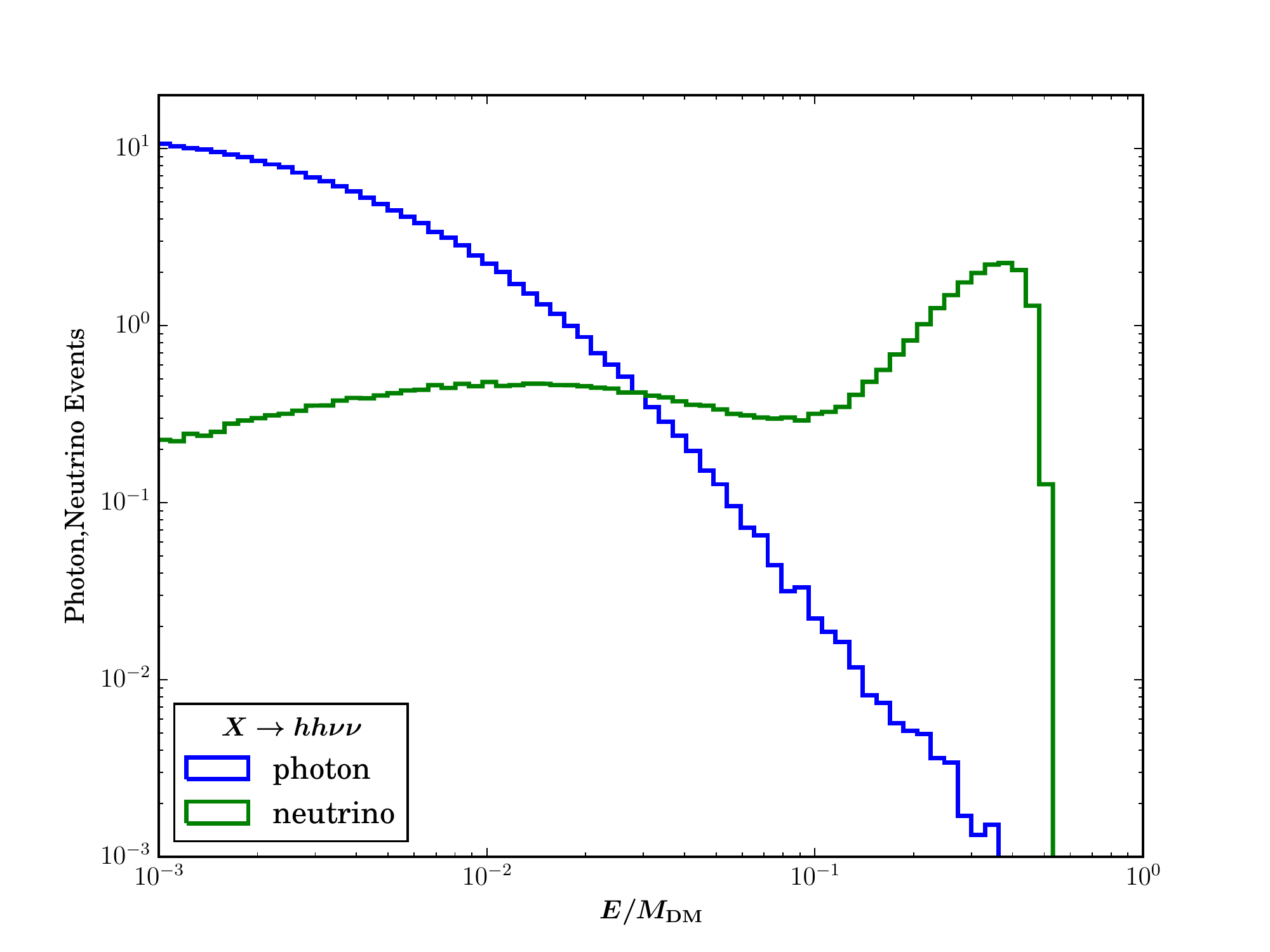}
\caption{A comparison of the photon and neutrino spectra for two representative decaying dark matter models: $\chi \rightarrow h+\nu$ (left panel) and $X \rightarrow 2h + 2\nu$ (right panel). The bin size is chosen to be 10\% of the corresponding energy.
\label{fig:DM-photon} }
\vspace{0.5cm}
\includegraphics[width=0.6\textwidth,clip=true,viewport= 10 0 540 420]{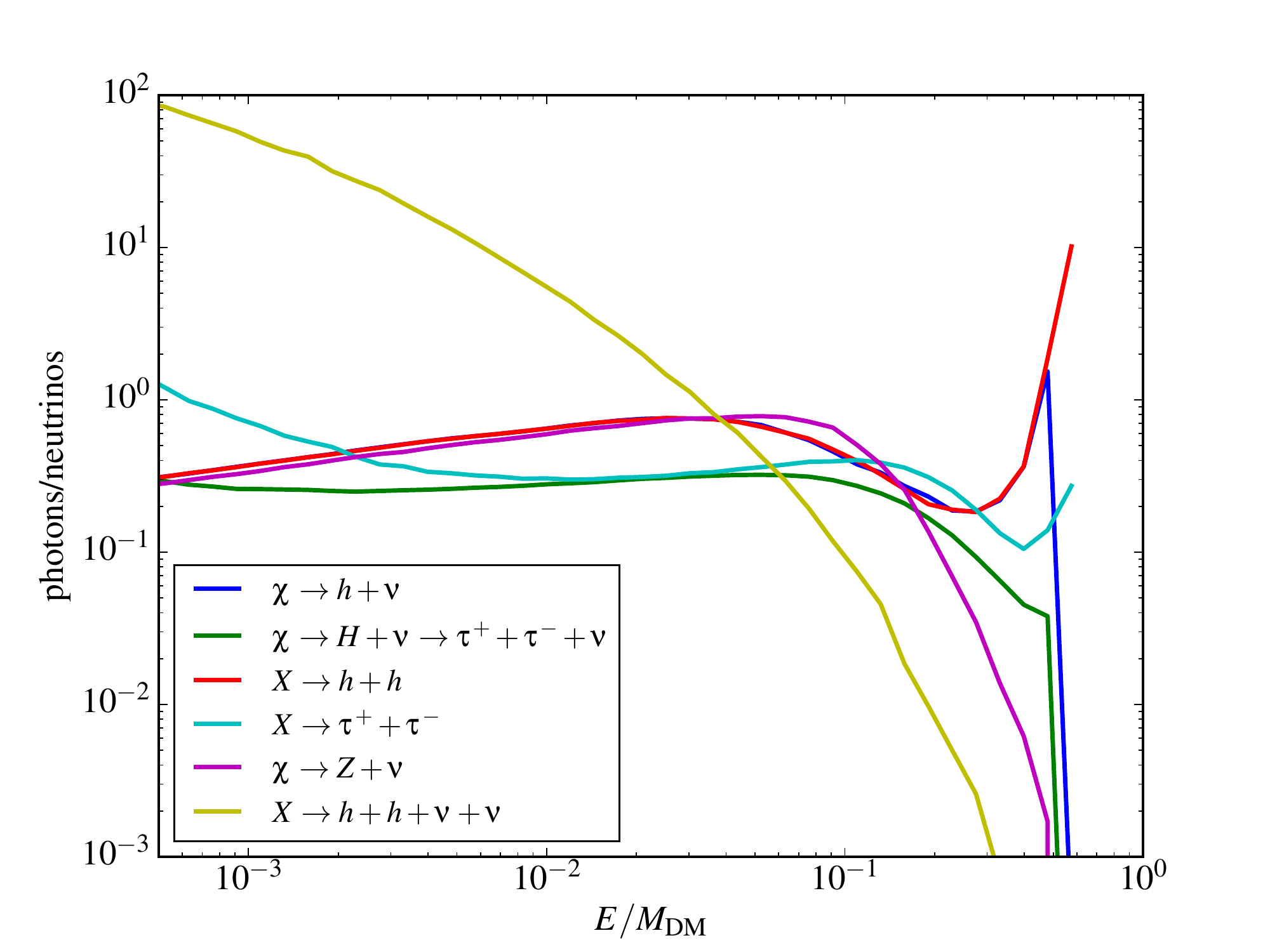}
\caption{The ratio of photon over neutrino fluxes for different dark matter decay modes.
\label{fig:DM-spectrum-ratio} }
\end{figure*}

For decaying dark matter models, several models of the energy spectra can provide a reasonable fit to the IceCube observed neutrino spectrum~\cite{Bai:2013nga}. As an example, we choose two representative models of a DM fermion $\chi$ and boson $X$ with coupling $\chi H \bar{L}$ and $X (H \bar{L})^2$, respectively, before electroweak symmetry breaking. The former operator has the dark matter decay as $\chi \rightarrow h+ \nu$, while the latter operator has the main decay mode of $X \rightarrow 2h + 2\nu$ (see also Ref.~\cite{Dudas:2014bca} for dark matter four-body decay). We show a comparison of photon and neutrino spectra for the two representative dark matter models in Fig.~\ref{fig:DM-photon} with the energy bin size to be 10\% of the corresponding energy. Compared to the observed IceCube neutrino flux, the associated $\gamma$-ray flux can be 10\% or above for a wide range of dark matter models. For more dark matter models, we show the ratios of photon over neutrino fluxes for different dark matter decay modes in Fig.~\ref{fig:DM-spectrum-ratio}.

One could have a neutrino-philic model with suppressed $\gamma$-ray fluxes, although it may not provide a good fit to the IceCube observed spectrum. One example is to have the dark matter particle to be the electric-neutral component of, $X$, $3_1$ under $SU(2)_L\times U(1)_Y$. The coupling like $\bar{L}_L^c X L_L$ can induce a  two-body decay of the dark matter particle $X^0$ to two neutrinos~\cite{Feldstein:2013kka}. The suppressed three-body decay like $X^0 \rightarrow \nu + e^+ + W^-$ can generate a small $\gamma$-ray flux.

\section{Background and Signal Sampling}\label{app2}

The background samples of the high-energy starting event (HESE) analysis are produced in the following way. The expected zenith distribution of isotropic signal events and atmospheric backgrounds above a deposited energy of $60$~TeV is shown in Fig.~5 of the supplementary material of the official IceCube publication\cite{Aartsen:2014gkd}. In each of the ten declination bands (equivalent to zenith bands, since $\theta_{\rm zen} = \pi/2+\delta$) we generate the expected number of events $\mu$ for conventional atmospheric neutrinos (``atmo $\nu$''), atmospheric muons (``atmo $\mu$'') and the best-fit isotropic astrophysical signal (``astro'') following an $E^{-2.3}$ spectrum. The zenith angle $\theta$ of each event is uniformly sampled over the $\cos\theta$ range of the particular bin and the right ascension angle is randomly chosen in $[0,2\pi)$. 

For each of the three subsets we generate track ($\diamond$) and cascade ($\circ$) events with a track fraction $x_\diamond = 0.9$ for atmospheric muons, $x_\diamond = 0.71$ for conventional atmospheric neutrinos and $x_\diamond = 0.19$ for the astrophysical signal following the values of Table IV of Ref.~\cite{Aartsen:2014gkd}. The fraction of cascades is denoted by $x_\circ = 1-x_\diamond$. The median angular error $\alpha_{50\%}$ of each track and cascade is randomly selected from the actual values reconstructed for the 4 tracks and 16 cascades in the data set above $60$~TeV (Tab.~IV in Ref.~\cite{Aartsen:2014gkd}). The background events are then re-sampled from their original distribution following a von-Mises-distribution. The angular distance $\alpha$ of events from the true position is sampled as $\cos\alpha = 1+ \sigma^2\ln(1 - x(1-\exp(-2/\sigma^2))$, where $x$ is a random number in $[0,1)$ and the azimuthal direction of the scatter is a random angle in $[0,2\pi)$. The angular uncertainty $\sigma$ is related to the median provided in Table I in Ref.~\cite{Aartsen:2014gkd} via $\cos\alpha_{50\%} = 1+ \sigma^2\ln(1 - 0.5(1-\exp(-2/\sigma^2))$.

The signal samples are generated under the assumption that a fraction $1-f_{\rm iso}$ of the best-fit astrophysical contribution is due to a Galactic contribution. The first step of the sample generation is the same as for the background accept that the isotropic astrophysical expectation is re-scaled in each zenith bin to a fraction $f_{\rm iso}$. For the Galactic distribution we sample events from the Galactic signal map $w_{\rm signal}$ which is re-weighted according to the zenith distribution of an isotropic component in Fig.~5 of Ref.~\cite{Aartsen:2014gkd} and then normalized to the expected Galactic event fraction $1-f_{\rm iso}$. The distribution between tracks and cascades and the res-sampling of positions in the sky via the reconstruction uncertainty is the same as in the case of the isotropic component. 

For the classical $\nu_\mu$ search we are looking for events below zenith angles of $85^\circ$ to avoid large backgrounds from atmospheric muons. The only background that we consider in this case is atmospheric $\nu_\mu$. As a benchmark value we expect 10 isotropic astrophysical events in one year with a muon energy larger than about $100$~TeV. This is comparable with the $60$~TeV energy cut used in the HESE analysis. During the same period we expect 10 conventional atmospheric $\nu_\mu$ events. Signal and background events are exponentially suppressed at large zenith angles due to Earth absorption~\cite{Gandhi:1995tf}. Here we account for this absorption via the charged and neutral current interaction of $100$~TeV neutrinos as they traverse the Earth. We use the preliminary Earth density model provided in Ref.~\cite{Gandhi:1995tf}. The conventional atmospheric $\nu_\mu$ is enhanced towards the horizon due to an increased atmospheric column depth. We use the parametrization of Ref.~\cite{Chirkin:2004ic} to account for this. 

As in the case of the HESE event sampling we generate Galactic signal events by a rescaling of the isotropic signal via the fraction $f_{\rm iso}$ and the corresponding Galactic emission accounting for an event fraction $1-f_{\rm Gal}$ sampled from the corresponding Galactic emission template. Note, that in this study we do not have the actual IceCube event distribution for three years of observation and we can only use these simulations to study the sensitivity.

\section{Likelihood Function}\label{app3}

For the HESE 3yr search the expectation values for tracks ($\diamond$) and cascades ($\circ$) are defined in the following way:
\begin{equation}
\mu^{\rm sig}_{\diamond/\circ}(\Omega) = \frac{1-f_{\rm iso}}{\mathcal{N}_{\rm Gal}}x^{\rm astro}_{\diamond/\circ}\mu^{\rm astro}(\Omega){w_{\rm signal}(\Omega)}\,,
\end{equation}
\begin{equation}
\mu_{\diamond/\circ}^{\rm bgr}(\Omega) = f_{\rm iso}x^{\rm astro}_{\diamond/\circ}\mu^{\rm astro}(\Omega)+ x^{\rm atmo\,\,\nu}_{\diamond/\circ}\mu^{\rm atmo\,\,\nu}(\Omega)+ x^{\rm atmo\,\,\mu}_{\diamond/\circ}\mu^{\rm atmo\,\,\mu}(\Omega)\,,
\end{equation}
where $x_{\diamond/\circ}$ corresponds to the fraction of tracks and cascades for the three separate components and $\mu$ are the expectation values extracted from Fig.5 and Tab.~IV in Ref.~\cite{Aartsen:2014gkd}. The factor $\mathcal{N}_{\rm Gal}$ renormalizes the Galactic flux such that the total number of expected events is conserved,
\begin{equation}\label{eq:norm}
\mathcal{N}_{\rm Gal} =\int{\rm d}\Omega\mu^{\rm astro}_{\diamond/\circ}(\Omega){w_{\rm signal}(\Omega)}\left(\int{\rm d}\Omega\mu^{\rm astro}_{\diamond/\circ}(\Omega)\right)^{-1} \,.
\end{equation}
Numerically, we find that $\mathcal{N}_{\rm Gal}$ is in the range 1.02 (Galactic diffuse) to 1.09 (DM decay) except for the {\it Fermi Bubble} template where $\mathcal{N}_{\rm Gal}\simeq 1.24$.

The three year expectation values of the isotropic astrophysical contribution ($\mu^{\rm astro}_{\diamond/\circ}$), the atmospheric neutrino ($\mu^{\rm atmo\,\,\nu}_{\diamond/\circ}$) and the atmospheric muon ($\mu^{\rm atmo\,\,\mu}_{\diamond/\circ}$) are known from Fig.~3 and Tab.~IV of Ref.~\cite{Aartsen:2014gkd}. Now, the angular uncertainty of the events can be accounted for by replacing the expectation values for signal and background by the weighted values
\begin{equation}
\tilde\mu_i = \int{\rm d}\Omega \,{\rm pdf}_i(\Omega)\,\mu(\Omega) \,.
\end{equation}
The fraction $x_i$ indicates the contributions into tracks and cascades.
For normalized pdf's this implies that $\tilde\mu_{\rm tot} = \mu_{\rm tot}$ and the exponential terms in the likelihood drop out of the likelihood ratio in Eq.~(\ref{eq:LH}).

The case of the classical $\nu_\mu$ search follows the same prescription. However, due to the good angular resolution of tracks we choose an initial binning of the skymaps such that each pixel has an approximate diameter of $1^\circ$, corresponding to about twice the angular resolution of IceCube. In this case we do not need to resample the angular resolution from the true position. Also, we do not have to account for different event topologies in the simulation, {\it i.e.}~$x_\diamond =1$. The background and signal expectation are both reduced towards high zenith angles due to Earth absorption via charged and neutral current interactions. The conventional atmospheric $\nu_\mu$ flux is enhanced towards the horizon due to the scaling with the atmospheric depth. We already commented about this approximation in the previous section. The total expected astrophysical $\nu_\mu$ event rate is normalized to 10 events per year and the total expected conventional atmospheric $\nu_\mu$ event rate is also normalized to 10 events per year. This approximates the expected integrated event rate above muon energies of $100$~TeV~\cite{Karle_Arlington}. Due to the limited field of view with $\theta_{\rm zen}>85^\circ$ the normalization factor (\ref{eq:norm}) is smaller than unity: $0.79$ (PWN), $0.81$ (DM decay), and $0.86$ (SNR \& Galactic diffuse).

\section{Comparison of Galactic and Extragalactic Emission}\label{app4}

Any mechanism that provides a strong anisotropic Galactic neutrino emission is also expected to contribute via an isotropic emission from similar, but distant galaxies. If $\widehat{Q}_\nu(E)$ is the total spectral emission rate of our Galaxy then the contribution of extragalactic source can be written as the red-shift integral~\cite{Waxman:1998yy}
\begin{equation}
\phi^{\rm iso}_\nu(E_\nu) = \frac{c}{4\pi}\int \frac{{\rm d}z}{H(z)}\rho(z)\,\widehat{Q}_\nu[(1+z)E_\nu]\,,
\end{equation}
where $H(z)/H_0 = \sqrt{\Omega_\Lambda + \Omega_{\rm m}(1+z)^3}$ is the red-shift dependent Hubble constant in the matter-dominated era with $H_0 = 67.27\pm 0.66~\mbox{km}\,\mbox{s}^{-1}\,\mbox{Mpc}^{-1}$ and $\Omega_{\rm m} = 1 - \Omega_\Lambda = 0.3156\pm 0.0091$~\cite{Ade:2015xua}. The distribution of sources is taken to be its averaged value $\rho(0) \equiv \rho_0 \simeq (10^{-3}-10^{-2})\,{\rm Mpc}^{-3}$. Comparing to the average diffuse emission from the Galaxy can be expressed via Eq.~(\ref{eq:source-generated-neutrino}) and approximating the neutrino flux as power-law with index $\Gamma$, we have the relation
\beqa\label{eq:ratio}
\frac{\phi^{\rm iso}_\nu}{\phi^{\rm Gal}_\nu} = \int {{\rm d}z}\frac{c\,\rho(z)\,(1+z)^{-\Gamma}}{\widehat{J}H(z)} = \frac{c\,\rho_0\,\xi_z}{H_0\,\widehat{J}} \simeq 2\times10^{-3}\xi_z\left(\frac{\rho_0}{10^{-3}\,{\rm Mpc}^{-3}}\right)\left(\frac{(20\,{\rm kpc})^{-2}}{\widehat{J}}\right)\,,
\eeqa
with source evolution factor $\xi_z=\mathcal{O}(1)$ ($\xi_z \simeq 0.5$ for no evolution and $\Gamma\simeq 2$).

Note that the parameter $\widehat{J}$ is defined as $\langle({\bf r}-{\bf r}_\odot)^{-2}\rangle/(4\pi)$, averaged over the source distribution in the Galaxy. The SNR and PWN distributions in Eq.~(\ref{eq:Case}) are essentially 2-dimensional and have a maximal contribution at $r_{\rm max} = \alpha R_\odot/\beta $. For $r_{\rm max}<R_\odot$ the size of the $J$-factor is then expected to be of the order of $4\pi\widehat{J} \simeq \mathcal{O}[(R_\odot^2-r^2_{\rm max})^{-1}]$. However, this does not take into account the radial width of the distribution. Numerically, we find $4\pi\widehat{J} \simeq (4.6\,{\rm kpc})^{-2}$ for both, SNR and PWN distributions. The local Galactic emission is then typically much stronger than the corresponding extragalactic contribution.

On the other hand the dark matter distribution in Eq.~(\ref{eq:Einasto}) is 3-dimensional with a characteristic extension $r_s\gg R_\odot$. The dark matter profile is extended to a wider range compared to the SNR or PWN sources, and therefore has a larger value of $\widehat{J}^{-1/2}$. The characteristic length scale of the $J$-factor is of the order of $4\pi\widehat{J} \simeq \mathcal{O}(r^{-2}_s/3)$. This agrees well with the numerical value $4\pi\widehat{J} \simeq (34\,{\rm kpc})^{-2}$. For a flatter dark matter profile, {\it i.e.}~larger $r_s$, we also anticipate a larger value of $\widehat{J}^{-1/2}$ or a smaller value of $\widehat{J}$ and more extragalactic contributions. In the case of dark matter decay the local density in Eq.~(\ref{eq:ratio}) is given by $\rho_0\simeq\Omega_{\rm cdm}\rho_{\rm cr}/M^{\rm total}_{\rm DM} \simeq 10^{-2}~{\rm Mpc}^{-3}$~\cite{Ade:2015xua}. Therefore, up to uncertainties of the Galaxy density and the dark matter distribution, the isotropic emission from extragalactic dark matter is expected to be at a similar level as the Galactic emission~\cite{Esmaili:2013gha,Murase:2015gea}.

\end{appendix}


\begin{thebibliography}{100}

\bibitem{Hunter:1997}
S.~Hunter et~al., {\it {EGRET Observations of the Diffuse Gamma-Ray Emission
  from the Galactic Plane}},  {\em APJ} {\bf 481} (May, 1997) 205--240.

\bibitem{FermiLAT:2012aa}
{\bf Fermi-LAT} Collaboration, M.~Ackermann et~al., {\it {Fermi-LAT
  Observations of the Diffuse Gamma-Ray Emission: Implications for Cosmic Rays
  and the Interstellar Medium}},  {\em Astrophys.J.} {\bf 750} (2012) 3,
  [\href{http://arxiv.org/abs/1202.4039}{{\tt arXiv:1202.4039}}].

\bibitem{Ackermann:2013wqa}
{\bf Fermi-LAT} Collaboration, M.~Ackermann et~al., {\it {Detection of the
  Characteristic Pion-Decay Signature in Supernova Remnants}},  {\em Science}
  {\bf 339} (2013) 807, [\href{http://arxiv.org/abs/1302.3307}{{\tt
  arXiv:1302.3307}}].

\bibitem{Aartsen:2013jdh}
{\bf IceCube} Collaboration, M.~Aartsen et~al., {\it {Evidence for High-Energy
  Extraterrestrial Neutrinos at the IceCube Detector}},  {\em Science} {\bf
  342} (2013) 1242856, [\href{http://arxiv.org/abs/1311.5238}{{\tt
  arXiv:1311.5238}}].

\bibitem{Aartsen:2014gkd}
{\bf IceCube} Collaboration, M.~Aartsen et~al., {\it {Observation of
  High-Energy Astrophysical Neutrinos in Three Years of IceCube Data}},  {\em
  Phys.Rev.Lett.} {\bf 113} (2014) 101101,
  [\href{http://arxiv.org/abs/1405.5303}{{\tt arXiv:1405.5303}}].

\bibitem{Anchordoqui:2013dnh}
L.~A. Anchordoqui, V.~Barger, I.~Cholis, H.~Goldberg, D.~Hooper, et~al., {\it
  {Cosmic Neutrino Pevatrons: A Brand New Pathway to Astronomy, Astrophysics,
  and Particle Physics}},  {\em JHEAp} {\bf 1-2} (2014) 1--30,
  [\href{http://arxiv.org/abs/1312.6587}{{\tt arXiv:1312.6587}}].

\bibitem{Fox:2013oza}
D.~Fox, K.~Kashiyama, and P.~Meszaros, {\it {Sub-PeV Neutrinos from TeV
  Unidentified Sources in the Galaxy}},  {\em Astrophys.J.} {\bf 774} (2013)
  74, [\href{http://arxiv.org/abs/1305.6606}{{\tt arXiv:1305.6606}}].

\bibitem{Gonzalez-Garcia:2013iha}
M.~Gonzalez-Garcia, F.~Halzen, and V.~Niro, {\it {Reevaluation of the Prospect
  of Observing Neutrinos from Galactic Sources in the Light of Recent Results
  in Gamma Ray and Neutrino Astronomy}},  {\em Astropart.Phys.} {\bf 57-58}
  (2014) 39--48, [\href{http://arxiv.org/abs/1310.7194}{{\tt
  arXiv:1310.7194}}].

\bibitem{Anchordoqui:2014rca}
L.~A. Anchordoqui, H.~Goldberg, T.~C. Paul, L.~H.~M. da~Silva, and B.~J. Vlcek,
  {\it {Estimating the contribution of Galactic sources to the diffuse neutrino
  flux}},  {\em Phys.Rev.} {\bf D90} (2014), no.~12 123010,
  [\href{http://arxiv.org/abs/1410.0348}{{\tt arXiv:1410.0348}}].

\bibitem{Padovani:2014bha}
P.~Padovani and E.~Resconi, {\it {Are both BL Lacs and pulsar wind nebulae the
  astrophysical counterparts of IceCube neutrino events?}},  {\em
  Mon.Not.Roy.Astron.Soc.} {\bf 443} (2014), no.~1 474--484,
  [\href{http://arxiv.org/abs/1406.0376}{{\tt arXiv:1406.0376}}].

\bibitem{Razzaque:2013uoa}
S.~Razzaque, {\it {The Galactic Center Origin of a Subset of IceCube Neutrino
  Events}},  {\em Phys.Rev.} {\bf D88} (2013) 081302,
  [\href{http://arxiv.org/abs/1309.2756}{{\tt arXiv:1309.2756}}].

\bibitem{Ahlers:2013xia}
M.~Ahlers and K.~Murase, {\it {Probing the Galactic Origin of the IceCube
  Excess with Gamma-Rays}},  {\em Phys.Rev.} {\bf D90} (2014), no.~2 023010,
  [\href{http://arxiv.org/abs/1309.4077}{{\tt arXiv:1309.4077}}].

\bibitem{Lunardini:2013gva}
C.~Lunardini, S.~Razzaque, K.~T. Theodoseau, and L.~Yang, {\it {Neutrino Events
  at IceCube and the Fermi Bubbles}},  {\em Phys.Rev.} {\bf D90} (2014) 023016,
  [\href{http://arxiv.org/abs/1311.7188}{{\tt arXiv:1311.7188}}].

\bibitem{Lunardini:2015laa}
C.~Lunardini, S.~Razzaque, and L.~Yang, {\it {A multi-messenger study of the
  Fermi Bubbles: very high energy gamma rays and neutrinos}},
  \href{http://arxiv.org/abs/1504.07033}{{\tt arXiv:1504.07033}}.

\bibitem{Taylor:2014hya}
A.~M. Taylor, S.~Gabici, and F.~Aharonian, {\it {A Galactic Halo Origin of the
  Neutrinos Detected by IceCube}},  {\em Phys.Rev.} {\bf D89} (2014) 103003,
  [\href{http://arxiv.org/abs/1403.3206}{{\tt arXiv:1403.3206}}].

\bibitem{Bai:2014kba}
Y.~Bai, A.~Barger, V.~Barger, R.~Lu, A.~Peterson, et~al., {\it {Neutrino
  Lighthouse at Sagittarius A*}},  {\em Phys.Rev.} {\bf D90} (2014), no.~6
  063012, [\href{http://arxiv.org/abs/1407.2243}{{\tt arXiv:1407.2243}}].

\bibitem{Neronov:2013lza}
A.~Neronov, D.~Semikoz, and C.~Tchernin, {\it {PeV neutrinos from interactions
  of cosmic rays with the interstellar medium in the Galaxy}},  {\em Phys.Rev.}
  {\bf D89} (2014) 103002, [\href{http://arxiv.org/abs/1307.2158}{{\tt
  arXiv:1307.2158}}].

\bibitem{Guo:2014laa}
Y.~Guo, H.~Hu, and Z.~Tian, {\it {On the Contribution of "Fresh" Cosmic Rays to
  the Excesses of Secondary Particles}},
  \href{http://arxiv.org/abs/1412.8590}{{\tt arXiv:1412.8590}}.

\bibitem{Feldstein:2013kka}
B.~Feldstein, A.~Kusenko, S.~Matsumoto, and T.~T. Yanagida, {\it {Neutrinos at
  IceCube from Heavy Decaying Dark Matter}},  {\em Phys.Rev.} {\bf D88} (2013),
  no.~1 015004, [\href{http://arxiv.org/abs/1303.7320}{{\tt arXiv:1303.7320}}].

\bibitem{Esmaili:2013gha}
A.~Esmaili and P.~D. Serpico, {\it {Are IceCube neutrinos unveiling PeV-scale
  decaying dark matter?}},  {\em JCAP} {\bf 1311} (2013) 054,
  [\href{http://arxiv.org/abs/1308.1105}{{\tt arXiv:1308.1105}}].

\bibitem{Bai:2013nga}
Y.~Bai, R.~Lu, and J.~Salvado, {\it {Geometric Compatibility of IceCube TeV-PeV
  Neutrino Excess and its Galactic Dark Matter Origin}},
  \href{http://arxiv.org/abs/1311.5864}{{\tt arXiv:1311.5864}}.

\bibitem{Bhattacharya:2014vwa}
A.~Bhattacharya, M.~H. Reno, and I.~Sarcevic, {\it {Reconciling neutrino flux
  from heavy dark matter decay and recent events at IceCube}},  {\em JHEP} {\bf
  1406} (2014) 110, [\href{http://arxiv.org/abs/1403.1862}{{\tt
  arXiv:1403.1862}}].

\bibitem{Esmaili:2014rma}
A.~Esmaili, S.~K. Kang, and P.~D. Serpico, {\it {IceCube events and decaying
  dark matter: hints and constraints}},  {\em JCAP} {\bf 1412} (2014), no.~12
  054, [\href{http://arxiv.org/abs/1410.5979}{{\tt arXiv:1410.5979}}].

\bibitem{Cherry:2014xra}
J.~F. Cherry, A.~Friedland, and I.~M. Shoemaker, {\it {Neutrino Portal Dark
  Matter: From Dwarf Galaxies to IceCube}},
  \href{http://arxiv.org/abs/1411.1071}{{\tt arXiv:1411.1071}}.

\bibitem{Murase:2015gea}
K.~Murase, R.~Laha, S.~Ando, and M.~Ahlers, {\it {Testing the Dark Matter
  Scenario for PeV Neutrinos Observed in IceCube}},
  \href{http://arxiv.org/abs/1503.04663}{{\tt arXiv:1503.04663}}.

\bibitem{Stecker:1978ah}
F.~Stecker, {\it {Diffuse Fluxes of Cosmic High-Energy Neutrinos}},  {\em
  Astrophys.J.} {\bf 228} (1979) 919--927.

\bibitem{Domokos:1991tt}
G.~Domokos, B.~Elliott, and S.~Kovesi-Domokos, {\it {Cosmic neutrino production
  in the Milky Way}},  {\em J.Phys.} {\bf G19} (1993) 899--912.

\bibitem{Berezinsky:1992wr}
V.~Berezinsky, T.~Gaisser, F.~Halzen, and T.~Stanev, {\it {Diffuse radiation
  from cosmic ray interactions in the galaxy}},  {\em Astropart.Phys.} {\bf 1}
  (1993) 281--288.

\bibitem{Bertsch:1993}
D.~L. {Bertsch}, T.~M. {Dame}, C.~E. {Fichtel}, S.~D. {Hunter}, P.~{Sreekumar},
  J.~G. {Stacy}, and P.~{Thaddeus}, {\it {Diffuse Gamma-Ray Emission in the
  Galactic Plane from Cosmic-Ray, Matter, and Photon Interactions}},  {\em APJ}
  {\bf 416} (Oct., 1993) 587.

\bibitem{Ingelman:1996md}
G.~Ingelman and M.~Thunman, {\it {Particle production in the interstellar
  medium}},  \href{http://arxiv.org/abs/hep-ph/9604286}{{\tt hep-ph/9604286}}.

\bibitem{Evoli:2007iy}
C.~Evoli, D.~Grasso, and L.~Maccione, {\it {Diffuse Neutrino and Gamma-ray
  Emissions of the Galaxy above the TeV}},  {\em JCAP} {\bf 0706} (2007) 003,
  [\href{http://arxiv.org/abs/astro-ph/0701856}{{\tt astro-ph/0701856}}].

\bibitem{Joshi:2013aua}
J.~C. Joshi, W.~Winter, and N.~Gupta, {\it {How Many of the Observed Neutrino
  Events Can Be Described by Cosmic Ray Interactions in the Milky Way?}},
  \href{http://arxiv.org/abs/1310.5123}{{\tt arXiv:1310.5123}}.

\bibitem{Kachelriess:2014oma}
M.~Kachelrie{\ss} and S.~Ostapchenko, {\it {Neutrino yield from Galactic cosmic
  rays}},  {\em Phys.Rev.} {\bf D90} (2014), no.~8 083002,
  [\href{http://arxiv.org/abs/1405.3797}{{\tt arXiv:1405.3797}}].

\bibitem{Strong:1998pw}
A.~Strong and I.~Moskalenko, {\it {Propagation of cosmic-ray nucleons in the
  galaxy}},  {\em Astrophys.J.} {\bf 509} (1998) 212--228,
  [\href{http://arxiv.org/abs/astro-ph/9807150}{{\tt astro-ph/9807150}}].

\bibitem{CREAM}
Y.~Yoon, H.~Ahn, P.~Allison, M.~Bagliesi, J.~Beatty, et~al., {\it {Cosmic-Ray
  Proton and Helium Spectra from the First CREAM Flight}},  {\em Astrophys.J.}
  {\bf 728} (2011) 122, [\href{http://arxiv.org/abs/1102.2575}{{\tt
  arXiv:1102.2575}}].

\bibitem{KASCADE}
{\bf KASCADE} Collaboration, T.~Antoni et~al., {\it {KASCADE measurements of
  energy spectra for elemental groups of cosmic rays: Results and open
  problems}},  {\em Astropart.Phys.} {\bf 24} (2005) 1--25,
  [\href{http://arxiv.org/abs/astro-ph/0505413}{{\tt astro-ph/0505413}}].

\bibitem{KASCADE-Grande}
W.~Apel, J.~Arteaga-Vel{\'a}zquez, K.~Bekk, M.~Bertaina, J.~Bl{\"u}mer, et~al.,
  {\it {KASCADE-Grande measurements of energy spectra for elemental groups of
  cosmic rays}},  {\em Astropart.Phys.} {\bf 47} (2013) 54--66,
  [\href{http://arxiv.org/abs/1306.6283}{{\tt arXiv:1306.6283}}].

\bibitem{Gaisser:2013bla}
T.~K. Gaisser, T.~Stanev, and S.~Tilav, {\it {Cosmic Ray Energy Spectrum from
  Measurements of Air Showers}},  {\em Front.Phys.China} {\bf 8} (2013)
  748--758, [\href{http://arxiv.org/abs/1303.3565}{{\tt arXiv:1303.3565}}].

\bibitem{Hillas:2005cs}
A.~Hillas, {\it {Can diffusive shock acceleration in supernova remnants account
  for high-energy galactic cosmic rays?}},  {\em J.Phys.} {\bf G31} (2005)
  R95--R131.

\bibitem{Blum:2014ewa}
K.~Blum, A.~Hook, and K.~Murase, {\it {High energy neutrino telescopes as a
  probe of the neutrino mass mechanism}},
  \href{http://arxiv.org/abs/1408.3799}{{\tt arXiv:1408.3799}}.

\bibitem{Fong:2014bsa}
C.~S. Fong, H.~Minakata, B.~Panes, and R.~Z. Funchal, {\it {Possible
  Interpretations of IceCube High-Energy Neutrino Events}},  {\em JHEP} {\bf
  1502} (2015) 189, [\href{http://arxiv.org/abs/1411.5318}{{\tt
  arXiv:1411.5318}}].

\bibitem{Chen:2014gxa}
C.-Y. Chen, P.~S.~B. Dev, and A.~Soni, {\it {A Possible Two-component Flux for
  the High Energy Neutrino Events at IceCube}},
  \href{http://arxiv.org/abs/1411.5658}{{\tt arXiv:1411.5658}}.

\bibitem{Palomares-Ruiz:2015mka}
S.~Palomares-Ruiz, A.~C. Vincent, and O.~Mena, {\it {Spectral analysis of the
  high-energy IceCube neutrinos}},  \href{http://arxiv.org/abs/1502.02649}{{\tt
  arXiv:1502.02649}}.

\bibitem{Kamada:2015era}
A.~Kamada and H.-B. Yu, {\it {Coherent Propagation of PeV Neutrinos and the Dip
  in the Neutrino Spectrum at IceCube}},
  \href{http://arxiv.org/abs/1504.00711}{{\tt arXiv:1504.00711}}.

\bibitem{Aartsen:2014qna}
{\bf IceCube} Collaboration, M.~Aartsen et~al., {\it {Development of a General
  Analysis and Unfolding Scheme and its Application to Measure the Energy
  Spectrum of Atmospheric Neutrinos with IceCube}},  {\em Eur.Phys.J.} {\bf
  C75} (2015), no.~3 116, [\href{http://arxiv.org/abs/1409.4535}{{\tt
  arXiv:1409.4535}}].

\bibitem{Gondolo:1995fq}
P.~Gondolo, G.~Ingelman, and M.~Thunman, {\it {Charm production and high-energy
  atmospheric muon and neutrino fluxes}},  {\em Astropart.Phys.} {\bf 5} (1996)
  309--332, [\href{http://arxiv.org/abs/hep-ph/9505417}{{\tt hep-ph/9505417}}].

\bibitem{Honda:2006qj}
M.~Honda, T.~Kajita, K.~Kasahara, S.~Midorikawa, and T.~Sanuki, {\it
  {Calculation of atmospheric neutrino flux using the interaction model
  calibrated with atmospheric muon data}},  {\em Phys.Rev.} {\bf D75} (2007)
  043006, [\href{http://arxiv.org/abs/astro-ph/0611418}{{\tt
  astro-ph/0611418}}].

\bibitem{Enberg:2008te}
R.~Enberg, M.~H. Reno, and I.~Sarcevic, {\it {Prompt neutrino fluxes from
  atmospheric charm}},  {\em Phys.Rev.} {\bf D78} (2008) 043005,
  [\href{http://arxiv.org/abs/0806.0418}{{\tt arXiv:0806.0418}}].

\bibitem{Gaisser:2014bja}
T.~K. Gaisser, K.~Jero, A.~Karle, and J.~van Santen, {\it {Generalized
  self-veto probability for atmospheric neutrinos}},  {\em Phys.Rev.} {\bf D90}
  (2014), no.~2 023009, [\href{http://arxiv.org/abs/1405.0525}{{\tt
  arXiv:1405.0525}}].

\bibitem{Casse:2001be}
F.~Casse, M.~Lemoine, and G.~Pelletier, {\it {Transport of cosmic rays in
  chaotic magnetic fields}},  {\em Phys. Rev.} {\bf D65} (2002) 023002,
  [\href{http://arxiv.org/abs/astro-ph/0109223}{{\tt astro-ph/0109223}}].

\bibitem{Effenberger:2012jc}
F.~Effenberger, H.~Fichtner, K.~Scherer, and I.~Busching, {\it {Anisotropic
  diffusion of galactic cosmic ray protons and their steady-state azimuthal
  distribution}},  {\em Astron. Astrophys.} {\bf 547} (2012) A120,
  [\href{http://arxiv.org/abs/1210.1423}{{\tt arXiv:1210.1423}}].

\bibitem{Evoli:2008dv}
C.~Evoli, D.~Gaggero, D.~Grasso, and L.~Maccione, {\it {Cosmic-Ray Nuclei,
  Antiprotons and Gamma-rays in the Galaxy: a New Diffusion Model}},  {\em
  JCAP} {\bf 0810} (2008) 018, [\href{http://arxiv.org/abs/0807.4730}{{\tt
  arXiv:0807.4730}}].

\bibitem{Gaggero:2013rya}
D.~Gaggero, L.~Maccione, G.~Di~Bernardo, C.~Evoli, and D.~Grasso, {\it
  {Three-Dimensional Model of Cosmic-Ray Lepton Propagation Reproduces Data
  from the Alpha Magnetic Spectrometer on the International Space Station}},
  {\em Phys. Rev. Lett.} {\bf 111} (2013) 021102,
  [\href{http://arxiv.org/abs/1304.6718}{{\tt arXiv:1304.6718}}].

\bibitem{Kissmann:2014sia}
R.~Kissmann, {\it {PICARD: A novel code for the Galactic Cosmic Ray propagation
  problem}},  {\em Astropart. Phys.} {\bf 55} (2014) 37--50,
  [\href{http://arxiv.org/abs/1401.4035}{{\tt arXiv:1401.4035}}].

\bibitem{Werner:2014sya}
M.~Werner, R.~Kissmann, A.~W. Strong, and O.~Reimer, {\it {Spiral Arms as
  Cosmic Ray Source Distributions}},  {\em Astropart. Phys.} {\bf 64} (2014)
  18--33, [\href{http://arxiv.org/abs/1410.5266}{{\tt arXiv:1410.5266}}].

\bibitem{Gaggero:2015xza}
D.~Gaggero, D.~Grasso, A.~Marinelli, A.~Urbano, and M.~Valli, {\it {The
  gamma-ray and neutrino sky: a consistent picture of Fermi-LAT, H.E.S.S.,
  Milagro, and IceCube results}},  \href{http://arxiv.org/abs/1504.00227}{{\tt
  arXiv:1504.00227}}.

\bibitem{Moskalenko:2005ng}
I.~V. Moskalenko, T.~A. Porter, and A.~W. Strong, {\it {Attenuation of vhe
  gamma rays by the milky way interstellar radiation field}},  {\em
  Astrophys.J.} {\bf 640} (2006) L155--L158,
  [\href{http://arxiv.org/abs/astro-ph/0511149}{{\tt astro-ph/0511149}}].

\bibitem{Aartsen:2014cva}
{\bf IceCube} Collaboration, M.~Aartsen et~al., {\it {Searches for Extended and
  Point-like Neutrino Sources with Four Years of IceCube Data}},
  \href{http://arxiv.org/abs/1406.6757}{{\tt arXiv:1406.6757}}.

\bibitem{Blasi:2011fi}
P.~Blasi and E.~Amato, {\it {Diffusive propagation of cosmic rays from
  supernova remnants in the Galaxy. I: spectrum and chemical composition}},
  {\em JCAP} {\bf 1201} (2012) 010, [\href{http://arxiv.org/abs/1105.4521}{{\tt
  arXiv:1105.4521}}].

\bibitem{Berezhko:2000vy}
E.~G. Berezhko and H.~J. V{\"o}lk, {\it {Galactic gamma-ray background
  radiation from supernova remnants}},  {\em Astrophys. J.} {\bf 540} (2000)
  923--929, [\href{http://arxiv.org/abs/astro-ph/0004353}{{\tt
  astro-ph/0004353}}].

\bibitem{Berezhko:2004qb}
E.~G. Berezhko and H.~J. Volk, {\it {The Contribution of different supernova
  populations to the Galactic gamma-ray background}},  {\em Astrophys. J.} {\bf
  611} (2004) 12--19, [\href{http://arxiv.org/abs/astro-ph/0404307}{{\tt
  astro-ph/0404307}}].

\bibitem{Kelner:2006tc}
S.~Kelner, F.~A. Aharonian, and V.~Bugayov, {\it {Energy spectra of gamma-rays,
  electrons and neutrinos produced at proton-proton interactions in the very
  high energy regime}},  {\em Phys.Rev.} {\bf D74} (2006) 034018,
  [\href{http://arxiv.org/abs/astro-ph/0606058}{{\tt astro-ph/0606058}}].

\bibitem{Block:2011vz}
M.~M. Block and F.~Halzen, {\it {Experimental Confirmation that the Proton is
  Asymptotically a Black Disk}},  {\em Phys.Rev.Lett.} {\bf 107} (2011) 212002,
  [\href{http://arxiv.org/abs/1109.2041}{{\tt arXiv:1109.2041}}].

\bibitem{Blondin1998}
J.~M. {Blondin}, E.~B. {Wright}, K.~J. {Borkowski}, and S.~P. {Reynolds}, {\it
  {Transition to the Radiative Phase in Supernova Remnants}},  {\em
  Astrophys.J.} {\bf 500} (June, 1998) 342--354.

\bibitem{Case:1998qg}
G.~L. Case and D.~Bhattacharya, {\it {A new sigma-d relation and its
  application to the galactic supernova remnant distribution}},  {\em
  Astrophys.J.} {\bf 504} (1998) 761,
  [\href{http://arxiv.org/abs/astro-ph/9807162}{{\tt astro-ph/9807162}}].

\bibitem{Lorimer:2006qs}
D.~Lorimer, A.~Faulkner, A.~Lyne, R.~Manchester, M.~Kramer, et~al., {\it {The
  Parkes multibeam pulsar survey: VI. Discovery and timing of 142 pulsars and a
  Galactic population analysis}},  {\em Mon.Not.Roy.Astron.Soc.} {\bf 372}
  (2006) 777--800, [\href{http://arxiv.org/abs/astro-ph/0607640}{{\tt
  astro-ph/0607640}}].

\bibitem{Abdo:2007ad}
A.~Abdo, B.~T. Allen, D.~Berley, S.~Casanova, C.~Chen, et~al., {\it {TeV
  Gamma-Ray Sources from a Survey of the Galactic Plane with Milagro}},  {\em
  Astrophys.J.} {\bf 664} (2007) L91--L94,
  [\href{http://arxiv.org/abs/0705.0707}{{\tt arXiv:0705.0707}}].

\bibitem{Aliu:2014xra}
E.~Aliu, S.~Archambault, T.~Aune, B.~Behera, M.~Beilicke, et~al., {\it
  {Investigating the TeV Morphology of MGRO J1908+06 with VERITAS}},  {\em
  Astrophys.J.} {\bf 787} (2014) 166,
  [\href{http://arxiv.org/abs/1404.7185}{{\tt arXiv:1404.7185}}].

\bibitem{Ioka:2009dh}
K.~Ioka and P.~Meszaros, {\it {Hypernova and Gamma-Ray Burst Remnants as TeV
  Unidentified Sources}},  {\em Astrophys.J.} {\bf 709} (2010) 1337--1342,
  [\href{http://arxiv.org/abs/0901.0744}{{\tt arXiv:0901.0744}}].

\bibitem{Su:2010qj}
M.~Su, T.~R. Slatyer, and D.~P. Finkbeiner, {\it {Giant Gamma-ray Bubbles from
  Fermi-LAT: AGN Activity or Bipolar Galactic Wind?}},  {\em Astrophys.J.} {\bf
  724} (2010) 1044--1082, [\href{http://arxiv.org/abs/1005.5480}{{\tt
  arXiv:1005.5480}}].

\bibitem{Crocker:2010dg}
R.~M. Crocker and F.~Aharonian, {\it {The Fermi Bubbles: Giant,
  Multi-Billion-Year-Old Reservoirs of Galactic Center Cosmic Rays}},  {\em
  Phys.Rev.Lett.} {\bf 106} (2011) 101102,
  [\href{http://arxiv.org/abs/1008.2658}{{\tt arXiv:1008.2658}}].

\bibitem{Mertsch:2011es}
P.~Mertsch and S.~Sarkar, {\it {Fermi gamma-ray `bubbles' from stochastic
  acceleration of electrons}},  {\em Phys.Rev.Lett.} {\bf 107} (2011) 091101,
  [\href{http://arxiv.org/abs/1104.3585}{{\tt arXiv:1104.3585}}].

\bibitem{Yang:2013kca}
H.~Y.~K. Yang, M.~Ruszkowski, and E.~Zweibel, {\it {The Fermi Bubbles:
  Gamma-ray, Microwave, and Polarization Signatures of Leptonic AGN Jets}},
  \href{http://arxiv.org/abs/1307.3551}{{\tt arXiv:1307.3551}}.

\bibitem{Fermi-LAT:2014sfa}
{\bf Fermi-LAT} Collaboration, M.~Ackermann et~al., {\it {The Spectrum and
  Morphology of the Fermi Bubbles}},  {\em Astrophys.J.} (2014)
  [\href{http://arxiv.org/abs/1407.7905}{{\tt arXiv:1407.7905}}].

\bibitem{Doro:2012xx}
{\bf CTA} Collaboration, M.~Doro et~al., {\it {Dark Matter and Fundamental
  Physics with the Cherenkov Telescope Array}},  {\em Astropart.Phys.} {\bf 43}
  (2013) 189--214, [\href{http://arxiv.org/abs/1208.5356}{{\tt
  arXiv:1208.5356}}].

\bibitem{Pierre:2014tra}
M.~Pierre, J.~M. Siegal-Gaskins, and P.~Scott, {\it {Sensitivity of CTA to dark
  matter signals from the Galactic Center}},  {\em JCAP} {\bf 1406} (2014),
  no.~10 024, [\href{http://arxiv.org/abs/1401.7330}{{\tt arXiv:1401.7330}}].

\bibitem{Abeysekara:2013tza}
A.~Abeysekara, R.~Alfaro, C.~Alvarez, J.~{\'A}lvarez, R.~Arceo, et~al., {\it
  {Sensitivity of the High Altitude Water Cherenkov Detector to Sources of
  Multi-TeV Gamma Rays}},  {\em Astropart.Phys.} {\bf 50-52} (2013) 26--32,
  [\href{http://arxiv.org/abs/1306.5800}{{\tt arXiv:1306.5800}}].

\bibitem{Abeysekara:2014ffg}
{\bf HAWC} Collaboration, A.~Abeysekara et~al., {\it {Sensitivity of HAWC to
  high-mass dark matter annihilations}},  {\em Phys.Rev.} {\bf D90} (2014),
  no.~12 122002, [\href{http://arxiv.org/abs/1405.1730}{{\tt
  arXiv:1405.1730}}].

\bibitem{Ema:2013nda}
Y.~Ema, R.~Jinno, and T.~Moroi, {\it {Cosmic-Ray Neutrinos from the Decay of
  Long-Lived Particle and the Recent IceCube Result}},  {\em Phys.Lett.} {\bf
  B733} (2014) 120--125, [\href{http://arxiv.org/abs/1312.3501}{{\tt
  arXiv:1312.3501}}].

\bibitem{Ema:2014ufa}
Y.~Ema, R.~Jinno, and T.~Moroi, {\it {Cosmological Implications of High-Energy
  Neutrino Emission from the Decay of Long-Lived Particle}},  {\em JHEP} {\bf
  1410} (2014) 150, [\href{http://arxiv.org/abs/1408.1745}{{\tt
  arXiv:1408.1745}}].

\bibitem{Graham:2006ae}
A.~W. Graham, D.~Merritt, B.~Moore, J.~Diemand, and B.~Terzic, {\it {Empirical
  Models for Dark Matter Halos. II. Inner profile slopes, dynamical profiles,
  and $\rho/\sigma^3$}},  {\em Astron.J.} {\bf 132} (2006) 2701--2710,
  [\href{http://arxiv.org/abs/astro-ph/0608613}{{\tt astro-ph/0608613}}].

\bibitem{Agashe:2014kda}
{\bf Particle Data Group} Collaboration, K.~Olive et~al., {\it {Review of
  Particle Physics}},  {\em Chin.Phys.} {\bf C38} (2014) 090001.

\bibitem{Cowan:2010js}
G.~Cowan, K.~Cranmer, E.~Gross, and O.~Vitells, {\it {Asymptotic formulae for
  likelihood-based tests of new physics}},  {\em Eur.Phys.J.} {\bf C71} (2011)
  1554, [\href{http://arxiv.org/abs/1007.1727}{{\tt arXiv:1007.1727}}].

\bibitem{Braun:2008bg}
J.~Braun, J.~Dumm, F.~De~Palma, C.~Finley, A.~Karle, et~al., {\it {Methods for
  point source analysis in high energy neutrino telescopes}},  {\em
  Astropart.Phys.} {\bf 29} (2008) 299--305,
  [\href{http://arxiv.org/abs/0801.1604}{{\tt arXiv:0801.1604}}].

\bibitem{Karle_Arlington}
A.~Karle,
  ``{\href{https://docushare.icecube.wisc.edu/dsweb/Get/Document-69617/4-Karle_NGIC_Arlington.pdf}{A
  next-generation IceCube}}.'' {Talk presented at the workshop {\it Neutrinos
  Beyond IceCube}, Arlington, Virginia}, April, 2014.

\bibitem{Prodanovic:2006bq}
T.~Prodanovic, B.~D. Fields, and J.~F. Beacom, {\it {Diffuse gamma rays from
  the galactic plane: probing the gev excess and identifying the TeV excess}},
  {\em Astropart.Phys.} {\bf 27} (2007) 10--20,
  [\href{http://arxiv.org/abs/astro-ph/0603618}{{\tt astro-ph/0603618}}].

\bibitem{Gupta:2013xfa}
N.~Gupta, {\it {Galactic PeV Neutrinos}},  {\em Astropart.Phys.} {\bf 48}
  (2013) 75--77, [\href{http://arxiv.org/abs/1305.4123}{{\tt
  arXiv:1305.4123}}].

\bibitem{Anchordoqui:2013qsi}
L.~A. Anchordoqui, H.~Goldberg, M.~H. Lynch, A.~V. Olinto, T.~C. Paul, et~al.,
  {\it {Pinning down the cosmic ray source mechanism with new IceCube data}},
  {\em Phys.Rev.} {\bf D89} (2014), no.~8 083003,
  [\href{http://arxiv.org/abs/1306.5021}{{\tt arXiv:1306.5021}}].

\bibitem{Learned:1994wg}
J.~G. Learned and S.~Pakvasa, {\it {Detecting tau-neutrino oscillations at PeV
  energies}},  {\em Astropart.Phys.} {\bf 3} (1995) 267--274,
  [\href{http://arxiv.org/abs/hep-ph/9405296}{{\tt hep-ph/9405296}}].

\bibitem{Athar:2000yw}
H.~Athar, M.~Jezabek, and O.~Yasuda, {\it {Effects of neutrino mixing on
  high-energy cosmic neutrino flux}},  {\em Phys.Rev.} {\bf D62} (2000) 103007,
  [\href{http://arxiv.org/abs/hep-ph/0005104}{{\tt hep-ph/0005104}}].

\bibitem{Aartsen:2015ivb}
{\bf IceCube} Collaboration, M.~Aartsen et~al., {\it {Flavor Ratio of
  Astrophysical Neutrinos above 35 TeV in IceCube}},  {\em Phys.Rev.Lett.} {\bf
  114} (2015), no.~17 171102, [\href{http://arxiv.org/abs/1502.03376}{{\tt
  arXiv:1502.03376}}].

\bibitem{Palladino:2015zua}
A.~Palladino, G.~Pagliaroli, F.~Villante, and F.~Vissani, {\it {Which is the
  flavor of cosmic neutrinos seen by IceCube?}},  {\em Phys.Rev.Lett.} {\bf
  114} (2015), no.~17 171101, [\href{http://arxiv.org/abs/1502.02923}{{\tt
  arXiv:1502.02923}}].

\bibitem{Beacom:2002vi}
J.~F. Beacom, N.~F. Bell, D.~Hooper, S.~Pakvasa, and T.~J. Weiler, {\it {Decay
  of high-energy astrophysical neutrinos}},  {\em Phys.Rev.Lett.} {\bf 90}
  (2003) 181301, [\href{http://arxiv.org/abs/hep-ph/0211305}{{\tt
  hep-ph/0211305}}].

\bibitem{Majumdar:2006px}
D.~Majumdar and A.~Ghosal, {\it {Probing deviations from tri-bimaximal mixing
  through ultra high energy neutrino signals}},  {\em Phys.Rev.} {\bf D75}
  (2007) 113004, [\href{http://arxiv.org/abs/hep-ph/0608334}{{\tt
  hep-ph/0608334}}].

\bibitem{Baerwald:2012kc}
P.~Baerwald, M.~Bustamante, and W.~Winter, {\it {Neutrino Decays over
  Cosmological Distances and the Implications for Neutrino Telescopes}},  {\em
  JCAP} {\bf 1210} (2012) 020, [\href{http://arxiv.org/abs/1208.4600}{{\tt
  arXiv:1208.4600}}].

\bibitem{Laha:2013lka}
R.~Laha, J.~F. Beacom, B.~Dasgupta, S.~Horiuchi, and K.~Murase, {\it
  {Demystifying the PeV Cascades in IceCube: Less (Energy) is More (Events)}},
  {\em Phys.Rev.} {\bf D88} (2013) 043009,
  [\href{http://arxiv.org/abs/1306.2309}{{\tt arXiv:1306.2309}}].

\bibitem{Barger:2014iua}
V.~Barger, L.~Fu, J.~Learned, D.~Marfatia, S.~Pakvasa, et~al., {\it {Glashow
  resonance as a window into cosmic neutrino sources}},  {\em Phys.Rev.} {\bf
  D90} (2014) 121301, [\href{http://arxiv.org/abs/1407.3255}{{\tt
  arXiv:1407.3255}}].

\bibitem{Moskalenko:1997gh}
I.~Moskalenko and A.~Strong, {\it {Production and propagation of cosmic ray
  positrons and electrons}},  {\em Astrophys.J.} {\bf 493} (1998) 694--707,
  [\href{http://arxiv.org/abs/astro-ph/9710124}{{\tt astro-ph/9710124}}].

\bibitem{Strong:1998fr}
A.~W. Strong, I.~V. Moskalenko, and O.~Reimer, {\it {Diffuse continuum
  gamma-rays from the galaxy}},  {\em Astrophys.J.} {\bf 537} (2000) 763--784,
  [\href{http://arxiv.org/abs/astro-ph/9811296}{{\tt astro-ph/9811296}}].

\bibitem{Strong:2004de}
A.~W. Strong, I.~V. Moskalenko, and O.~Reimer, {\it {Diffuse galactic continuum
  gamma rays. A Model compatible with EGRET data and cosmic-ray measurements}},
   {\em Astrophys.J.} {\bf 613} (2004) 962--976,
  [\href{http://arxiv.org/abs/astro-ph/0406254}{{\tt astro-ph/0406254}}].

\bibitem{Yusifov:2004fr}
I.~Yusifov and I.~Kucuk, {\it {Revisiting the radial distribution of pulsars in
  the galaxy}},  {\em Astron.Astrophys.} {\bf 422} (2004) 545--553,
  [\href{http://arxiv.org/abs/astro-ph/0405559}{{\tt astro-ph/0405559}}].

\bibitem{Bronfman:2000tw}
L.~Bronfman, S.~Casassus, J.~May, and L.~Nyman, {\it {The radial distribution
  of ob star formation in the galaxy}},  {\em Astron.Astrophys.} {\bf 358}
  (2000) 521, [\href{http://arxiv.org/abs/astro-ph/0006104}{{\tt
  astro-ph/0006104}}].

\bibitem{GALPROPv54}
GALPROPv54-r2504. {\url{http://sourceforge.net/projects/galprop}}.

\bibitem{Huang:2007wk}
C.-Y. Huang and M.~Pohl, {\it {Production of Neutrinos and Secondary Electrons
  in Cosmic Sources}},  {\em Astropart.Phys.} {\bf 29} (2008) 282--289,
  [\href{http://arxiv.org/abs/0711.2528}{{\tt arXiv:0711.2528}}].

\bibitem{Dudas:2014bca}
E.~Dudas, Y.~Mambrini, and K.~A. Olive, {\it {Monochromatic neutrinos generated
  by dark matter and the seesaw mechanism}},  {\em Phys.Rev.} {\bf D91} (2015),
  no.~7 075001, [\href{http://arxiv.org/abs/1412.3459}{{\tt arXiv:1412.3459}}].

\bibitem{Gandhi:1995tf}
R.~Gandhi, C.~Quigg, M.~H. Reno, and I.~Sarcevic, {\it {Ultrahigh-energy
  neutrino interactions}},  {\em Astropart.Phys.} {\bf 5} (1996) 81--110,
  [\href{http://arxiv.org/abs/hep-ph/9512364}{{\tt hep-ph/9512364}}].

\bibitem{Chirkin:2004ic}
D.~Chirkin, {\it {Fluxes of atmospheric leptons at 600-GeV - 60-TeV}},
  \href{http://arxiv.org/abs/hep-ph/0407078}{{\tt hep-ph/0407078}}.

\bibitem{Waxman:1998yy}
E.~Waxman and J.~N. Bahcall, {\it {High-energy neutrinos from astrophysical
  sources: An Upper bound}},  {\em Phys.Rev.} {\bf D59} (1999) 023002,
  [\href{http://arxiv.org/abs/hep-ph/9807282}{{\tt hep-ph/9807282}}].

\bibitem{Ade:2015xua}
{\bf Planck} Collaboration, P.~Ade et~al., {\it {Planck 2015 results. XIII.
  Cosmological parameters}},  \href{http://arxiv.org/abs/1502.01589}{{\tt
  arXiv:1502.01589}}.

\end{thebibliography}
\providecommand{\href}[2]{#2}\begingroup\raggedright\endgroup

\end{document}